\documentclass[fleqn,usenatbib]{mnras}

\usepackage[T1]{fontenc}
\usepackage{ae,aecompl}
\usepackage{bm}

\usepackage{graphicx}	
\usepackage{amsmath}	
\usepackage{amssymb}	
\usepackage{mathrsfs}
\usepackage{hyperref}
\usepackage[dvipsnames]{xcolor}
\usepackage{newtxtext,newtxmath}
\usepackage{widetext}

\newcommand{\LCDM}{$\Lambda$CDM }
\newcommand{\MSun}{\, \rm M_{\odot}}
\newcommand{\K}{\, \rm K}

\DeclareMathOperator\arcsinh{arcsinh}

\def\citejap#1{\citeauthor{#1}\ \citeyear{#1}}

\title[Impact of $\Lambda$ on star formation history]{The impact of the cosmological constant on past and future star formation}

\author[D. Sorini, J. A. Peacock \& L. Lombriser]{%
Daniele Sorini$^{1,\, 2, \, 3}$\thanks{E-mail: daniele.sorini@durham.ac.uk},
John A. Peacock$^2$
and Lucas Lombriser$^3$
\\
$^{1}$Institute for Computational Cosmology, Department of Physics, Durham University, South Road, Durham DH1 3LE, United Kingdom\\
$^2$Institute for Astronomy, University of Edinburgh, Royal Observatory, Edinburgh, EH9 3HJ, United Kingdom\\
$^3$D\'epartement de Physique Th\'eorique, Universit\'e de Gen\`eve, 24 quai Ernest Ansermet, 1211 Gen\`eve 4, Switzerland
}

\pubyear{2024}

\begin{document}
\label{firstpage}
\pagerange{\pageref{firstpage}--\pageref{lastpage}}
\maketitle

\begin{abstract}
We present an extended analytic model for cosmic star formation, with the aim of investigating the impact of cosmological parameters on the star formation history within the \LCDM paradigm. Constructing an ensemble of flat \LCDM models where the cosmological constant varies between $\Lambda = 0$ and $10^5$ times the observed value, $\Lambda_{\rm obs}$,
we find that the fraction of cosmic baryons that are converted into stars over the entire history of the universe peaks at $\sim$\,27\% for $0.01 \lesssim \Lambda/\Lambda_{\rm obs} \lesssim 1$. We explain, from first principles, that the decline of this asymptotic star-formation efficiency for lower and higher values of $\Lambda$ is driven respectively by the astrophysics of star formation, and by the suppression of cosmic structure formation. However, the asymptotic efficiency declines slowly as $\Lambda$ increases, falling below 5\% only for $\Lambda>100 \, \Lambda_{\rm obs}$. Making the minimal assumption that the probability of generating observers is proportional to this efficiency, and following Weinberg in adopting a flat prior on $\Lambda$, the median posterior value of $\Lambda$ is $539 \, \Lambda_{\rm obs}$. Furthermore, the probability of observing $\Lambda\leq \Lambda_{\rm obs}$ is only $0.5\%$. Although this work has not considered recollapsing models with $\Lambda<0$, the indication is thus that $\Lambda_{\rm obs}$ appears to be unreasonably small compared to the predictions of the simplest multiverse ensemble. This poses a challenge for anthropic reasoning as a viable explanation for cosmic coincidences and the apparent fine-tuning of the universe: either the approach is invalid, or more parameters than $\Lambda$ alone must vary within the ensemble.
\end{abstract}




\section{Introduction}

Galaxies are the most obvious tracers of large-scale structure in the universe, and understanding how they emerge and evolve throughout cosmic history has been a longstanding goal of cosmology and astrophysics. The \LCDM paradigm approaches this via the gravitational collapse of dark matter haloes from an initial quasi-homogeneous universe. This process is well understood, thanks to early analytical models of halo assembly \citep{LaceyCole1993}, which were subsequently validated numerically with full N-body simulations \citep{Springel_2005, Klypin_2011, Angulo_2012, Fosalba_2015}. However, many of the detailed baryonic processes that govern the build up of galaxies, such as gas accretion, star formation and outflows, are not yet fully understood. A successful theory of galaxy formation needs to account for these processes in order to predict the main observed properties of galaxies. In particular,
reproducing the observed efficiency of star formation, both locally within individual galaxies and globally over a cosmologically representative volume, constitutes a crucial test for any model of galaxy formation \citep[see the review by][]{Madau_rev}. In this paper, we consider predictions for this efficiency, and how it depends on cosmological parameters, specifically the cosmological constant.

Early fully analytic models of star formation were based on simple prescriptions for the cooling time for the gas within galaxies and the typical time scale for converting it into stars \citep[e.g.][]{HS03}, while following the growth of structures with well established analytical forms for the halo mass function \citep{PS74, ST99, ST02}. However, these insightful models neglected more sophisticated mechanisms such as feedback processes driven by stars or active galactic nuclei (AGN). With a slightly different approach, \cite{White_1991} implemented the effect of sub-galactic physics through a series of approximate formulae, while keeping the treatment of structure formation nearly analytical. This seminal work set the stage for a sequence of refinements of the modelling \citep[e.g.][]{Kauffmann1993, Cole_1994, Guideroni_1998, Kauffmann_1999, Cole2000}. Further extensions included the assembly of the central black hole, which enabled a description of the co-evolution of galaxies and quasars \citep[e.g.][]{Kauffman2000, Somerville2008, Henriques_2015, Lacey_2016}. Other semi-analytical techniques followed the formation  of dark matter haloes in full N-body simulations, coupled with analytical recipes for the baryonic physics \citep[e.g.][]{Croton2006, LGalaxies2020}.

Full hydrodynamical simulations incorporate several physical processes into the modelling and follow the evolution of baryons and dark matter from first principles. Several cosmological simulations \citep[e.g.][]{OWLS,  Dubois_2014, FIRE2014, Illustris_V2014, Lukic_2015, EAGLE_Schaye2015, Mufasa, McCarthy_2017, IllustrisTNG2018, Simba_Dave2019}, often based on different numerical approaches \citep{Springel_2005,Springel_Gadget_2001, Almgren_2013, Enzo_Bryan2014, Gizmo, Gadget4}, generally managed to reproduce a plethora of observations to a satisfactory level of accuracy. 
However, feedback processes are implemented via various different numerical `subgrid' prescriptions, whose parameters are tuned to reproduce certain observations. It is therefore important to ask if the predictions of these calculations are robust, or whether they are fine-tuned to our universe and have a high sensitivity to model parameters. While a substantial body of literature has focused on the effect on star formation history of varying the 
subgrid parameters \citep[see the review by][]{feedback_review}, the impact of changing the cosmological parameters has historically received less attention. This may partially reflect the tremendous progress in constraining the parameters of the \LCDM paradigm (e.g. \citealt{Planck18}; but see also \citealt{LCDM_tensions} for a discussion on recent tensions). But studying the predictions of hydrodynamical simulations regarding the star formation history in counter-factual cosmological models would constitute an interesting `stress-test' of our understanding of galaxy formation.

There is also a more fundamental reason to explore this question. Although the \LCDM model is highly successful in many ways, it has theoretical issues that are hard to ignore \citep[see e.g. the review by][]{LCDM_problems}. While the cosmological constant $\Lambda$ is required to explain the accelerating expansion of the universe, there is no consensus on its physical meaning. If $\Lambda$ is a manifestation of the energy of the quantum vacuum, then one can estimate its value by integrating over the zero-point energy of all possible modes. Adopting a cutoff energy $E_{\rm v}$ yields a vacuum density of $E_{\rm v}^4$ in natural units; but the observed $\Lambda$ requires a cutoff at rather low energies, $E_{\rm v}\sim 1\,{\rm meV}$, in gross conflict with the scale of new physics at perhaps 10\,TeV or above \citep{CC_problem_review}. This common calculation is in fact deeply flawed, as it is non-relativistic and yields the wrong vacuum equation of state \citep{Koksma_2011}: a more sophisticated calculation yields a vacuum density of order $M^4$, in terms of the particle mass. But since particles exist with $M$ near to the TeV scale (0.17\, TeV for the top quark), the discrepancy in the estimated $\Lambda$ is much the same as in the naive approach. For a more detailed discussion of this
`cosmological constant problem' see e.g.\cite{Weinberg_2000_talk} or \cite{Abel_2002}. 

Many attempts have been made to move beyond the idea of $\Lambda$ representing a simple vacuum density, and to hypothesise some more general `dark energy' contributing to an effective $\Lambda$ that may vary with time.
\cite{Ratra_1988} proposed that a dynamical scalar field could cause the accelerating expansion of the universe.  While the scalar field proposed by \cite{Ratra_1988} was not motivated by fundamental physics, later models tried to connect it to extensions of the standard model of particle physics. In practice, the kinetic or potential terms of the Lagrangian of the scalar field depend on some fundamental mass scale \citep[e.g.][]{Zlatev_1999, Armendariz-Picon_2001}. An alternative view is that  the cosmic acceleration in fact shows the need for some modified theory of gravity \citep[see e.g.][]{MG_rev}. Other approaches include mechanisms that prevent vacuum energy from gravitating \citep{Kaloper_2014}. However, all these models effectively move the problem of the value of the $\Lambda$ to the fine-tuning of some other parameter of the theory, such as the mass scale associated with the scalar field \citep[see e.g.][]{Amendola_2010}. There is also a more radical position asserting that the debate on the physical nature of the cosmological constant is moot, and that $\Lambda$ is simply a fundamental constant emerging within the theory of General Relativity \citep{Bianchi_Rovelli_2010}.

Regardless of the physical interpretation of the cosmological constant, the oddly small non-null value of $\Lambda$ gives rise to a number of coincidences that cry out for an explanation. Perhaps the most well-known of these is that the universe became $\Lambda$-dominated at $z\approx 0.39$, near to the formation time of the Sun. We therefore appear to live near the unique era when $\Lambda$ transitions from being negligible to dominating the universe: this is the `why-now problem' \citep{why_now}.  A further time coincidence is that recombination occurred at the same time of baryon-radiation equality. In addition, \cite{Lombriser_2017} pointed out that the equality between $\Lambda$ and radiation occurs around the midpoint of cosmic reionisation. All these timescale puzzles are in principle distinct from the cosmological constant problem described earlier \citep[but see also the discussion in][]{Lombriser_2023}.

A possible explanation for the `why now' problem was suggested by \cite{Weinberg_1987}. He noted that if the cosmological constant had been much larger than observed, the accelerating expansion of the universe would have set in at earlier times, freezing out the growth of structure before galaxy-scale haloes had been able to form. Thus the star formation in galaxies that is necessary for the creation of observers would not occur if $\Lambda$ was substantially larger than the observed value. 
Such an argument is an example of anthropic reasoning \citep{Carter_1974}, which effectively considers the existence of observers (such as ourselves) as a `data point' and explores the implications for the cosmological parameters conditional on this information.

To make this Bayesian argument (probability of $\Lambda$ given that it is observed), we need there to be some physical mechanism that allows $\Lambda$ to vary. It is also common to invoke a multiverse: an ensemble of different universes. The probability calculus is the same whether or not the members of the ensemble actually exist, or merely have the potential to do so. But the idea of a concrete multiverse underlying Weinberg's argument was given stronger motivation by the theory of inflation. Here, a multiverse of causally disconnected `bubble universes' arises in models of stochastic inflation where inflation proceeds eternally \citep{Vilenkin_1983, Linde_1986, Guth2007,Freivogel2011}. All causally disconnected bubbles evolve as independent universes, each characterised by a different set of constants, including $\Lambda$. The advantage of this picture is that for a sufficiently large ensemble there are guaranteed to exist universes suitable for the formation of structure. Thus, that would explain the existence of our universe, no matter how atypical it is within the ensemble.

Anthropic arguments often encounter significant resistance, with many physicists arguing that efforts should be focused on finding solutions to cosmological puzzles from first principles \citep[e.g.][]{Kane_2002}. And of course such efforts should always be pursued. But we can note that anthropic approaches pervade other fields of astronomy, without generating controversy. A prime example is the concept of circumstellar `habitable zone', which is defined based on the conditions that can sustain life on a planet \citep[see e.g.][]{Kasting_1993}. Of course, unlike with exoplanets, there is only one universe that can actually be observed \citep[although see][]{Aguirre_2013, Wainwright_2014, Johnson_2016}. As such, anthropic arguments cannot be tested in the Galilean sense ingrained in the scientific method. But they can still be tested, since we can in principle predict the probability distribution of observed values of cosmological parameters over the ensemble. This prediction can be compared with the single `data point' of our universe, and a sufficiently large deviation from the average serves to rule out the model on which the prediction was based.

After the initial formulation given by \cite{Weinberg_1987}, investigations of the anthropic approach became progressively more refined. \cite{Weinberg_1989} extended his argument by noting that observer selection does not require $\Lambda$ to be exactly zero, and thus a small non-zero (in principle even negative) value of $\Lambda$ would be anthropically predicted. A critical element of this argument is the idea that the prior distribution of $\Lambda$ should be flat because $\Lambda=0$ is not a special point, and therefore any continuous prior must be treatable as a constant in a narrow range near zero. \cite{Efstathiou_1995} revisited the argument by making more detailed estimates of the abundance of galaxies in universes with $\Lambda \geq 0$, subject to the constraint that the observed temperature of the Cosmic Microwave Background (CMB) is equal to $2.73 \, \rm K$. He concluded that anthropic arguments may explain a value of $\Lambda$ close to the one that is observed. Later works allowed for more than one cosmological parameter to vary. \cite{Garriga_1999} considered simultaneous variations of $\Lambda$ and the density contrast at the time of recombination. \cite{Peacock_2007} explored anthropic arguments treating both the CMB temperature and $\Lambda$ as free parameters, and further allowing $\Lambda$ to assume negative values. \cite{Bousso_2010} also considered the variation of multiple cosmological parameters, as well as negative values of $\Lambda$. In their work, the generation of observers is tied to the global efficiency of star formation in the different members of the multiverse ensemble. The star formation history is predicted with a semi-analytic model, under the assumption that the astrophysics of star formation does not vary throughout the multiverse \citep{Bousso_2009}. Other semi-analytic work by \cite{Sudoh_2017} showed that incorporating the astrophysics of galaxy formation in anthropic reasoning can affect the range of the anthropically favoured values of $\Lambda$ by almost an order of magnitude.

More recently, progress in numerical power has enabled the testing of anthropic reasoning with full hydrodynamical simulations, exploring the simulated past and future star formation history for a wide range of $\Lambda$ (\citealt{Barnes_2018}; \citealt{Salcido_2018}; \citealt{BK_2022}). While such simulations include several astrophysical processes (albeit in an approximate parameterised form), it is hard to probe a wide parameter space, or to explore the far future of the universe (beyond $\sim$\,100\,Gyr cosmic time), without a massive commitment of computational resources.

Analytic models of star formation therefore represent an attractive complementary approach, as they are not subject to the same limitations as hydrodynamical simulations. While inevitably simplified in terms of astrophysics, they are an efficient technique that can offer a more intuitive picture of the evolution of star formation \citep[e.g.][]{Rasera_2006, Dave_2012, Ikea, Salcido_2020, Fukugita_2022}. Recently, \cite{SP21} generalised the \cite{HS03} model of cosmic star formation such that it can be applied to arbitrarily large times. In this work, we further adapt the \cite{SP21} formalism to explore the impact of different values of $\Lambda$ on past and future star formation. The main result of our study will be that values of $\Lambda$ in the range $0.01\lesssim \Lambda/ \Lambda_{\rm obs} \lesssim \Lambda_{\rm obs}$ maximise the global efficiency of star formation. However, the interesting question is how effectively star formation is truncated by large values of $\Lambda$. If this suppression is not effective until $\Lambda$ is vastly beyond the observed value, then Weinberg's flat prior will mean that the observed universe risks being an implausibly rare outlier. The main aim of this paper is to quantify just how unusual our universe is in this respect.

This manuscript is organised as follows. In \S\,\ref{sec:formalism} we give an overview of the formalism in \cite{SP21}, and improve it by introducing extra features. In \S\,\ref{sec:change_L} we explain how we adapt it to \LCDM models with arbitrary non-negative values of $\Lambda$ and discuss the impact of changing the cosmological constant on the halo mass function and on the efficiency of star formation within haloes. In \S\,\ref{sec:results} we present our predictions for the long-term efficiency of star formation for different values of $\Lambda$. In \S\,\ref{sec:discussion} we discuss the implications of our results for anthropic reasoning. We mainly focus on the cosmological constant problem, but also marginally consider the why-now problem. We also compare our results with previous literature on the subject and discuss the limitations of our model. Finally, in \S\,\ref{sec:conclusions} we summarise the main conclusions of our work and discuss the future developments of our line of research.

Throughout this work, unless otherwise indicated, units of distance are understood to be proper units. Co-moving units are designated with a `c' prefix (e.g. ckpc, cMpc).

\section{Formalism}
\label{sec:formalism}

In this work, we aim to exploit the analytic model for cosmic star formation developed by \cite{SP21} -- hereafter SP21. We therefore summarise the main aspects of the formalism in \S\,\ref{sec:summary}. In \S\,\ref{sec:generalise} we will show how we extended the SP21 model to obtain greater accuracy in the predictions of the late-time behaviour of star formation. As we will explain, this generalisation will be crucial in answering the main scientific questions addressed in this work.

\subsection{Summary of the SP21 model}
\label{sec:summary}

The SP21 model predicts the cosmic star formation rate density (CSFRD) from first principles, generalising the seminal work by \cite{HS03} -- hereafter HS03. The basic idea of the formalism is that the CSFRD is obtained by integrating the star formation rate (SFR) in all haloes within a given comoving volume, weighted by the halo multiplicity function:
\begin{equation}
\label{eq:CSFRD}
    \dot{\rho}_* (z)= \bar{\rho}_0 \int s(M, \, z) \frac{dF(M, \, z)}{d\ln M} d\ln M \, ,
\end{equation}
where the $\bar{\rho}_0$ is the comoving mean matter density of the universe and $dF(M, \, z)/d\ln M$ is the halo multiplicity function, with $F(M, \, z)$ being the collapsed mass fraction in haloes with total mass $>M$. These quantities encode the impact of background cosmology and of the growth of large-scale structure on the CSFRD. The specifically astrophysical component of the CSFRD is encapsulated in the term $s(M, \, z)={\rm SFR}/M$, which represents the average star formation rate over a population of haloes of mass $M$ at redshift $z$, normalised by the \textit{total} halo mass $M$. We will refer to $s(M, \, z)$ as `normalised SFR' (nSFR).

In this work, we follow SP21's choice of modelling $F(M, \, z)$ via the Sheth--Tormen formalism \citep{ST99, ST02}. We also adopt the same definitions of the virial radius, mass and temperature as in SP21. The virial radius $R$ of a halo of virial mass $M$ at redshift $z$ is defined as the radius of the sphere that contains a matter density equal to $\Delta \, \rho_{\rm c}(z)$, where $\rho_{\rm c}(z)$ is the critical density of the universe and $\Delta$ a suitable multiplying factor:
\begin{equation}
\label{eq:Mvir}
    M = \Delta \frac{4}{3} \pi R^3 \rho_{\rm c}(z) \, ;
\end{equation}
SP21 adopted $\Delta=200$. One can then define a characteristic virial velocity
\begin{equation}
    V^2 = \frac{GM}{R} \, ,
\end{equation}
where $G$ is the universal gravitational constant. We additionally \emph{define} the virial temperature $T$ such that
\begin{equation}
\label{eq:Tvir}
    V^2 = \frac{2k_{\rm B} T}{\mu} \, ,
\end{equation}
where $k_{\rm B}$ is the Boltzmann constant and $\mu$ is the mean molecular weight. We assume that $\mu \approx 0.6 m_{\rm p}$, which is a valid approximation for a fully ionised plasma of primordial composition. From equations~\eqref{eq:Mvir}--\eqref{eq:Tvir}, it follows that 
\begin{equation}
\label{eq:Mvir_T}
    M = \sqrt{\frac{2}{\Delta}} \frac{1}{GH(z)} \left( \frac{2 k_{\rm B} T}{\mu} \right)^{\frac{3}{2}} \, .
\end{equation}

Although the CSFRD in equation~\eqref{eq:CSFRD} is expressed as an integral over the halo mass, it may be at times more convenient to switch the integration variable to the virial temperature. In particular, within the HS03 formalism, the nSFR is more naturally expressed for a population of haloes at given $T$, as we shall now explain.

At any given $z$, the nSFR is set by a characteristic time scale. At high redshift, haloes are denser and hence gas cooling is more efficient. The bottleneck of star formation is therefore represented by the typical time scale that governs the conversion of cool gas into stars. This `average gas consumption time scale', $\langle t_* \rangle$, is assumed to be a constant, with no dependence on the virial temperature of the halo or on the cosmological epoch. This assumption was originally introduced by HS03 following the results of full hydrodynamical cosmological simulations \citep{SH03_sims, SH03_SFR}, but really asserts that $\langle t_* \rangle$ is a timescale set by the local microphysics of molecular clouds. The same prescription was adopted by SP21, who determined a physically reasonable range of $\langle t_* \rangle$ by matching their predictions of the Kennicutt--Schmidt law \citep{Kennicutt_1998} with measurements by \cite{Genzel_2010}. The value of $\langle t_* \rangle$ that best reproduces the observations was found to be $2.39\,\rm Gyr$, but values in the range $(1.63 - 3.87)\,\rm Gyr$ are consistent with the data within $3\sigma$. SP21 showed that values of $\langle t_* \rangle$ within this range can predict the CSFRD at high redshift within a factor of two, and we adopt the same prescription for the average high-redshift nSFR of a halo population at given virial temperature:
\begin{equation}
\label{eq:nSFR_high}
    s_{\rm high} (T, \, z) = \frac{(1-\beta) x f_{\rm gas}(T, \, z) }{\langle t_* \rangle} \, ,
\end{equation}
where $x$ is the fraction of cold gas clouds, $\beta$ is the mass fraction of massive ($>8 \MSun$) short-lived stars, and $f_{\rm gas}(T,\,z)$ is the mass fraction of gas within the haloes considered. Following the same choice as in HS03 and SP21, we set $x=0.95$. This value is again motivated by the sub-grid model for star formation by \cite{SH03_SFR}, which was utilised in hydrodynamical simulations by \cite{SH03_sims}. As in SP21, we set $\beta = 0.21$. This is determined by assuming a \cite{Chabrier_2003} initial mass function (IMF) over a stellar mass range between $0.1 \MSun$ and $80 \MSun$. 

At low redshift, haloes are less dense and gas cooling thus proceeds more slowly. The nSFR is then no longer dictated by an internal timescale, but is limited by the supply of new cold gas, which is set by the cooling time scale $t_{\rm cool}$. This quantity depends on the local gas density, which is assumed to follow a spherically symmetric power-law profile:
\begin{equation}
\label{eq:dens_prof}
	\rho _{\rm gas} (r) = (3-\eta) \frac{M_{\rm gas}}{4 \pi R^{3} } \left(\frac{R}{r} \right)^{\eta}  \, ,
\end{equation} 
where $M_{\rm gas}$ is the total gas mass enclosed in the halo, and the slope $\eta$ is a free parameter of the model. Clearly, one must have $\eta<3$ to prevent the halo gas mass from diverging. We also assume that $\eta>0$, so that density falls with radius. The assumption of a power-law gas density profile is generally supported by full cosmological simulations (e.g. \citealt{Sorini_2024_Simba}), but more complex models may provide a more realistic physical picture \citep[see e.g.][]{mNFW_2017, mNFW_2019, Khrykin_2024, Sorini_2024_MTNG}.

The cooling time at distance $r$ from the centre of the halo is given by
\begin{equation}
\label{eq:tcool}
	t_{\rm cool} = \frac{3 k_{\rm B} T \rho_{\rm gas} (r)}{2 \mu n_{\rm H}(r)^2 \mathcal{C} (T)} \, ,
\end{equation}
where $n_{\rm H}$ is the number density of hydrogen, and $\mathcal{C}(T)$ is the cooling function. As in HS03 and SP21, we assume a primordial cooling function \citep{Sutherland_1993}, effectively neglecting metal cooling. But this limitation is expected to alter the CSFRD at low redshift in a manner that does not significantly affect conclusions regarding the long-term star formation history (see the discussion in SP21 and HS03), so that metallicity evolution is unimportant for the scope of this work. The cooling function depends purely on temperature, so there is some advantage in considering the evolution of the nSFR for haloes with given $T$ rather than given $M$.

The cooling rate of gas within haloes of virial temperature $T$ at a given time is then estimated by following the expansion of a cooling front from the centre of the halo outwards. At any time $t$, the cooling front reaches the cooling radius $r_{\rm cool}(t)$, defined by the criterion $t_{\rm cool}(r_{\rm cool}(t))=t$. The gas mass $M_{\rm cool}$ within $r_{\rm cool}$ cools down and remains cool thereafter. The cooling rate is then determined by solving the equation
\begin{equation}
    \label{eq:cool_rate}
	\frac{d M_{\rm cool} }{d t} = 4 \pi \rho _{\rm gas} (r_{\rm cool}) r^2_{\rm cool} \frac{d r_{\rm cool}}{dt} \, .
\end{equation}
Here, the meaning of the time $t$ needs to be defined quite carefully. In principle, this would be the time since the formation of the halo, i.e. since the last major merger. In the matter-dominated era, low-mass haloes have a life span comparable to the age of the universe, whereas high-mass haloes beyond the exponential cutoff survive for shorter times. But massive haloes are rare, so one can effectively make the reasonable assumption that $t$ (and $t_{\rm cool}$) are comparable to the cosmic time. However, it has been suggested that $t_{\rm cool}$ should be of the order of the dynamical time of the halo $t_{\rm dyn} = R/V$, as it is on this time scale that the gas profile reacts to pressure losses, and hence should set the extent of the cooling radius \citep{Springel_2001}. Because this prescription provides good agreement with simulations \citep{Yoshida_2002}, HS03 solved equation~\eqref{eq:cool_rate} imposing $t_{\rm cool} = t_{\rm dyn}$. However, SP21 noted that this assumption breaks down in the far future, when the universe becomes $\Lambda$-dominated. In this regime, merging eventually ceases, and haloes are isolated. Therefore, haloes can in principle cool down for arbitrarily large times. Thus, SP21 adopted the prescription
\begin{equation}
   \label{eq:tcool_smooth}
    t_{\rm cool} (t) = f_{\rm dyn} t_{\rm dyn} \left[ 1 - E + \left(\frac{t}{f_{\rm dyn} t_{\rm dyn}}\right)^m \right]^{\frac{1}{m}} \, ,
\end{equation}
where
\begin{equation}
    E = \left(\frac{2}{3 f_{\rm dyn}} \sqrt{\frac{\Delta}{2}} \right)^m \, .
\end{equation}
By construction, equation~\eqref{eq:tcool_smooth} yields $t_{\rm cool} \approx f_{\rm dyn} t_{\rm dyn}$ for $t \ll t_{\rm dyn}$, where $f_{\rm dyn}$ is a constant of order unity, and $t_{\rm cool} \approx t$ for $t \gg t_{\rm dyn}$. SP21 showed that the values of the parameters $f_{\rm dyn}$ and $m$ have minimal impact on the predicted CSFRD. We will then adopt their fiducial values, $f_{\rm dyn}=1$ and $m=2$.

In a flat \LCDM universe, the correspondence between redshift and cosmic time can be obtained analytically outside the radiation-dominated era:
\begin{equation}
\label{eq:a_t}
a(t) = \left(\frac{\Omega_{\rm m}}{\Omega_{\Lambda}}\right)^{\frac{1}{3}} \left[ \sinh \left(\frac{3}{2} \sqrt{\Omega_{\Lambda}} H_0 t \right) \right]^{\frac{2}{3}} \, ,
\end{equation}
where $H_0$, $\Omega_{\rm m}$ and $\Omega_{\Lambda}$  are the Hubble parameter and density parameters of matter and $\Lambda$ at $z=0$, respectively. A principal interest of the present paper is the impact of changing the value of $\Lambda$, which then alters the expansion history of the universe and changes all cosmological parameters. We discuss below how to handle these changes in a consistent fashion. Combining equation~\eqref{eq:a_t} with equations~\eqref{eq:cool_rate} and \eqref{eq:tcool_smooth}, one can then obtain the cooling rate as a function of the virial temperature and redshift. Using the cooling rate as a proxy for the low-redshift nSFR, SP21 found
\begin{multline}
\label{eq:nSFR_low}
      s_{\rm low}(T,\,z) = \tilde{S}(T) \left( \frac{H(z)}{H_0} \frac{f_{\rm gas}(T, \, z)}{f_{\rm b}} \right)^{\frac{3}{\eta}} 
    \left( 1- E + A(z)^m \right)^{\frac{3-\eta}{m \eta}-1} \\
    \left[A(z)^{m-1} + (1-E)  \frac{3 f_{\rm dyn}}{2} \sqrt{\frac{2}{\Delta}} \left(\frac{H_0}{H(z)}\right)^2 \Omega_{\rm m} (1+z)^3 \right]  \, ,
\end{multline}
where
\begin{equation}
    A(z) = \frac{2}{3 f_{\rm dyn}} \sqrt{\frac{\Delta}{2\, \Omega_{\Lambda}}} \frac{H(z)}{H_0} \arcsinh \left(\sqrt{\frac{\Omega_{\Lambda}}{\Omega_{\rm m} (1+z)^3}} \right) 
\end{equation}
and $\tilde{S}(T)$ is a temperature-dependent proportionality factor with the units of a star formation rate per unit mass. 

Clearly, to determine the nSFR both at high and low redshift, it is crucial to have an expression for $f_{\rm gas}(T, \, z)$. SP21 adopt the approximation $f_{\rm gas}(T,\,z)\approx f_{\rm b,\, halo} (T,\,z)$, where $f_{\rm b,\, halo} (T,\,z)$ is the baryon mass fraction in haloes of temperature $T$ at redshift $z$. This is in turn determined by studying the balance between the gravitational force and outward pressure exerted by supernovae-driven winds on a gas parcel within the virial radius, as a function of its distance from the centre. One can then define a critical radius within which baryons are bound to the halo and escape otherwise. The question is whether such a critical radius falls within the cooling radius, which is the region of the halo that can produce stars in the simplified picture considered by SP21 and HS03. SP21 show that above a certain critical virial temperature $T_{\rm crit}(z)$ the critical radius falls outside the cooling radius, hence all haloes retain their cosmic share of baryons. Below $T_{\rm crit}(z)$, this approach yields a baryon mass fraction enclosed in the halo that scales as a power of its virial temperature:
\begin{gather}
\label{eq:fbh_T}
    f_{\rm b, \, \, halo} (T, \, z) =   
        \begin{cases}
         \left(  \frac{T}{T_{\rm crit}(z)}\right) ^{\frac{3-\eta}{2(\eta - 1)}} f_{\rm b} &  \textrm{if} \;\; T<T_{\rm crit}(z) \\
            f_{\rm b} & \rm otherwise
        \end{cases} \, .
\end{gather}
The transition between the two temperature regimes is smoothed with a suitable analytic function. Equation~\eqref{eq:fbh_T} effectively provides a prediction for the baryonic Tully-Fisher relationship (bTFR), an empirical correlation between the baryonic mass and total mass of haloes. Comparing the predicted bTFR with  observations of the bTFR by \cite{Lelli_2016}, SP21 conclude that values of $\eta$ in the range $1.9-2.4$ are compatible with the data within $3 \sigma$. The same values are simultaneously compatible, within a similar level of precision, with the \cite{Genzel_2010} observations of the Kennicutt-Schmidt relationship.

\begin{figure*}
    \centering
    \includegraphics[width=0.33\textwidth]{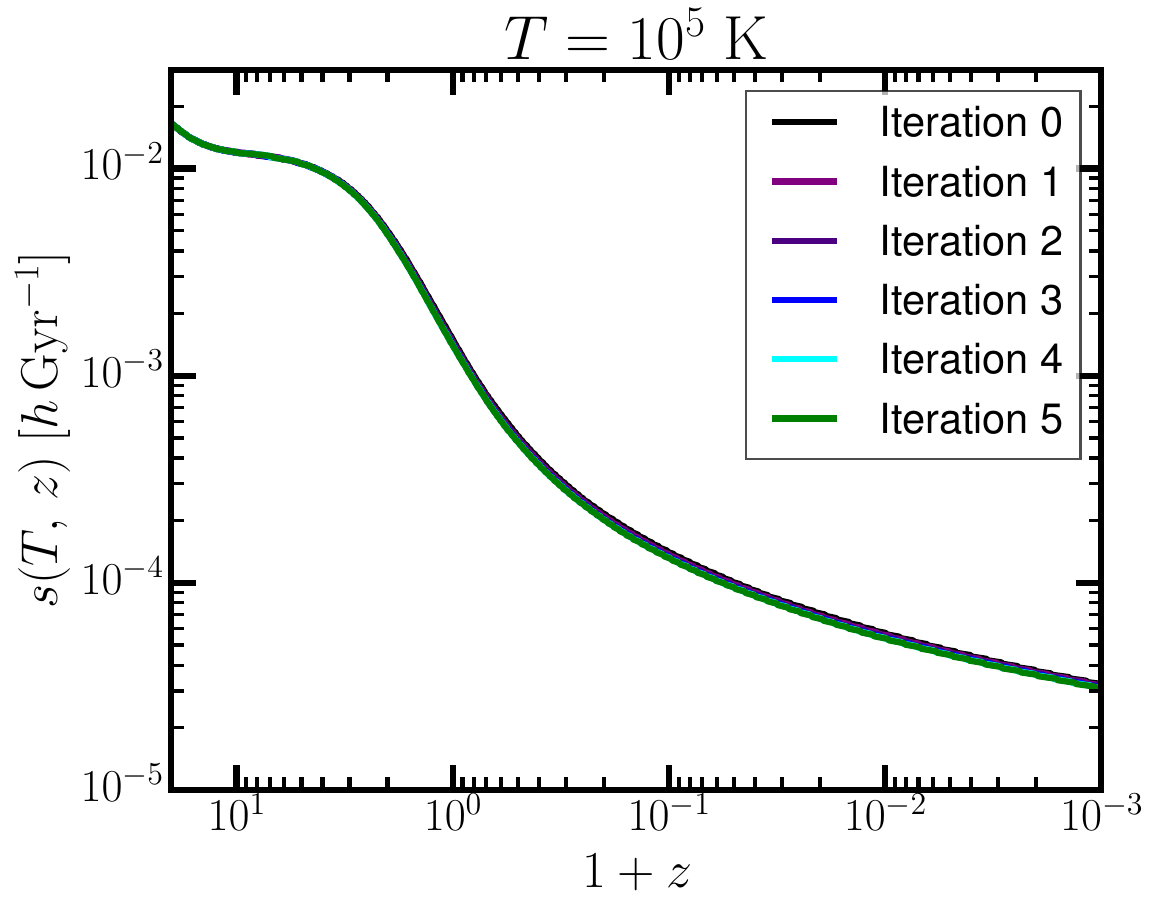}
    \hfill
    \includegraphics[width=0.33\textwidth]{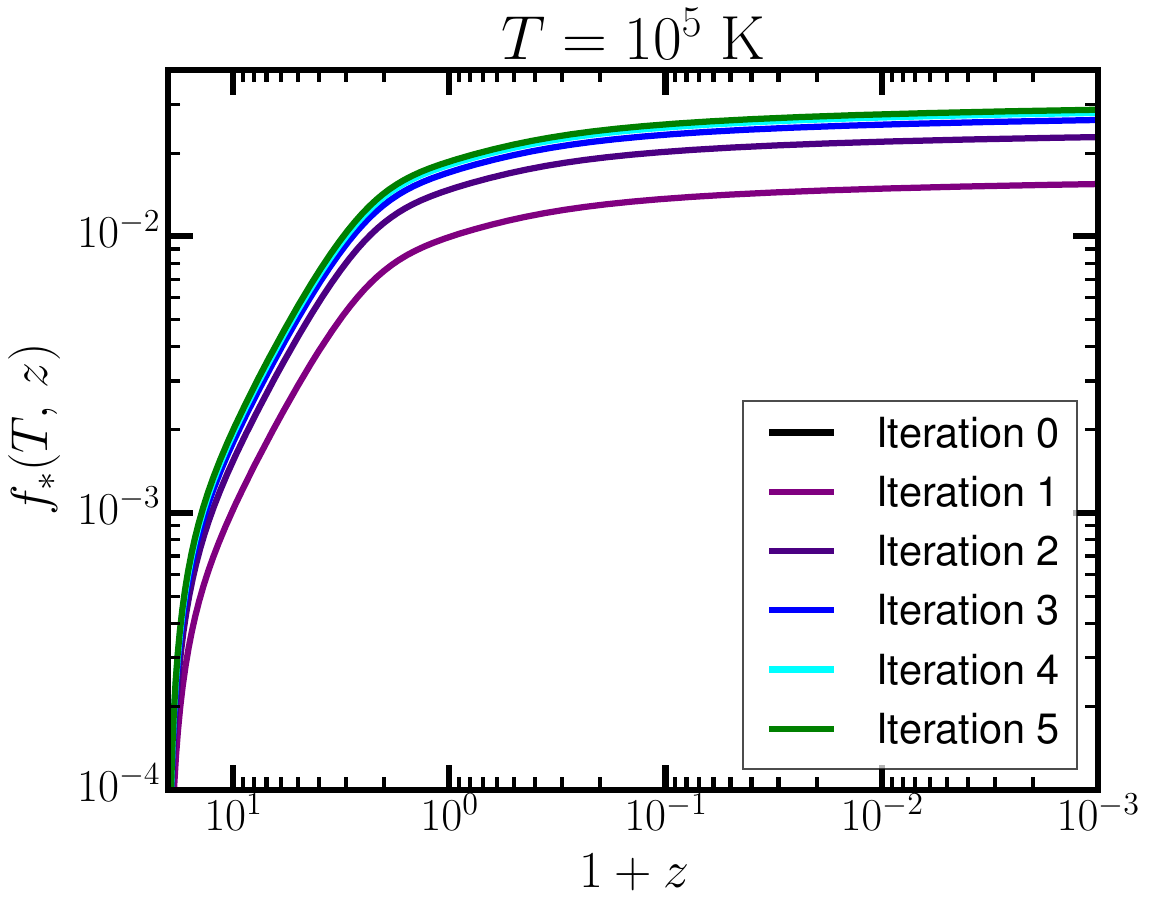}
    \hfill
    \includegraphics[width=0.325\textwidth]{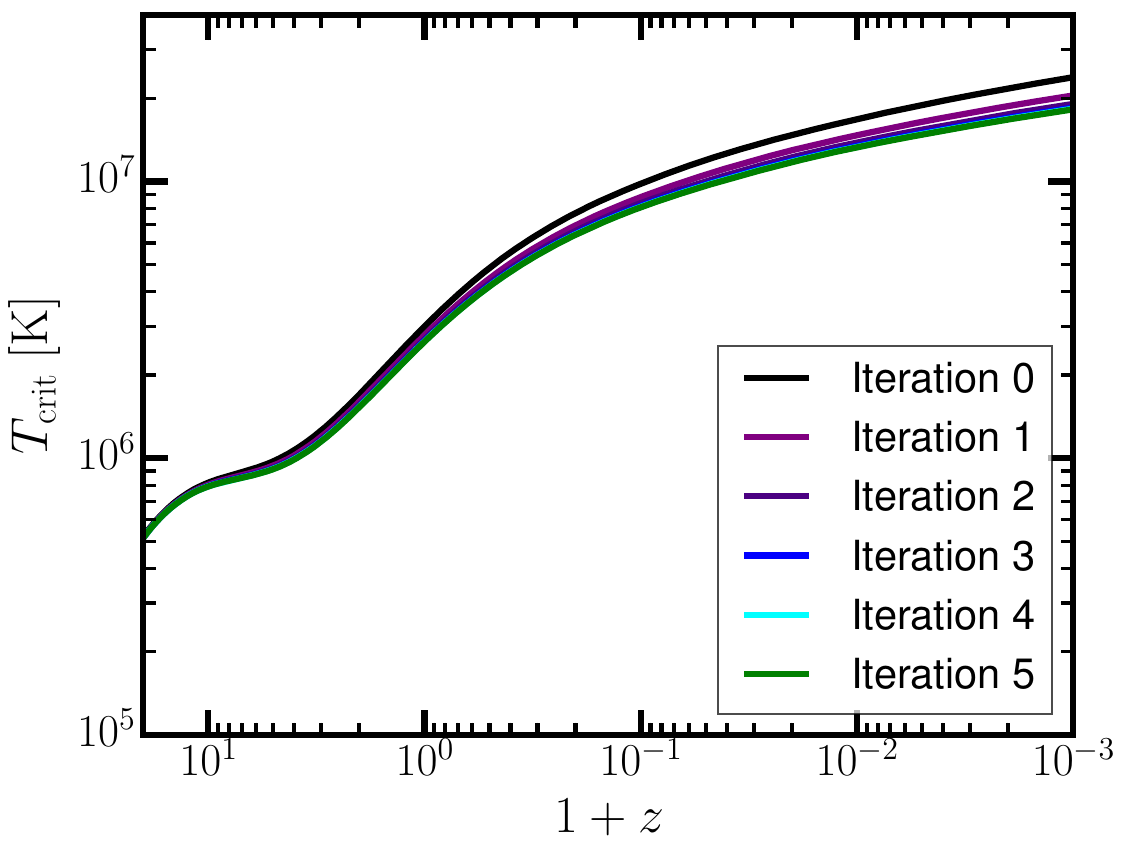}
    \caption{Redshift evolution of the nSFR (\textit{left panel}) and stellar mass fraction (\textit{middle panel}) for a halo with virial temperature $T=10^5 \K$, as predicted at every iteration of our updated version of \protect\cite{SP21} described in \S\,\ref{sec:generalise}. The \textit{right panel} shows the prediction for the redshift evolution of the critical temperature.  The black lines labelled ``iteration 0'' refer to the predictions of the original SP21 model with $f_*(T,\, z)=0$; thus no black line appears in the middle panel, which adopts a logarithmic scale on the $y$-axis. For all quantities plotted in this figure, full convergence is achieved after five iterations.}
    \label{fig:convergence}
\end{figure*}

The final expression of the nSFR as a function of time is then obtained by simply interpolating the high-redshift (equation~\ref{eq:nSFR_high}) and low-redshift solutions (equation~\ref{eq:nSFR_low}) with a smooth function, such that
\begin{equation}
\label{eq:nSFR_smooth}
    s(T,\,z) = \frac{s_{\rm high}(T,\,z) s_{\rm low}(T,\,z)}{(s_{\rm high}(T,\,z)^m+s_{\rm low}(T,\,z)^m)^{\frac{1}{m}}} \, .
\end{equation}
SP21 set $m=2$, although they showed that the impact of this parameter on the predicted CSFRD is minimal. Equation~\eqref{eq:nSFR_smooth}, combined with the Sheth--Tormen formalism for the halo multiplicity function, allows one to obtain an analytic expression for the CSFRD given by equation~\eqref{eq:CSFRD}.

\subsection{Improving the accuracy of the SFR at late times}
\label{sec:generalise}

As explained in \S\,\ref{sec:summary}, one simplifying assumption adopted in SP21 is that the baryon mass fraction in haloes is roughly equal to the gas mass fraction. This approximation can be easily justified at high redshift, shortly after the onset of star formation. But this assumption can be too strong at later times, when the stellar mass fraction of haloes is not negligible with respect to their gas mass fraction \citep[e.g.][]{McGaugh_2010}. This could become an even more important issue when considering the future of the universe. Since the major focus of this work is understanding the impact of $\Lambda$ on both the past and future star formation history, we need to first extend the SP21 formalism by keeping $f_{\rm gas}(T,\,z)$ and $f_{\rm b,\, halo} (T,\,z)$ distinct.

One would be tempted to compute $f_{\rm gas}$ from equation~\eqref{eq:fbh_T}, replacing the cosmic baryon mass fraction with $f_{\rm b}-f_*(T, \, z)$, where $f_*(T, \, z)$ is the stellar mass fraction for a halo with virial temperature $T$ at redshift $z$. The stellar mass fraction is related to the time integral of the nSFR provided by equation~\eqref{eq:nSFR_smooth}. The trouble is that the nSFR depends on $f_{\rm gas}$, while $f_{\rm gas}$ in turn depends on the nSFR.

To break this circularity, we adopt the following iterative method to calculate the nSFR:
\begin{enumerate}
    \item we assume that, at any fixed virial temperature, $f_{\rm gas}(T,\,z)\approx f_{\rm b,\, halo} (T,\,z)$ at a sufficiently high redshift $z_{\rm in}$, and compute the corresponding nSFR following the original SP21 model;
    \item \label{pt:iteration} we obtain the stellar mass fraction as a function of redshift by integrating the nSFR over time, or, in terms of redshift:
    \begin{equation}
    \label{eq:fstar}
        f_*(T,\, z) = \frac{1}{M(T, \, z)} \int ^{z_{\rm in}} _{z} \frac{M(T,\, z') \, s(T, \, z')}{(1+z') H(z')} dz' \, ;
    \end{equation}
    \item we compute the cooling radius as a function of redshift from the criterion $t_{\rm cool}(r_{\rm cool}(t))=t$ explained in \S\,\ref{sec:summary}; the cooling time is given by equation~\eqref{eq:tcool}, with the gas density given by equation~\eqref{eq:dens_prof}, where we further impose that $M_{
    \rm gas}(T, \, z) = (f_{\rm b} - f_*(T,\, z)) M$. We then compare the evolution of the cooling radius with that of the critical radius described in the previous section to determine the critical temperature $T_{\rm crit} (z)$ (see the appendix in SP21 for more details);
    \item we update $f_{\rm gas}(T,\,z)$ via equation~\eqref{eq:fbh_T}, where we now replace $f_{\rm b}$ with $f_{\rm b}-f_*(T, \, z)$;
    \item we use the new gas mass fraction to re-calculate the nSFR via equations~\eqref{eq:nSFR_high}, \eqref{eq:nSFR_low} and \eqref{eq:nSFR_smooth};
    \item we repeat the above protocol restarting from point~\ref{pt:iteration}.
\end{enumerate}

To ensure that the iterative procedure described above produces converging results, we examined the nSFR, stellar mass fraction, and critical temperature computed at each iteration. We plot the evolution of these quantities in Fig.~\ref{fig:convergence}. The nSFR and stellar mass fraction refer to a halo with virial temperature $T=10^5 \K$ in our \LCDM universe. For all quantities plotted, convergence is achieved very quickly -- within five iterations. As expected, a non-null stellar mass fraction reduces the gas available for star formation, so that the nSFR is reduced. This in turns makes the cooling radius smaller: haloes that were at the critical temperature in the SP21 work are therefore now above the critical temperature so that the critical temperature is lowered with respect to the original formalism.

As we can see in the middle panel of Fig.~\ref{fig:convergence}, the stellar mass fraction asymptotes to a constant in the far future. That is a consequence of the decay in the nSFR, which tends to zero in the limit of infinitely large cosmic time. We will discuss this in detail in \S\,\ref{sec:future}. The stellar mass fraction is a monotonic function of time, because the nSFR is always positive. There is no explicit mechanism in the model that removes stars that have gone supernova or became stellar remnants. The contribution of short-lived and massive stars to the nSFR is removed through the $\beta$ factor in equations~\eqref{eq:nSFR_high} and \eqref{eq:nSFR_low}, but all other stars are `eternal' once formed. This is a reasonably good approximation, since the time scale for white dwarfs to turn into black dwarfs has been estimated to be $\sim$\,$10^5\,\rm Gyr$ \citep{Dyson_1979, Adams_1997, Caplan_2020}; we will discuss the subsequent implications for anthropic reasoning in \S\,\ref{sec:anthropics}.

We verified that convergence is achieved in the range of interest for this study, that is $10^4 \, \mathrm{K} < T < 10^{10} \K$ (as explained in SP21, the contribution of haloes with $T>10^{10} \K$ to the CSFRD is negligible; we verified that this holds also for the cosmological models considered in this work). At higher temperatures, convergence is faster: at $T=10^7 \, \rm K$, two iterations are sufficient. For all quantities shown in Fig.~\ref{fig:convergence}, we chose $z_{\rm in}=20$ as the initial redshift, but we verified that the exact value is unimportant as far as our analysis is concerned (see \S\,\ref{sec:future} for a more detailed discussion).

To assess whether the stellar mass fractions predicted by our revised SP21 method are physically sensible, let us consider the mass of the Milky Way. Recent estimates point to a stellar mass of $(6.08 \pm 1.14) \times 10^{10} \MSun$ \citep{Licquia_2015} and a virial mass of $1.54_{-0.44}^{+0.75} \times 10^{12} \MSun$ \citep{Watkins_2019}. Therefore, the stellar-to-total mass ratio is expected to be in the range $2.2-6.6\%$. Applying equation~\eqref{eq:Mvir_T}, we can derive the virial temperatures corresponding to the virial mass found by \cite{Watkins_2019}, and then compute the corresponding stellar mass fraction at $z=0$ via equation~\eqref{eq:fstar}, obtaining $6.7\%$. Thus, the predictions of our model are compatible with the upper bound of the stellar-to-total halo mass range provided by the observations. Considering the simplifications made in the formalism, this is a reassuring result.

\begin{figure*}
    \centering
    \includegraphics[width=\textwidth]{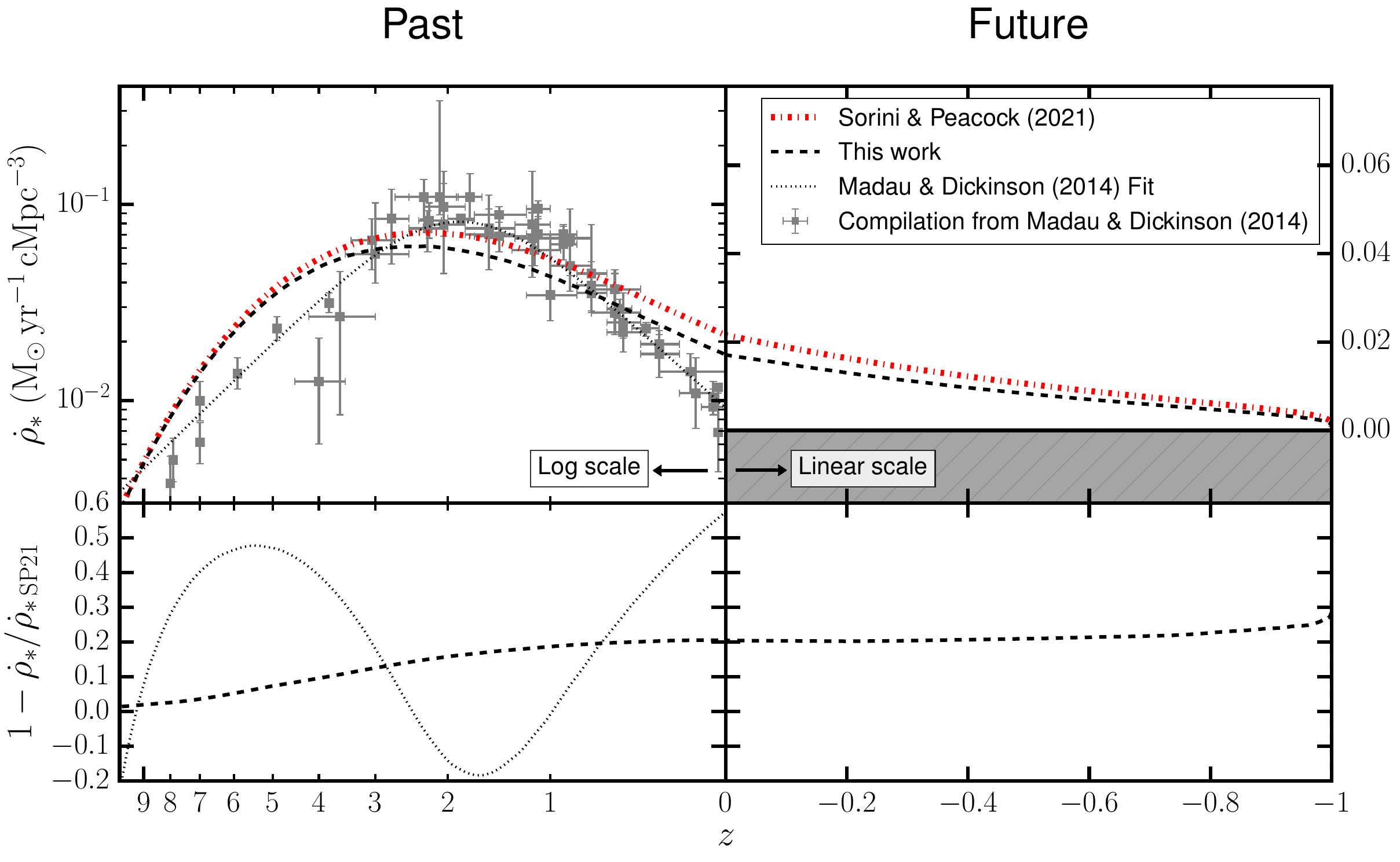}
    \caption{\textit{Upper panels}: Cosmic star formation rate density, as computed with our extension of the SP21 model (dashed black line) and as given by the original SP21 model (dot-dashed red line). In the left panel, we further show with the empirical fit (dotted black line) to a compilation of observational data provided by \protect\cite{Madau_rev} (grey data points). The right panel shows the behaviour of the SP21 model and our extended formalism in the future of the universe. As future times correspond to negative redshifts, and the cosmic star formation rate density becomes asymptotically null at arbitrarily large times, both axes switch from logarithmic to linear scale when moving from the left to the right panel. The grey hatched shaded area in the right panel excludes the non-physical region corresponding to $\dot{\rho}_*<0$. \textit{Lower panels}: Relative difference between the two models showed above (dashed black line), and between our formalism and the fit from \protect\cite{Madau_rev} (dotted black line). The latter comparison is available only for positive redshifts (i.e., past cosmic times), as the fit represents a purely empirical fit to the data, and not a predictive theoretical model. The SP21 model agrees well with the improved method introduced in this work at high redshift, but overestimates the cosmic star formation rate density at lower redshift and in the future. In the redshift range $0<z<10$, both the original SP21 model and our updated formalism agree with the empirical fit within a factor of $\sim\,2.$}
    \label{fig:CSFRD_comp}
\end{figure*}

The key quantity under study in this paper is the CSFRD, so it is important to see how this is affected by the changes in normalised star-formation rate shown in Fig.~\ref{fig:convergence}, which result from our generalisation of the SP21 model. In the upper-left panel of Fig.~\ref{fig:CSFRD_comp} we show the CSFRD in the redshift range $0<z<10$, computed both with the original and revised formalism. It is clear that our procedure yields a slightly lower CSFRD at low redshift, with little change in the slope at late times. In the upper-right panel, we show the behaviour of the two models in the future of the universe. As future times correspond to negative redshifts, and the CSFRD approaches zero as time tends to infinity, the scale of both axes switches from logarithmic to linear when moving from the upper-left to the upper-right panel. We quantify the relative difference between the original SP21 model and our updated formalism in the lower panels of Fig.~\ref{fig:CSFRD_comp} (dashed black line). The differences are at the percent level for $z>5$, and then grow progressively, reaching $\sim$\,10\% at the peak of star formation and $\sim$\,20\% at $z=0$. The discrepancy grows slowly but steadily in the far future of the universe, stopping short of $30\%$ in the limit $t\rightarrow \infty$. Such differences are sub-dominant with respect to the uncertainties of the parameters of the model. The original SP21 model is thus still suitable for rapidly predicting the past star formation history of our universe, but our generalisation becomes indispensable for the future star formation history. This is even more important when one considers larger values of the cosmological constant: we verified that the relative difference with respect to the SP21 model grows as $\Lambda$ increases. 

In the upper-left panel of Fig.~\ref{fig:CSFRD_comp}, we also show the compilation of observational data provided by \citealt{Madau_rev}, alongside their empirical fit to the data. Both data and fitting function have been properly rescaled from a \cite{Salpeter_1955} to a \cite{Chabrier_2003} IMF, for consistency with the SP21 formalism and the model presented in this work (the data points were not rescaled in this way in the SP21 paper). The relative difference between the SP21 model and the \cite{Madau_rev} fit are quantified in the lower-left panel, confirming that the SP21 predictions recover the fitting function within a factor of $\sim 2$ for $0<z<10$. Since our updated formalism deviates from SP21 by at most $20\%$ in the same redshift range, we can conclude that our model also matches the \cite{Madau_rev} fit within a factor of $\sim$\,2. This level of agreement is remarkable, considering the simplicity of the SP21 model. As a reference, sophisticated hydrodynamic cosmological simulations, while providing an overall better agreement with the data, exhibit discrepancies of a factor of $\sim$\,2 or larger around cosmic noon \citep{McCarthy_2017, Simba_Dave2019} and within a factor of $\sim$\,1.6 at $z\lesssim 3$ \citep{Salcido_2018}. However, other state-of-the-art cosmological simulations, such as IllustrisTNG \citep{IllustrisTNG2018}, achieve a significantly more accurate match with observations of the CSFRD \citep{Weinberger_2017}. Thus, whereas the SP21 model manages to broadly reproduce both observed and simulated trends of the CSFRD \citep{Scharre_2024}, there is certainly room for further improvement. A detailed discussion on the limitations of the SP21 model and the resulting impact on the predicted CSFRD can be found in \cite{SP21}.

It is also worth noting that the data in the \cite{Madau_rev} compilation might be affected by biases introduced by the parametric models underlying the estimation of the CSFRD from observations of galaxy spectral energy distributions \citep{Carnall_2019}. Additionally, subsequent estimates of the CSFRD deviate from the \cite{Madau_rev} fit, both at high \citep{Gruppioni_2015, Rowan-Robinson_2016} and low \citep{Gruppioni_2015} redshifts. Other measurements exhibited a lower normalisation for the CSFRD at the peak of star formation \citep{Liu_2018}, or a plateau at cosmic noon \citep{Traina_2024}. The spread of some of the data points from these more recent measurements is generally comparable to, or even larger than, the discrepancy between the SP21 model and the \cite{Madau_rev} fit. 
Given this range of predictions, the precision of the CSFRD given by the SP21 model seems satisfactory for the purpose of the present investigation.

As for the performance of the model in predicting the future star formation history, we note that the shape of the future CSFRD is essentially dictated by the evolution of the cooling radius and of the critical temperature (see 
Appendix~\ref{app:math} for a detailed derivation of the asymptotic scaling of the CSFRD). The normalisation of the CSFRD for $z<0$ will follow the overall normalisation of the CSFRD at earlier times. Thus, given a factor of $\sim$\,2 uncertainty in the past star formation history at $z\approx 0$, we can reasonably expect a similar systematic imprecision in the future CSFRD. As it will become apparent in \S~\ref{sec:future}, the future behaviour of the CSFRD may contribute significantly to the overall production of stars in the entire history of the universe. But as we will show, the differences in the total stellar mass produced in the universe for different values of the cosmological constant can span several orders of magnitude, thus our main conclusions will be largely unaffected by an uncertainty of a factor $2$ in the CSFRD.

We also highlight that our analysis is relative, i.e. based on the ratio of the star formation histories produced in different cosmologies, and will not depend on the absolute value of the star formation efficiency in a single cosmology. Of course, our posterior for $\Lambda$  will still be subject to accepting the underlying simplified astrophysical model for cosmic star formation.  We discuss the impact of these limitations of our study, and prospects for their future amelioration, in \S~\ref{sec:comparison}.

\section{Impact of changing $\Lambda$}
\label{sec:change_L}

Having introduced the basic formalism in the previous section, we will now explain how varying $\Lambda$ would affect the star formation history. Changing the cosmological constant obviously affects the evolution of large-scale structure; this is the subject of \S\,\ref{sec:cosmo}. But a different value of $\Lambda$ can also directly affect the cooling rate within haloes. The astrophysical impact of changing $\Lambda$ will be discussed in \S\,\ref{sec:astro}.

\subsection{Cosmological impact}
\label{sec:cosmo}

\begin{figure}
    \centering
    \includegraphics[width=\columnwidth]{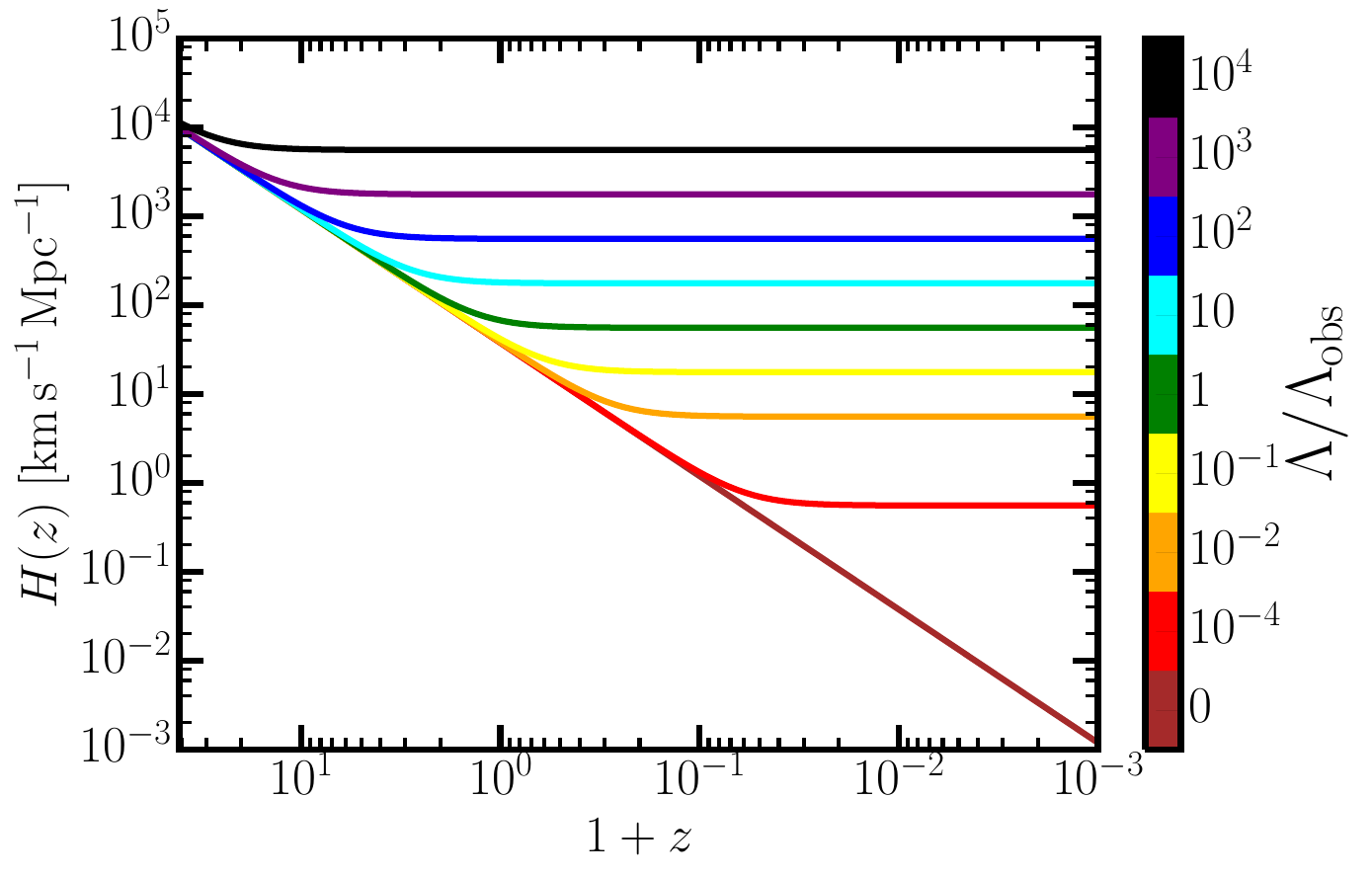}
    \caption{Evolution of the Hubble parameter $H(z)$ for different re-scalings of the cosmological constant. For larger values of $\Lambda$, the asymptotic value of the Hubble constant in the far future of the universe (negative redshift) increases. In an EdS universe, the Hubble constant becomes indefinitely small as the cosmic time tends to infinity.}
    \label{fig:hubble}
\end{figure}

To begin with, we need to understand how the empirical $z=0$ cosmological parameters such as $H_0$ are altered when we vary $\Lambda$. We will restrict our discussion to arbitrary \textit{flat} \LCDM universes only -- implicitly assuming an inflationary multiverse. In addition, all the dimensionless parameters of the \LCDM model are assumed to be unchanged: the horizon-scale amplitude $A_s$; the ratio of baryonic and CDM densities; the ratio of photon and baryon number densities. The consequence is that all model universes under study are indistinguishable copies of our own universe at very high redshifts where $\Lambda$ is dynamically unimportant. But in endowing them with different values of $\Lambda$, the histories of these copies become diverse at later times. The reason for this restriction is the measure problem: for $\Lambda$ we can appeal to Weinberg's argument for a flat prior, but if other parameters were to vary then we have little idea what the appropriate prior would be. Therefore, we concentrate on this simplest ensemble.

It is convenient to quantify these alternative models using the standard set of parameters for the \LCDM model: $\Omega_{\rm m}$, $\Omega_{\Lambda}$, etc. We will add the subscript `ref' when we indicate the corresponding parameters in our own universe. This requires a definition of the `present', which we take to be the point at which the CMB temperature equals the standard value ($T_0=2.7255 \K$; see \citealt{Planck18}). We normalise the scale factor such that $a=1$ at this point, so that the radiation density in all models scales $\propto a^{-4}$, with the identical constant of proportionality in all cases. However, changing the value of $\Lambda$ means that the time corresponding to $T=T_0$ and the Hubble parameter at that point take values that are different from those in the observed universe, as we now explain.

We start by noting that the Friedmann equation for a flat universe applies in all cases, so at times relevant for star formation we have
\begin{equation}
\label{eq:H_def}
    H^2 = H_0^2 (\Omega_{\rm m} a^{-3} + \Omega_\Lambda)\, .
\end{equation}
We preserve the matter density and re-scale the vacuum density by a factor $\alpha_{\Lambda} = \Lambda/\Lambda_{\rm ref}$, where $\Lambda_{\rm ref}$ is the cosmological constant in the reference cosmology, so that $\rho_{\rm m} = \rho_{\rm m, \, ref}$ and $\rho_{\Lambda} = \alpha_{\Lambda} \rho_{\Lambda, \, \rm ref}$ in all universes. The Hubble parameter in the generic universe of the ensemble can be then be re-written as:
\begin{equation}
\label{eq:H_rescale}
    H^2 = H_{0,\rm ref}^2 \Omega_{\rm m, ref} a^{-3} + \alpha_\Lambda H_{0,\rm ref}^2  \Omega_{\Lambda, \rm ref}\, .
\end{equation}
Writing the ratio of the two terms on the right hand side of equations~\eqref{eq:H_def}-\eqref{eq:H_rescale}, and remembering that flatness always ensures $\Omega_\Lambda =1-\Omega_m$, we deduce
\begin{equation}
    \frac{\Omega_{\rm m}}{1-\Omega_{\rm m}} = \frac{ \Omega_{\rm m, \, ref}}{\alpha_{\Lambda} \, (1-\Omega_{\rm m, \, \rm ref})}\, .
\end{equation}
But the unchanged matter density tells us that
$H_0^2=\Omega_{\rm m,\ ref} H_{0,\rm ref}^2/ \Omega_{\rm m}$, so that
\begin{equation}
\label{eq:H_alt}
    H_0^2= H_{0,\rm ref}^2 [\alpha_{\Lambda} + (1-\alpha_{\Lambda})\Omega_{\rm m,\ ref}] \, .
\end{equation}
Hence we readily obtain the new Hubble parameter and density parameter at $a=1$, for a given scaling of $\Lambda$. The cosmic baryon mass fraction is taken as unchanged. Similar reasoning was given by \cite{BK_2022}.

This argument applies for any sign of $\alpha_\Lambda$, but rescaling to negative values complicates things. The expansion ceases at some maximum value of $a$ and the universe subsequently undergoes recollapse. In principle the SP21 code should be able to handle such changes, although in practice it will require amendment in order to cope with a non-monotonic $a(t)$ relation. But in any case, the physical effect on the CSFRD of a negative $\Lambda$ is qualitatively different to that of a positive value. In the latter case, a large value of $\Lambda$ is expected to suppress star formation because structure growth ceases before galaxy-scale haloes can be generated. The main aim of the present paper is to make a quantitative estimate of this suppression. In recollapsing models, however, growth does not freeze out and the relevant question is how much star formation can occur in the limited time before the big crunch. We aim to consider this question elsewhere.

In Fig.~\ref{fig:hubble}, we show the resulting redshift evolution of the Hubble parameter for different values of $\alpha_{\Lambda}$. As expected, for larger values of $\Lambda$ the Hubble parameter reaches its asymptotic value $H\to H_0 \sqrt{\Omega_{\Lambda}}$ at earlier redshifts, and this value increases with $\Lambda$.
This reflects the fact that the early phase of all models is
Einstein--de Sitter (EdS), with $H\propto a^{-3/2}$, while $H$ freezes out at the value of the EdS model at the point where $\Lambda$ comes to dominate.

Calculating the new value of $\sigma_8$ and its scaling with $\alpha_{\Lambda}$ is more complicated: $\sigma_8$ is affected both by the new $\Omega_{\rm m}$, which changes the power spectrum shape and evolution, and by the new Hubble constant, which changes the scale $8 \, h^{-1} \, \rm cMpc$. These effects can be allowed for by recalling that conditions at very high $z$ are the same in all models, so that $\sigma(8\, h_{\rm ref}^{-1}{\rm cMpc})$ is known at $z\gg1$. We then compute the scale-dependent correction numerically, using the code \texttt{CAMB} \citep{CAMB_paper, CAMB_code}. This dependence of $\sigma_8$ on $\Lambda$ was studied by \cite{BK_2022}, who showed that as the universe approaches an EdS solution, $\sigma_8$ exhibits an asymptotic behaviour (see their figure 1). For a Planck-2018 cosmology, the asymptotic value  of $\sigma_8$ for low values of $\Lambda$ is 0.6786 (\citealt{BK_2022}). If one increases the value of $\Lambda$ with respect to the fiducial one, $\sigma_8$ reaches a maximum of $\sigma_8 \sim 0.9$ at around $\alpha_{\Lambda}\sim 8$. For larger values of $\Lambda$, it keeps declining. As a reference, for $\alpha_{\Lambda}=1000$ one obtains $\sigma_8 \approx 0.5$.

\begin{figure}
    \centering
    \includegraphics[width=\columnwidth]{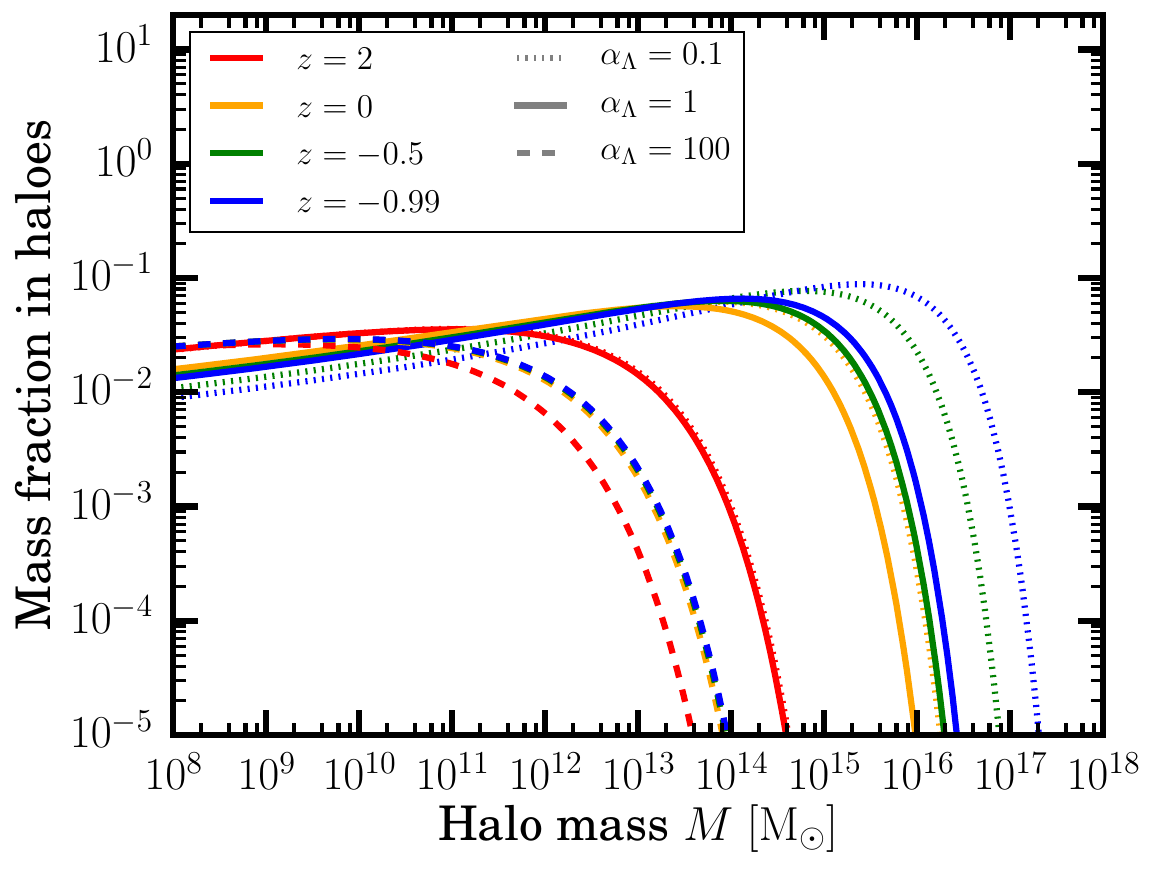}
    \caption{Sheth--Tormen halo mass function at different redshift, represented by different colours, and for different values of $\Lambda$ (different line styles). For a given redshift, larger values of $\Lambda$ result in a mass cut at smaller halo masses. In an EdS universe, the mass cut grows indefinitely in the future ($z<0$). This is not the case for $\Lambda>0$, where the mass cut approaches an asymptotic value as $z\rightarrow -1$ (i.e., $t\rightarrow\infty$). This is a consequence of the freeze out of structure formation in the \LCDM model. }
    \label{fig:hmf}
\end{figure}

We are now fully equipped to study the impact of $\Lambda$ on the CSFRD. The most obvious effect of altering $\Lambda$ is a modification of the cosmological term in the integrand of equation~\eqref{eq:CSFRD}, i.e. the halo mass function. A larger cosmological constant causes the $\Lambda$-dominated  accelerating phase to begin earlier. Consequently, the freeze-out of structure formation would also occur earlier, thus shifting the truncation of the halo mass function to smaller halo masses, and decreasing the merger rate. This is exactly what we observe in Fig.~\ref{fig:hmf}, which plots the halo multiplicity function for different values of $\alpha_{\Lambda}$  and redshift. For a cosmological constant 100 times larger than the observed value (dashed lines), the cut-off in the HMF drops by two orders of magnitude at $z=0$. For reduced values of $\Lambda$, the trend is opposite. At any given redshift, models with $\alpha_{\Lambda}<1$ overproduce massive haloes relative to the reference cosmology. In the limit of an EdS universe (not plotted in the figure), there is no freeze-out and the halo mass cut-off becomes arbitrarily large at sufficiently great times.

To summarise, lower positive values of $\Lambda$ promote structure formation. However, it would be premature to assume that the probability of generating observers is solely dependent on the collapsed mass fraction in the universe, and thus to conclude that the EdS model is anthropically favoured. Rather, we must consider the effect of $\Lambda$ on the astrophysical processes that govern cosmological star formation. This will be the focus of the next section.

\subsection{Astrophysical impact}
\label{sec:astro}

Altering $\Lambda$ affects the astrophysics of star formation because both the gas mass fraction in haloes and the gas cooling rate depend on the virial quantities of haloes, which in turn are sensitive to the background cosmology via the Hubble parameter (see \S\,\ref{sec:formalism}). In this section, we will illustrate these points in detail.

In this work, we will fix the astrophysical parameters to the `best-fit' choice in SP21, i.e.: $\eta=1.9$, $\langle t_* \rangle = 3.87\,\rm Gyr$ and $T_{\rm min} = 10^{4.5} \K$.

\subsubsection{Cooling rate}
\label{sec:rcool}

\begin{figure}
    \centering
    \includegraphics[width=\columnwidth]{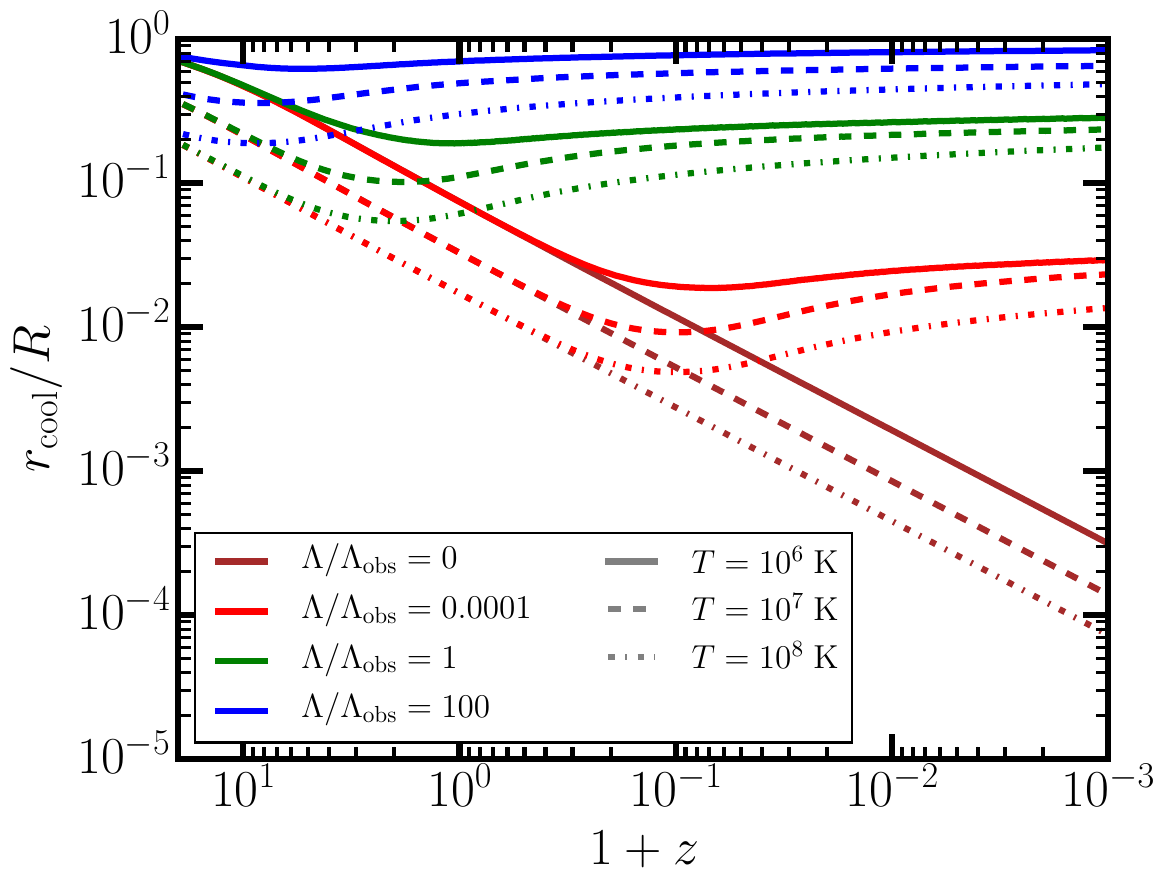}
    \caption{Redshift evolution of the cooling radius $r_{\rm cool}$, in units of the virial radius, at fixed virial temperature. Different colours refer to different values of $\Lambda$, while different line styles represent different virial temperatures. In an EdS universe, the cooling radius shrinks indefinitely in the far future. For $\Lambda>0$, it reaches a minimum value, and then slowly resumes increasing. This is a consequence of the freeze-out of structure formation: see \S\,\ref{sec:rcool} for further details. }\label{fig:rcool}
\end{figure}

As we recalled in \S\,\ref{sec:formalism}, within the SP21 model the low-redshift SFR is set by the gas cooling rate. This is in turn determined by the extent of the cooling radius $r_{\rm cool}$. We will now analyse how the virial temperature of the halo affects the cooling radius, and how this evolves depending on the cosmological model. A particular emphasis will be given to the asymptotic behaviour of the cooling radius for arbitrarily large cosmic times. 

We can express the cooling radius in terms of the cooling time by obtaining the gas density at $r_{\rm cool}$ via equation~\eqref{eq:dens_prof}, and then inserting it into equation~\eqref{eq:tcool}: 
\begin{equation}
    \left( \frac{r_{\rm cool}}{R} \right)^{\eta} = \frac{(3-\eta) \mu X^2 \mathcal{C}(T) M_{\rm gas}}{6\pi k_{\rm B}T m_{\rm H}^2 R^3}\, t_{\rm cool} \, .
\end{equation}
To obtain an explicit dependence of $r_{\rm cool}$ on redshift, we can simply replace $t_{\rm cool}$ with the prescription for the `effective cooling time' given by equation~\eqref{eq:tcool_smooth}, making use of the cosmic time-redshift relationship defined by equation~\eqref{eq:a_t}. We can further apply the definition $M_{\rm gas} = f_{\rm gas}(T,\, z) M$, and then express all virial quantities in terms of $T$ through equations~\eqref{eq:Mvir}--\eqref{eq:Mvir_T}. 
\begin{multline}
\label{eq:rcool_T}
    \left( \frac{r_{\rm cool}}{R} \right)^{\eta} = \sqrt{\frac{\Delta}{2}} \frac{(3-\eta)X^2\mu}{6\pi k_{\rm B} G m_{\rm H}^2 } \frac{\mathcal{C}(T)}{T} f_{\rm gas}(T, \, z) H(z) \\ 
    \times \left[1-E+\left(\frac{t(z)}{f_{\rm dyn} t_{\rm dyn}(z)}\right)^m \right]^{\frac{1}{m}} \, .
\end{multline}
We can therefore follow the redshift evolution of $r_{\rm cool}$; this is shown in Fig.~\ref{fig:rcool}
for different virial temperatures and different values of $\Lambda$.

For a fixed value of $\Lambda$, $r_{\rm cool}/R$ decreases with increasing virial temperature. In haloes above the critical temperature $T_{\rm crit}$ (see \S\,\ref{sec:formalism}), this behaviour can be readily understood from equation~\eqref{eq:rcool_T}. If $T>T_{\rm crit}$, then the gas mass fraction in haloes reaches its maximum value, and no longer depends on the virial temperature. The cooling radius is then simply proportional to $(\mathcal{C}(T)/T)^{1/\eta}$. For $T\gtrsim 10^6 \K$, a \cite{Sutherland_1993} cooling function for a H/He plasma with primordial abundances scales approximately as $\propto T^{1/2}$. Therefore, $r_{\rm cool}/R \propto T^{-1/2 \eta}$, so the cooling radius decreases with temperature in the regime considered. For $T<T_{\rm crit}$, there is an extra temperature-dependence carried by the $f_{\rm gas}(T, \, z)$ factor, but we verified that this does not qualitatively impact the scaling of $r_{\rm cool}$ with temperature.

We can now examine the redshift evolution of $r_{\rm cool}/R$ for a fixed virial temperature, for different cosmological models. To begin with, we restrict the discussion to universes with $\Lambda >0$. At high redshift, the effective cooling time is approximately equal to the dynamical time (see equation~\ref{eq:tcool_smooth}). Thus, the factor containing the cosmic time in equation~\eqref{eq:rcool_T} is very close to unity. Also, the critical temperature drops at sufficiently high redshift (see SP21 and Fig.~\ref{fig:Tcrit}). Therefore, at early times, we have that $f_{\rm gas} \approx f_{\rm b, \, halo} \approx f_{\rm b}$, so that the evolution of the cooling radius is dictated primarily by the Hubble parameter in equation~\eqref{eq:rcool_T}. In this regime, the universe is dominated by matter, thus the cooling radius scales as $H(z)^{1/\eta}  \propto (1+z)^{3/2\eta}$. Recalling that we set $\eta=1.9$, we would then expect that the redshift-evolution of $r_{\rm cool}/R$ to be a power law with index close to $3/4$ at high redshift, where all models are matter dominated. This is exactly what we observe in Fig.~\ref{fig:rcool}. 

Conversely, as the universe becomes  dominated by $\Lambda$, the effective cooling time given by equation~\eqref{eq:tcool_smooth} approaches the cosmic time, so that the cooling radius scales as $\propto (f_{\rm gas} \; t)^{1/\eta}$. To determine the evolution of the cooling radius, we then need to understand how $f_{\rm gas}$ scales with redshift or, equivalently, cosmic time. Because for $\Lambda>0$ the critical temperature is monotonically increasing with cosmic time after the point of $\Lambda$-domination (see Fig.~\ref{fig:Tcrit}), there is always a sufficiently late cosmic time after which the virial temperature of a halo drops below $T_{\rm crit}(z)$. Thereafter, the gas mass fraction within haloes of a given virial temperature scales with $T/T_{\rm crit}(z)$ as in equation~\eqref{eq:fbh_T}. 

The redshift evolution of $T_{\rm crit}(z)$ needs to be computed numerically, but we can find an analytic approximation by simplifying the \cite{Sutherland_1993} cooling function with a piece-wise power law. Under this assumption, the critical temperature itself scales as a power of cosmic time, hence so does the cooling radius (see Appendix~\ref{app:pos_L} for details). We verified that the index of the power law is positive and below unity, explaining why the cooling radius increases after the onset of $\Lambda$-domination, while its time derivative flattens out to zero. This is indeed the trend that we observe in Fig.~\ref{fig:rcool} at $z<0$ for all models with $\Lambda>0$. For a fixed redshift and virial temperature and after the point of $\Lambda$-domination, $r_{\rm cool}/R$ becomes larger as $\Lambda$ increases. To prevent the cooling radius from overflowing beyond the virial radius, we cap the ratio $r_{\rm cool}/R$ to unity with a suitably smooth function, similar to the one used for the nSFR in equation~\eqref{eq:nSFR_smooth}.

 The behaviour that we just described descends from the definition of the boundaries of haloes: for a given virial temperature, it follows from equations~\eqref{eq:Mvir}--\eqref{eq:Mvir_T} that both the virial radius and virial mass of a halo scale as $1/H(z)$. Thus, in a universe with a larger value of $\Lambda$, haloes of a given temperature become more concentrated (see Fig.~\ref{fig:hubble}). Therefore, the gas density becomes overall larger, and consequently cooling becomes more efficient. However, in the far future of the universe structure formation freezes out, and haloes become isolated. Eventually, the SP21 framework assumes that haloes will expel their baryon content through feedback processes, progressively becoming devoid of gas and unable to form stars (see Fig.~\ref{fig:convergence}).

To summarise, we saw that in a universe with $\Lambda>0$ the cooling radius decreases as a power of time at high redshift, but that it eventually switches to evolve in the opposite direction. The point of turnaround between these two regimes occurs at the transition to $\Lambda$-domination, when the effective cooling time in equation~\eqref{eq:tcool_smooth} switches from being approximated  by the dynamical time to tracking the cosmic time. As can be seen in Fig.~\ref{fig:rcool}, increasing $\Lambda$ moves the turnaround of the cooling radius to earlier redshifts. 
In contrast, there is no such turnaround in an EdS universe, and the cooling radius (in units of the virial radius) shrinks indefinitely as time goes by. Because there is no lower limit to the Hubble parameter, the size of haloes with a given virial temperature can become arbitrarily large. This makes the diffuseness of haloes and consequent inefficiency of gas cooling more dramatic than in any universe with positive $\Lambda$.

\subsubsection{Gas mass fraction in haloes}
\label{sec:fb_haloes}

\begin{figure}
    \centering
    \includegraphics[width=0.5\textwidth]{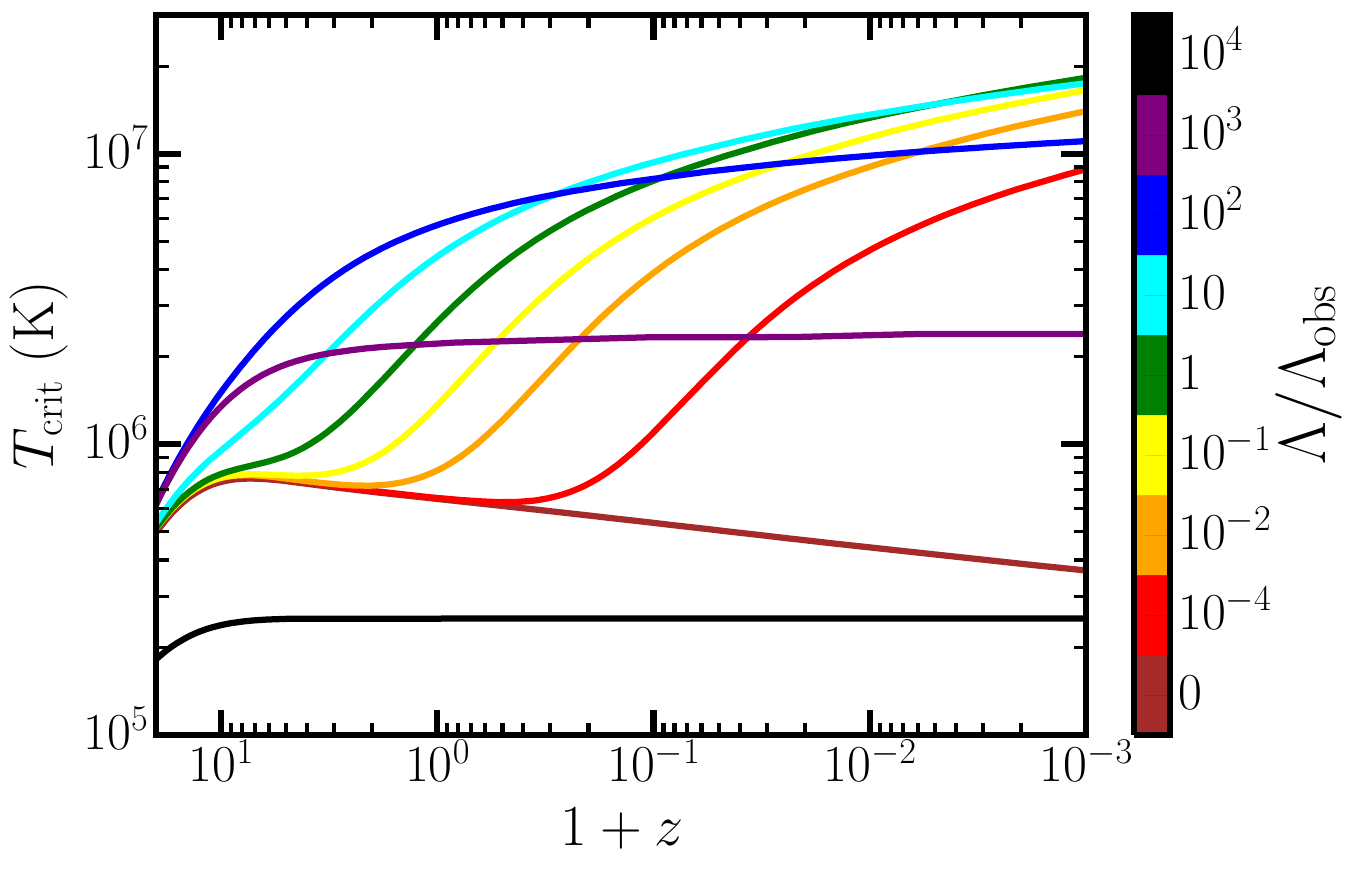}
    \caption{Redshift evolution of the critical virial temperature $T_{\rm crit}$ above which the baryon mass fraction in haloes saturates to the cosmic value $f_{\rm b}=\Omega_{\rm b}/\Omega_{\rm m}$, according to the SP21 model, for different values of the cosmological constant (see colour bar). The scaling of $T_{\rm crit}$ with redshift mirrors the relationship between cosmic time and redshift. In an EdS universe, the long-term  behaviour resembles a power law. As $\Lambda$ increases, $T_{\rm crit}(z)$ deviates from a power law at progressively earlier redshifts (see \S\,\ref{sec:fb_haloes} for further details).}\label{fig:Tcrit}
\end{figure}

\begin{figure*}
    \centering
    \includegraphics[width=\textwidth]{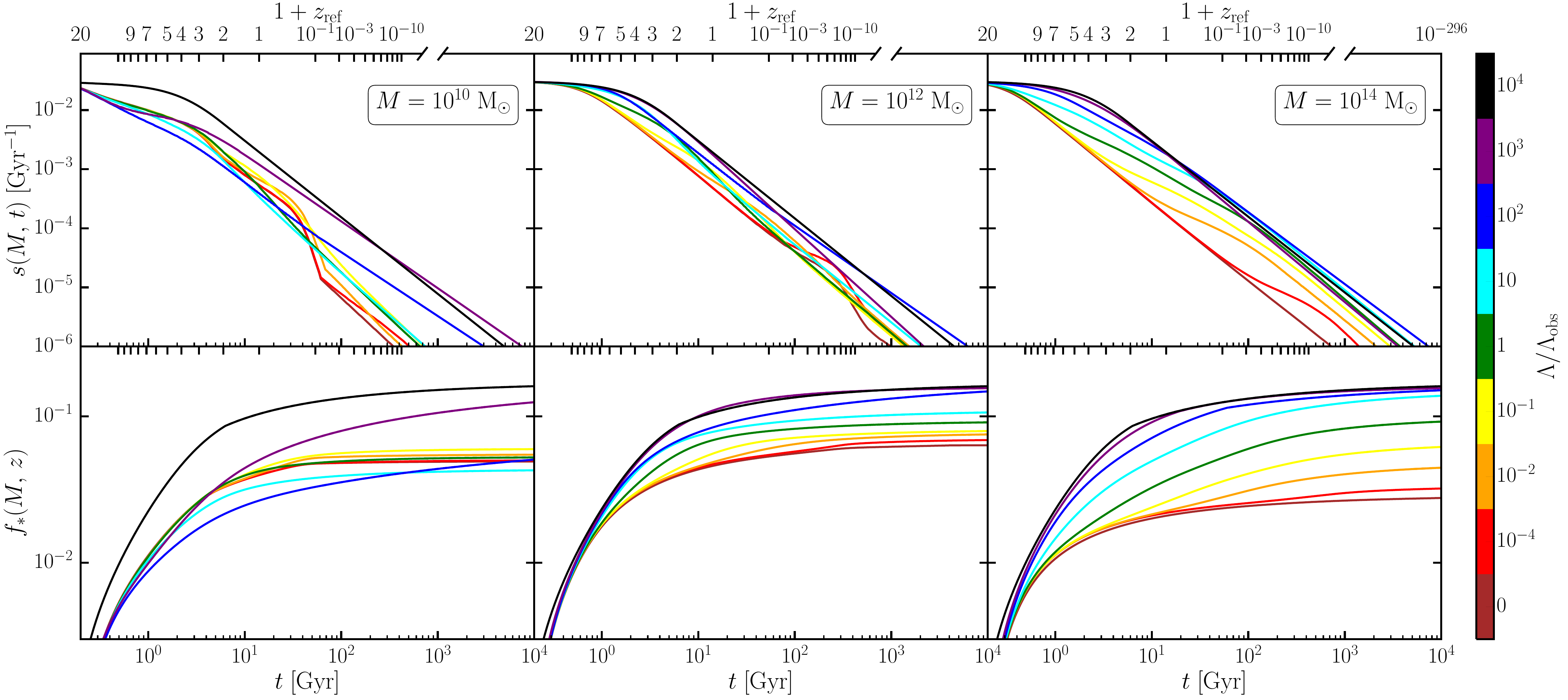}
    \caption{\textit{Upper panels}: Time evolution of the SFR, normalised by the total halo mass. Each panel refers to a population of haloes with fixed virial mass $M$, as annotated in the figure. Different colours refer to different values of $\Lambda$ in units of the observed value $\Lambda _{\rm obs}$, as reported in the colour bar. The lower $x$-axis reports the cosmic time, while the upper $x$-axis the corresponding redshift that would be observed in our universe ($\Lambda = \Lambda_{\rm obs}$).  The break in the upper $x$-axis indicates that the ticks are omitted for $z>-1+10^{-10}$, since the exponential growth of the scale factor in the future makes it increasingly hard to represent the redshift on a meaningful scale. Regardless of the halo mass, the normalised SFR tends to zero in the limit of $t\rightarrow \infty$. \textit{Lower panels}: Fraction of the stellar mass produced up to a certain cosmic time, with respect to the total halo mass, for haloes with fixed virial mass. This is the time integral of the quantity shown in the corresponding upper panels. The colour coding and the upper and lower $x$-axes are the same as in the upper panels. In all haloes and cosmological models considered, the cumulative stellar mass fraction reaches an asymptotic value for $t \rightarrow\infty$.
    }\label{fig:nSFR}
\end{figure*}

As explained in section~\ref{sec:formalism}, the SP21 model assumes that the baryon mass fraction in haloes scales with the virial temperature below a certain critical threshold $T_{\rm crit}$, and saturates to $f_{\rm b}$ above this value. In our improved variation laid out in \S\,\ref{sec:generalise}, the \textit{gas} mass in haloes above the critical temperature saturates to $f_{\rm b}-f_*(T, \, z)$, where $f_*(T, \, z)$ is the stellar mass fraction of a halo of virial temperature $T$ at redshift $z$. In either case, the physical significance of the critical temperature is the same.

In haloes with $T<T_{\rm crit}$ the gaseous component is gravitationally bound only within a certain critical radius $r_{\rm crit}$. Beyond this scale, the momentum injected by supernovae-driven winds into the surrounding gas is sufficient to overcome gravity and eject the gas. In practice then, $T_{\rm crit}$ is the temperature at which $r_{\rm crit}$ coincides with the boundary of the star-forming region within the halo. At high redshift, this is the entirety of the halo, so that $T_{\rm crit}$ is determined by the condition $r_{\rm crit}<R$. At low redshift, star formation is cooling-driven, and the relevant condition is $r_{\rm crit}<r_{\rm cool}$. 

Since the cosmological model affects both the virial radius and the cooling radius (see \S\,\ref{sec:rcool}), the critical temperature is also sensitive to the cosmological parameters. In particular, in Fig.~\ref{fig:Tcrit} we show the redshift evolution of $T_{\rm crit}$ for different values of $\alpha_{\Lambda}$. The late-time behaviour of the critical temperature can be understood starting from the condition $r_{\rm crit}=r_{\rm cool}$, which defines $T_{\rm crit}$. As discussed in the previous section, the cooling radius explicitly depends on the cooling time; in the far future of the universe, $r_{\rm cool}$ corresponds to the point at which the cooling time is approximately equal to the cosmic time.

In a universe with $\Lambda>0$, under the reasonable approximation of the cooling function as a piece-wise power law, the critical temperature scales as a power of cosmic time in the limit $t \rightarrow \infty$ (see Appendix~\ref{app:pos_L} for details). We verified that the index of this power law is positive in all physically relevant cases. Thus, for $\Lambda>0$, the critical temperature increases at arbitrarily large times. The mapping between the scale factor and cosmic time becomes exponential for $t\rightarrow \infty$ in models with $\Lambda>0$. Therefore, in the far future, the critical temperature scales with \textit{redshift} as a power of $-\ln (1+z)$, and that is why the increase of $T_{\rm crit}$ with redshift is rather slow (see Fig.~\ref{fig:Tcrit}).

In an EdS universe, we observe a qualitatively different behaviour for the critical temperature, which decreases rather than increasing at late times. This can be understood by considering that for a fixed virial temperature, the virial radius scales $\propto H(z)^{-1} = (1+z)^{-3/2}$, hence diverging for $z\rightarrow -1$. Therefore, all haloes of a given $T$ become arbitrarily large in the far future, so that their baryonic mass fraction must eventually match the cosmic value. Within the SP21 formalism, this means that all haloes must asymptotically exceed the critical temperature. This condition is achieved if $T_{\rm crit} \rightarrow 0$ for $t \rightarrow \infty$, in accord with our numerical results.

At the high-redshift end of Fig.~\ref{fig:Tcrit}, all models start out with the same $T_{\rm crit}(z)$ relation, because all universes are matter dominated at early times. As we saw in \S\,\ref{sec:rcool}, $r_{\rm cool}$ in the far future  increases for larger values of $\Lambda$ (see Fig.~\ref{fig:rcool}). The critical temperature increases accordingly for $\alpha_{\Lambda} \leq 100$. For $\alpha_{\Lambda}>100$, the critical temperature appears to invert this trend, and although it still increases at later times, its value is overall lower than in universes with a smaller cosmological constant. The reason is that in universes with $\alpha_{\Lambda} \gg 1$,
haloes reach overall higher densities: for a fixed halo mass $M$ (or, equivalently, for a fixed virial temperature at future times), the virial radius scales as \smash{$R \propto M^{1/3} H(z)^{-2/3} \propto M^{1/3} \alpha_{\Lambda}^{-1/3}$}. A higher gas density makes the cooling process more efficient (see equation~\ref{eq:tcool}), so that the cooling radius extends to the virial radius. This means that in practice the condition for determining the critical temperature, $r_{\rm crit} < r_{\rm cool}$, translates to $r_{\rm crit}<R$. It easily follows that $T_{\rm crit}(z) \propto H(z)^{-2}$, regardless of the value of $\eta$. At late times, $H(z)^2 \propto \alpha_{\Lambda}$, therefore the critical temperature diminishes as $\Lambda$ increases (see Fig.~\ref{fig:hubble}).

The behaviour of $T_{\rm crit}$ shown in Fig.~\ref{fig:Tcrit} directly affects the nSFR both at high and low redshift, as it affects the evolution of the baryon mass fraction in haloes. Piecing together the discussion in \S\,\ref{sec:rcool} and in the present section, we gain two important physical insights into the astrophysical impact of $\Lambda$ on the SFR. For small values of $\Lambda$, there is a larger gas mass fraction available in haloes at earlier times, but a progressively smaller fraction will cool down at later times. For larger values of $\Lambda$, there is an overall smaller mass fraction of gas that resides within haloes at later times, but a larger fraction will cool down as time goes by. These features will play a prominent role in shaping the dependence of the cosmic star formation efficiency as a function of $\Lambda$, as we will discuss in \S\,\ref{sec:future}-\ref{sec:efficiency}.

\section{Results}
\label{sec:results}

\subsection{Long-term efficiency of star formation}
\label{sec:future}

In this section, we will show our results for the star formation history for different values of $\Lambda$, focusing especially on the long-term behaviour of star formation. Determining the asymptotic behaviour of the nSFR will tell us whether there is a well defined final cumulative efficiency of cosmic star formation -- both in individual haloes and in the universe as a whole.

Let us then start with the evolution of the nSFR. In the original SP21 formalism, the nSFR is most naturally expressed for a fixed virial temperature, as that is the independent variable in the cooling function. Nevertheless, as we explained earlier in \S\,\ref{sec:formalism}, the Sheth--Tormen formalism for the clustering of haloes allows us to convert straightforwardly between the fixed-$T$ and the fixed-$M$ views, making use of the definitions expressed in equations~\eqref{eq:Mvir}--\eqref{eq:Mvir_T}. For the remainder of this paper, the fixed-$M$ view will be more convenient.

In the upper panels in Fig.~\ref{fig:nSFR} we show the time evolution of the nSFR for populations of haloes at given virial mass. From left to right, we show the results for a typical dwarf galaxy ($10^{10} \MSun$), a Milky-Way-like galaxy ($10^{12} \MSun$), and a galaxy cluster ($10^{14} \MSun$), respectively. The main independent variable in the plot is cosmic time; the upper $x$-axis shows the redshift that would be measured by an observer \textit{in the reference universe} for a given cosmic time. Therefore, it applies only to the case $\alpha_{\Lambda}=1$. 

We note that for all halo masses and all cosmologies, the nSFR tends to a universal value at sufficiently early times. This happens because, for a given virial mass, the virial temperature increases at earlier times, while the critical temperature is lower (see Fig.~\ref{fig:Tcrit}). There will thus always be a sufficiently early time when the baryon mass fraction within haloes saturates to $f_{\rm b}$. Equation~\eqref{eq:nSFR_high} then tells us that the nSFR reaches a fixed constant value in this regime. 

\begin{figure}
    \centering
    \includegraphics[width=\columnwidth]{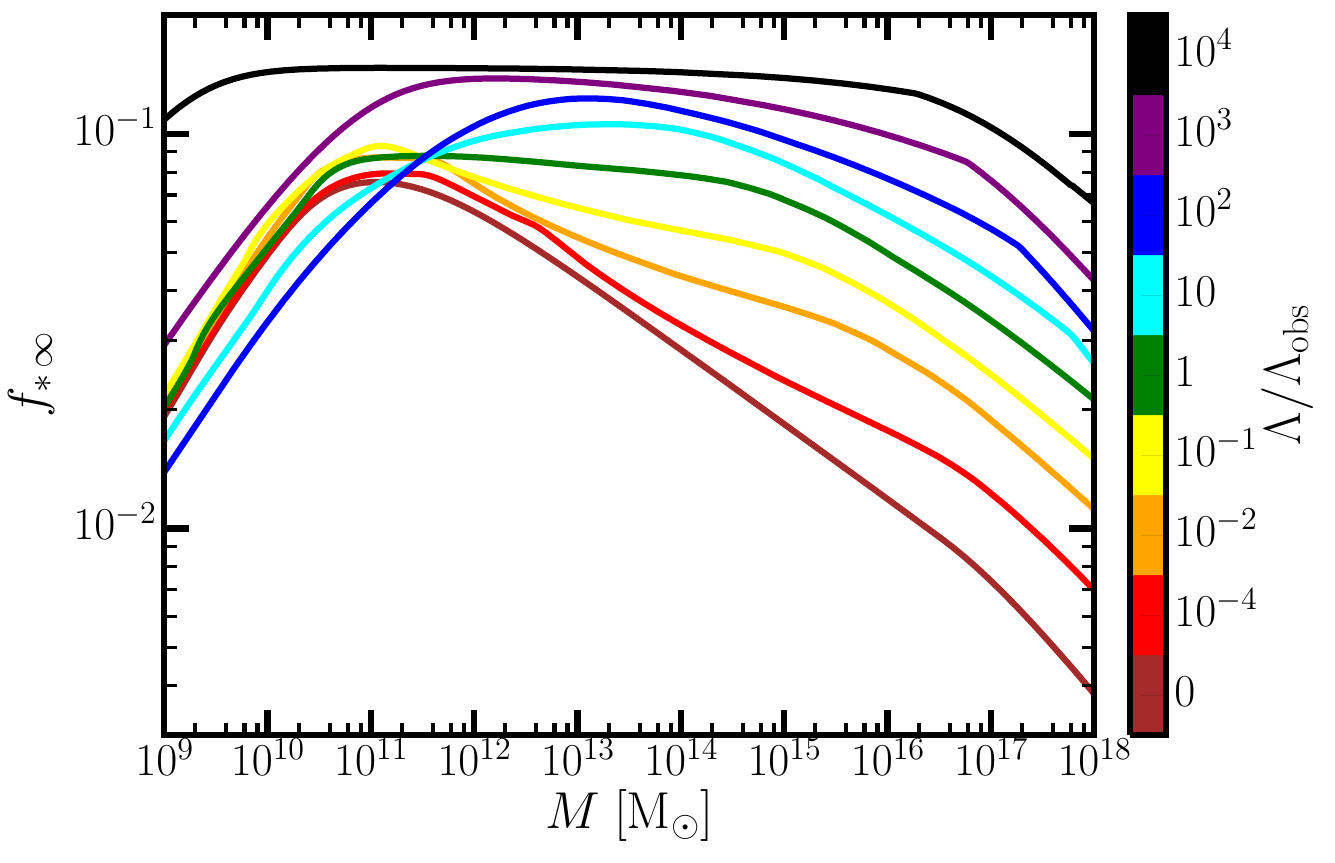}
    \caption{Asymptotic value (i.e., in the limit $t\rightarrow \infty$) of the stellar mass fraction for a halo population with given total mass $M$. Different colours refer to different values of $\Lambda$, as indicated in the colour bar. Even for large values of $\Lambda$, massive haloes can still be very efficient (see \S\,\ref{sec:future} for the detailed explanation).}\label{fig:efficiency}
\end{figure}

Since all universes in our ensemble are identical and matter-dominated at a sufficiently high redshift, they are naturally indistinguishable at early times. But for all halo masses considered, the nSFR drops steeply in the future, and the departure from the EdS solution occurs earlier for universes with a larger value of $\Lambda$. The asymptotic behaviour can be deduced by studying each factor in the r.h.s. of equation~\eqref{eq:nSFR_low}. As discussed in \S\,\ref{sec:astro}, the main challenge is understanding the late-time behaviour of the critical temperature, which needs to be determined numerically. However, under the reasonable assumption of a piece-wise cooling function, this asymptotic behaviour can be approximated analytically with a power of time. The index of the power law depends on the virial temperature of the halo and on the slope of the gas density profile. It is then possible to prove that the time integral of the nSFR at fixed mass (i.e., the stellar mass fraction) is convergent (although the argument is relatively detailed, and is presented in  Appendix~\ref{app:math}). This convergence holds both in an EdS universe and for $\Lambda>0$, and is seen in Fig.~\ref{fig:nSFR}, where the cumulative stellar mass fraction reaches a plateau at late times. We plot this asymptotic stellar mass fraction, $f_{*\, \infty}$, as a function of halo mass in Fig.~\ref{fig:efficiency}.

For all values of $\Lambda$ considered, we notice that $f_{*\, \infty}$ increases with mass for very low masses.
For $\alpha_{\Lambda} <1$, the efficiency then peaks at $M\approx10^{11} \MSun$, and then decreases for larger halo masses, with the drop being faster in an EdS universe. But even for still relatively small values of $\Lambda$, such as $\alpha_{\Lambda}=0.1$, the asymptotic efficiency decreases more slowly after reaching its maximum, and for $M\approx 10^{18} \MSun$ it is almost one order of magnitude larger than in an EdS universe. 
For $\Lambda$ at the observed level or above, the initial decline in efficiency above $M=10^{11} \MSun$ is rather slow, with a plateau of nearly constant efficiency, followed by an eventual sharper drop. In the case of $\alpha_{\Lambda}=10^4$, $f_{* \, \infty}$ exhibits little variation over the nine decades of  halo mass shown in Fig.~\ref{fig:efficiency}. In such an extreme cosmology, the freeze out of structure formation occurs very early and the critical temperature is so low (see Fig.~\ref{fig:Tcrit}) that most haloes retain their cosmic share of baryons; thus they are always efficient at forming stars, regardless of their mass.

The absence of a sharp peak in the stellar mass fraction and subsequent steep decline at the higher mass end might seem puzzling, at least for the case $\alpha =1$, as it appears to contradict the current paradigm of galaxy formation \citep[see e.g.][]{Behroozi_2019}. However, we stress that Fig.~\ref{fig:efficiency} shows the \textit{asymptotic} stellar mass fraction (i.e., in the limit $t\rightarrow \infty$), whereas the consensus on the mass dependence of the stellar mass fraction is based on observations and models of the past star formation history. We verified that our formalism would also predict a sharper peak and a steeper decline of the stellar mass fraction for $M\gtrsim 10^{12} \, M_{\odot}$ at $z=0$, in qualitative agreement with the accepted framework for galaxy formation. Therefore, Fig.~\ref{fig:efficiency} tells us that, according to our model, high-mass haloes produce a larger fraction of their asymptotic stellar mass in the future of the universe compared to low-mass haloes. Higher-mass haloes appear to be especially efficient at forming stars in cosmologies with larger values of $\Lambda$, where haloes of a given virial mass are overall denser, hence more efficient at cooling (see \S~\ref{sec:fb_haloes}). While this argument provides us with a basic understanding of the trend observed in Fig.~\ref{fig:rcool}, one should bear in mind that gas cooling is not the only relevant process for star formation, even at late times. While we incorporate the SP21 model of supernovae-driven winds (see their Appendix A) in our calculations, we do not include an explicit self-regulation mechanism due to AGN. This may have a noteworthy impact on the future star formation rate. We will discuss the subsequent implications for our conclusions in \S~\ref{sec:limitations}.

 \begin{figure*}
    \centering
    \includegraphics[width=\textwidth]{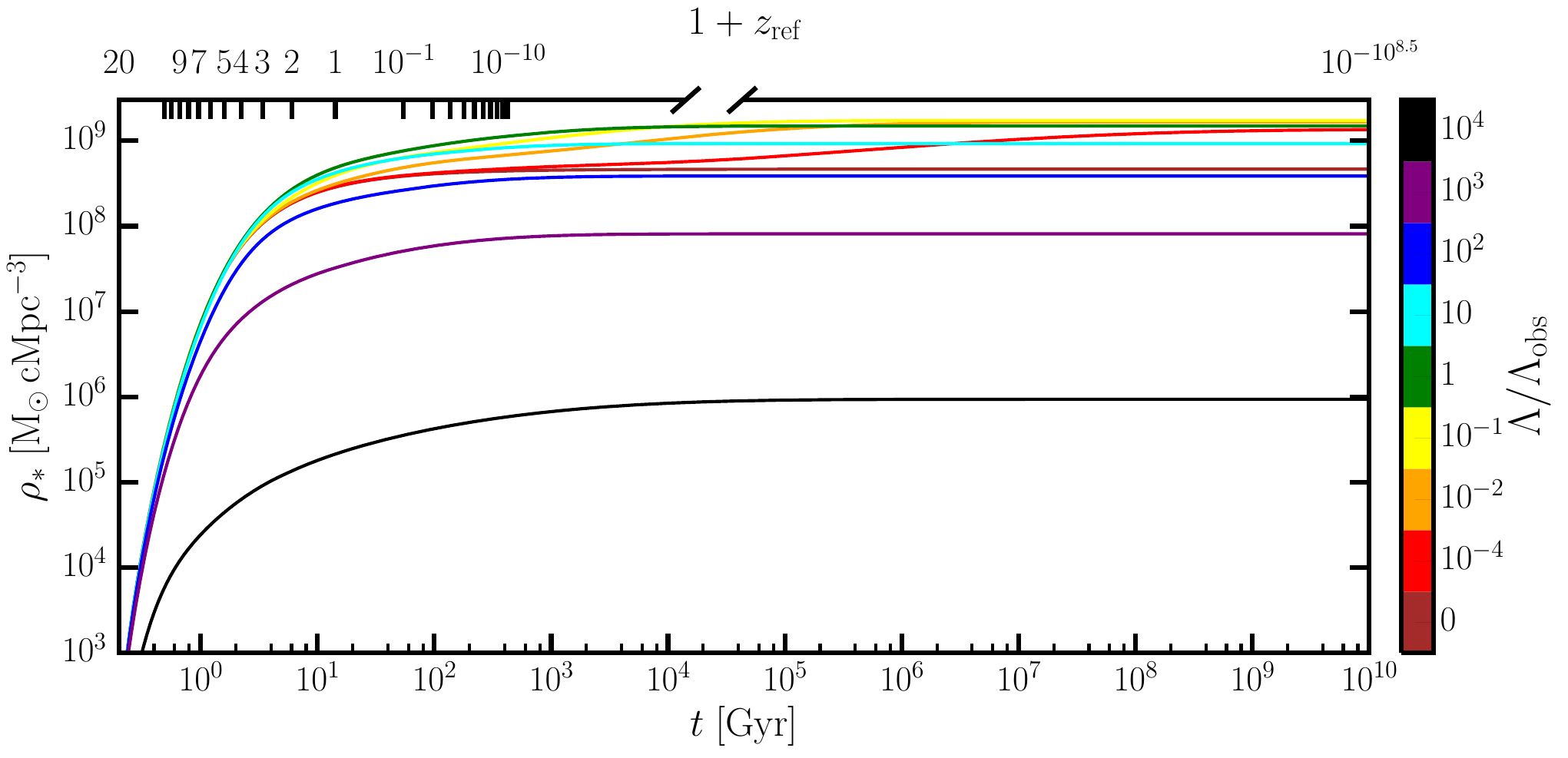}
    \includegraphics[width=\textwidth]{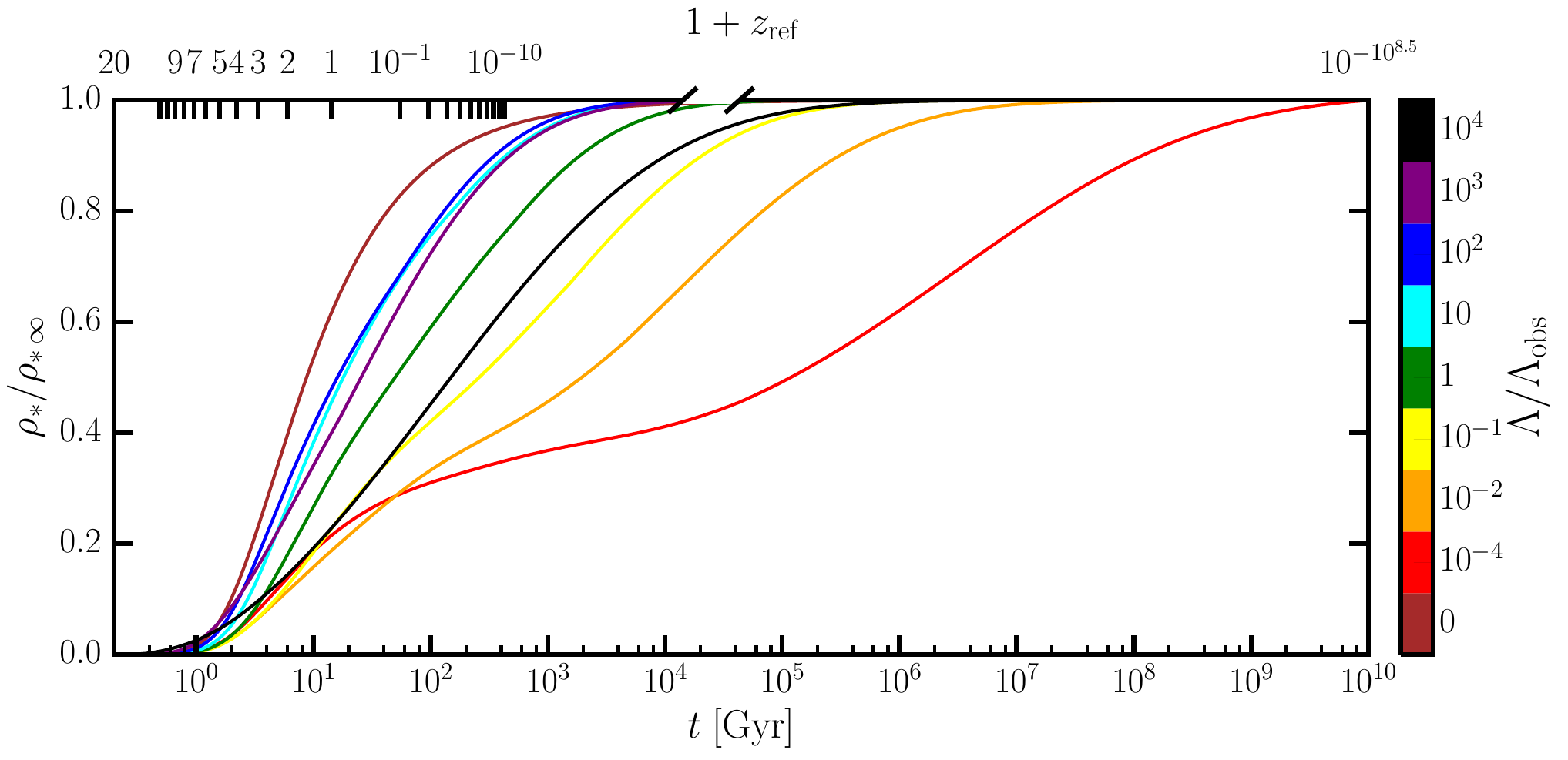}
    \caption{\textit{Upper panel}: Cumulative stellar mass produced in a unit comoving volume up to a certain cosmic time. Different lines correspond to different values of $\Lambda$, as indicated in the colour bar. The upper $x$-axis shows the redshift that corresponds to the cosmic time, for the reference cosmology. Therefore, the correspondence between the upper and lower $x$-axes applies only to the case $\Lambda = \Lambda_{\rm obs}$. The break in the upper $x$-axis indicates that the ticks are omitted for $z>-1+10^{-10}$, since the exponential growth of the scale factor in the future makes it increasingly hard to represent the redshift on a meaningful scale. \textit{Lower panel}: Same as in the upper panel, but normalised to the total stellar mass density produced over the entire history of the universe. The stellar mass fraction produced up to present time in our universe is $\sim$\, 32\%, hence we are typical observers. However, in general the mid-point of star formation history does not coincide with $\Lambda$-domination in other universes, meaning that the `why-now' problem is specific to $\Lambda = \Lambda_{\rm obs}$ (see discussion in \S\,\ref{sec:why-now}).    }\label{fig:rho*_frac}
\end{figure*}

It is important to bear in mind that Fig.~\ref{fig:efficiency} ignores the cosmology dependence of halo multiplicities. We then need to weight the efficiencies shown in Fig.~\ref{fig:efficiency} by the HMF. If we then further integrate over halo mass, that gives us the the stellar mass density (SMD) of all stars formed up to time $t$ from the time corresponding to the onset of star formation $t_{\rm in}$. This quantity is by definition the time integral of the CSFRD or, in terms of redshift:
\begin{equation}
\label{eq:rho*}
    \rho_* (z) = \int_z ^{z_{\rm in}} \frac{\dot{\rho}_*(z')}{(1+z')H(z')} dz' \,
\end{equation}
where $\dot{\rho}_*(z)$ is given by equation~\eqref{eq:CSFRD}. To study the asymptotic behaviour of $\rho_*(z)$, we should thus study the convergence of the integrand, which depends both on the nSFR and the HMF. We have already discussed the convergence of the nSFR for $\Lambda>0$ above, and we now focus on the behaviour of the HMF for $t\rightarrow \infty$. In a universe with $\Lambda>0$, the asymptotic behaviour of $\sigma(M, \, z)$ is a non-null constant, for any $M$. Therefore, the integral in equation~\eqref{eq:rho*} converges as long as the stellar mass fraction converges -- and we have already seen that this quantity does indeed asymptote to a finite constant in the far future. In an EdS model, 
the HMF continues to evolve indefinitely, but in fact the total stellar density still converges (see Appendix~\ref{app:zero_L} for the detailed argument).

The asymptotic study of the CSFRD described above also allows us to define physically motivated analytic approximations for how the CSFRD approaches its asymptotic value in the far future (see Appendix~\ref{app:math} for details). The results of this exercise are shown in the upper panel of Fig.~\ref{fig:rho*_frac}, plotting the stellar density as a function of time for various values of $\alpha_{\Lambda}$. 
In an EdS universe, the asymptotic SMD is reached at $t\sim 100\,\rm Gyr$; but as $\Lambda$ is increased from zero the
asymptotic SMD becomes larger, because now the cooling radius will not shrink indefinitely at late times, hence increasing the star formation efficiency of haloes. However, for $\alpha_{\Lambda} > 0.1$, the trend is reversed. This might seem somewhat at odds with Fig.~\ref{fig:efficiency}, which shows that the stellar mass fraction within haloes keeps increasing for $\alpha_{\Lambda} >  0.1$. However, Fig.~\ref{fig:efficiency} focuses on isolated haloes, without accounting for their multiplicity. Instead, the stellar production needs to be weighted by the HMF when computing the SMD. Thus, Fig.~\ref{fig:rho*_frac} tells us that the lower asymptotic SMD for large values of $\Lambda$ is a consequence of the suppression of structure due to the cutoff of the HMF occurring at lower masses.

In the lower panel of Fig.~\ref{fig:rho*_frac}, we normalise the SMD by the asymptotic value. For the observed value of $\Lambda$ (green line), about 32\% of the stellar mass that will ever exist in our universe has already been formed. In the original SP21 paper, this quantity was quoted to be close to 50\%, but in that case the SMD was computed differently. Namely, SP21 did not go through the detailed convergence study undertaken here, but simply integrated the CSFRD up to $z=-0.9999$, which corresponds to $t\sim 200\,\rm Gyr$ in our universe; Fig.~\ref{fig:rho*_frac} teaches us that this is not always sufficiently late to reach the true asymptotic stellar mass density.

\begin{figure*}
    \centering
    \includegraphics[width=\textwidth]{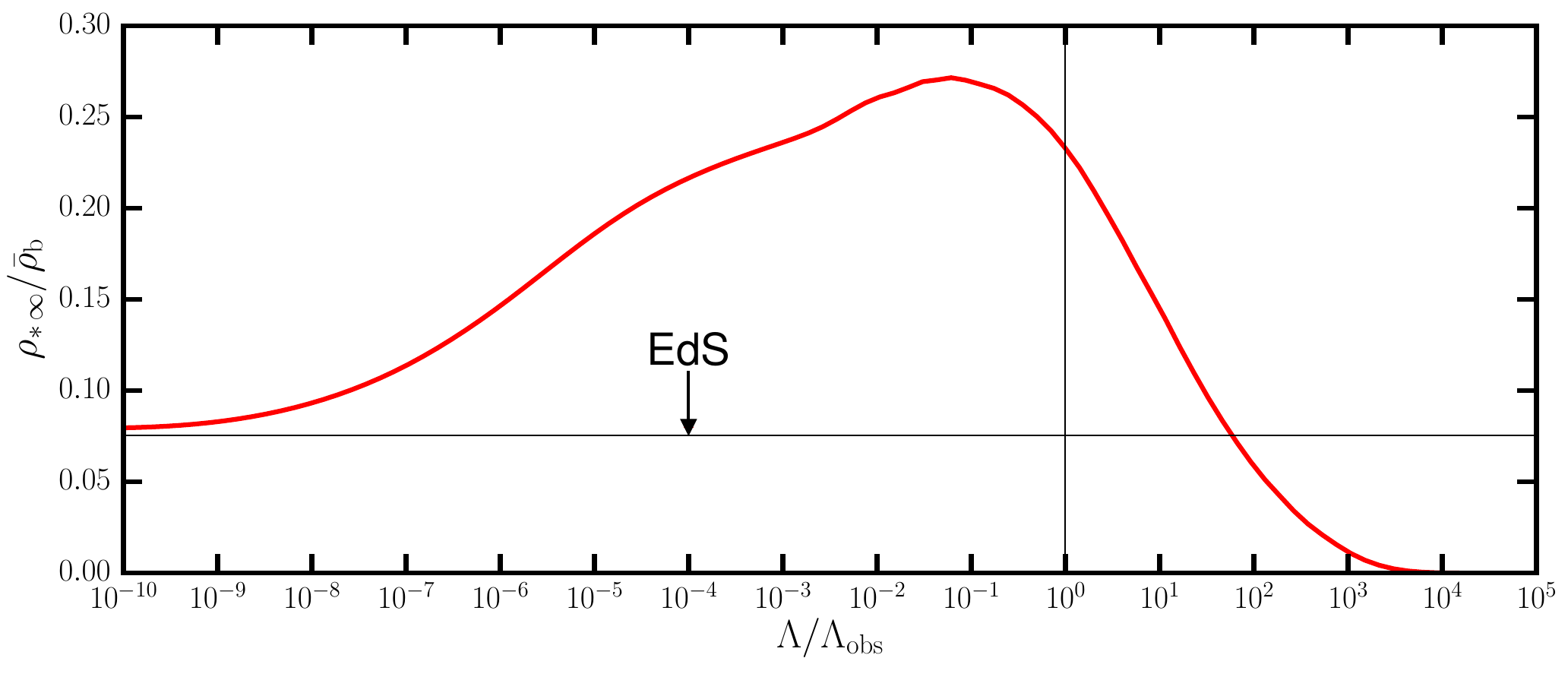}
    \caption{\textit{Right panel}: Fraction of baryonic mass (per unit comoving volume) that is converted into stars over the entire history of the universe, as a function of $\Lambda$. This global stellar efficiency peaks around one tenth of the observed value of $\Lambda$ (thin black vertical line), and becomes negligibly small for large values of $\Lambda$. This is due to the suppression of cosmological structure formation caused by the earlier and larger acceleration of the expansion of the universe. The horizontal thin black line represents the global stellar efficiency in an EdS universe, which is smaller than in our universe. The decrease in the cosmic star formation efficiency at small positive $\Lambda$ is driven by astrophysical rather than cosmological factors: haloes of a given virial mass become larger as $\Lambda$ decreases, and the consequent reduction in internal gas density diminishes the gas cooling rate, which in turn determines the star formation rate at low redshift (see discussion in \S~\ref{sec:efficiency}).
    } \label{fig:ms-mb}
\end{figure*}

For small positive $\Lambda$, the \textit{fractional} SMD in the lower panel of Fig.~\ref{fig:rho*_frac} departs from the EdS solution much earlier than for the SMD evolution shown in the upper panel, reflecting the larger asymptotic values reached for $0 <\Lambda <0.1$. Thus, similar values of the SMD at early times can still differ considerably in fractional terms. Instead, for large values of $\Lambda$, haloes are more efficient at producing stars, therefore the asymptotic SMD is reached earlier. Fig.~\ref{fig:efficiency} shows that for $\Lambda \lesssim 10^3$, star formation is driven by higher-mass haloes. For extreme cosmologies such as $\alpha_{\Lambda}=10^4$, haloes are at nearly full star formation efficiency irrespective of their mass, but the cutoff of the HMF at late times occurs at $M\sim 10^9 \MSun$. Recalling that we consider only haloes with $T>10^{4.5} \K$ as eligible for star formation, this means that stars are produced only by haloes in the narrow mass range $10^8 \, \mathrm{M}_{\odot} \lesssim M \lesssim 10^9 \MSun$. Therefore, it takes longer to reach the asymptotic SMD.

In conclusion, if $\Lambda$ is higher, halo formation ceases at early times: haloes are unable to attain masses as large as in our universe (see \S\,\ref{sec:cosmo}), and are less efficient at forming stars. The dominant fraction of stars is formed in the future (i.e., at negative redshift), when haloes are isolated and efficient at forming stars due to their larger cooling rate with respect to their counterparts in our own universe (see \S\,\ref{sec:rcool}). This long-term future of star formation is expected to play an important role in determining the asymptotic efficiency of star formation in universes with a large cosmological constant.

\subsection{Cosmic star formation efficiency}
\label{sec:efficiency}

In the previous section, we showed that the comoving stellar mass density produced in a universe with $\Lambda \geq 0$ asymptotes to a constant in the limit $t \rightarrow \infty$. We now want to understand how this asymptotic density, $\rho_{* \, \infty}$, varies as a function of $\Lambda$. In practice, we consider the cosmic star formation efficiency, defined as $\varepsilon = \rho_{* \, \infty}/\bar{\rho}_{\rm b}$, where $\bar{\rho}_{\rm b}$ is the mean comoving baryon density in the universe.  

We plot this cosmic stellar efficiency in Fig.~\ref{fig:ms-mb}, where we see the striking result that the peak of cosmic star formation efficiency occurs within one order of magnitude of $\Lambda_{\rm obs}$. At the peak efficiency, $\Lambda \approx 0.1 \Lambda_{\rm obs}$, the  model predicts that about 27\% of the baryonic mass in the universe will eventually be converted into stars. In our universe, this figure is 23\%. For larger values of $\Lambda$, the cosmic efficiency declines, becoming negligibly small as $\alpha_{\Lambda}$ approaches $10^5$. 

It might seem surprising that the cosmic stellar efficiency does not fall monotonically as $\Lambda$ increases. A larger cosmological constant would raise the acceleration of the cosmic expansion at earlier times, hence suppressing the gravitational collapse of galaxy-scale haloes; higher values of $\Lambda$ thus correspond to smaller fractions of matter in collapsed structures \cite[e.g.][]{Weinberg_1987, Martel_1998}. However, the picture is more complicated when considering the impact of $\Lambda$ on the fraction of \textit{baryonic} rather than \textit{total} matter within haloes \citep[see, e.g., the discussion in][]{Page_2011}. Furthermore, in Fig.~\ref{fig:ms-mb} we are concerned with the \textit{stellar} component only, which is subject to the complex interplay of astrophysical processes and the evolution of the large-scale structure of the universe. 

The suppression of the star formation efficiency at high $\Lambda$ cannot be attributed to astrophysical processes, since Fig.~\ref{fig:efficiency} shows that high-mass haloes remain highly efficient at producing stars, even for large values of $\Lambda$. Therefore, the small $\varepsilon$ found for large $\Lambda$ must be caused by the cutoff of the HMF at small halo masses (see \S\,\ref{sec:cosmo}). Ultimately, this is a consequence of the earlier freezeout that results from a large cosmological constant. In short, our model predicts that haloes are generally efficient at forming stars when $\Lambda$ is high, but their number density is then so low that the overall effect is a reduction in the cosmic star formation efficiency.

For small values of $\Lambda$, the opposite is true. There is now stronger clustering and haloes can attain larger masses. However, for a given virial mass, lower values of $\Lambda$ correspond to a larger virial radius, given that $R \propto M^{1/3} H_0^{-2/3}$ and $H_0$ is smaller for lower values of $\alpha_{\Lambda}$ (see equation~\ref{eq:H_alt} and discussion in \S~\ref{sec:fb_haloes}). Thus, a halo of a given virial mass has a lower internal density in a universe with a smaller cosmological constant. This reflects the continued progress of mergers, which dilute the internal density of haloes to a multiple of the cosmic density, up to the point where $\Lambda$ dominates. The reduced gas density results in a smaller cooling rate, making star formation less efficient at late times. The cosmic stellar efficiency at low values of $\Lambda$ in Fig.~\ref{fig:ms-mb} is therefore suppressed for astrophysical reasons.

These astrophysics-driven and cosmology-driven suppression effects on star formation balance out around $\alpha_\Lambda\sim0.1$. Fig.~\ref{fig:ms-mb} suggests that the observed value of $\Lambda$ is somewhat peculiar, in the sense that it is close to the optimal value for maximum cosmic star formation efficiency. However, this is not the same as saying that it is the most likely value of $\Lambda$ for observers to experience. In the next Section, we will address the implications of Fig.~\ref{fig:ms-mb} for such anthropic issues.

\begin{figure*}
    \centering
    \includegraphics[width=\columnwidth]{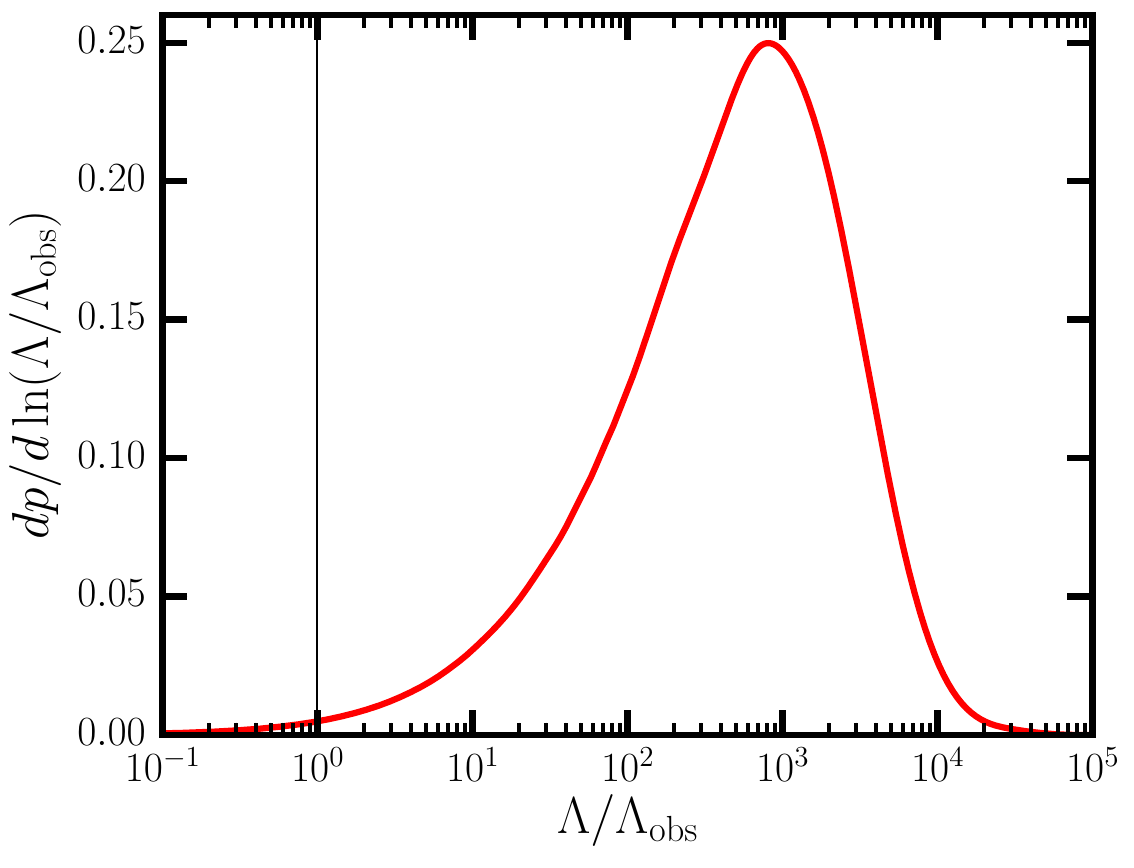}
        \includegraphics[width=\columnwidth]{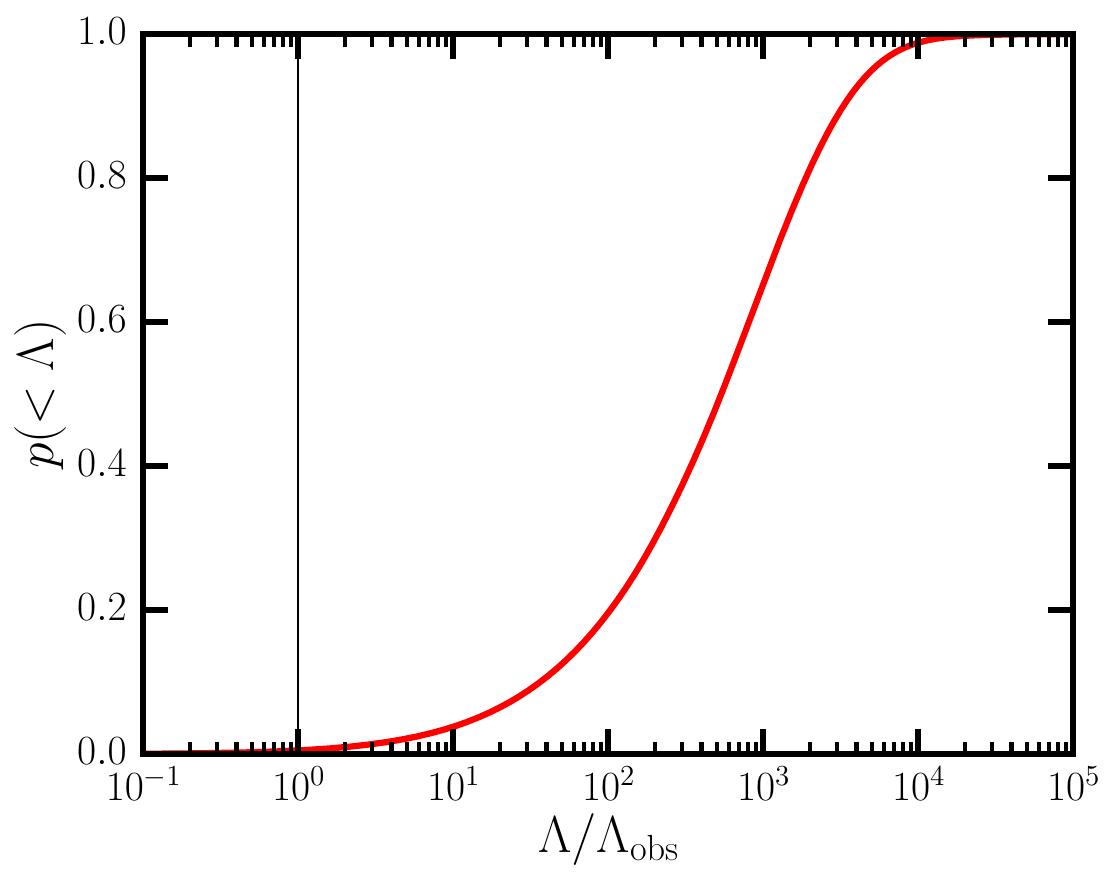}
    \caption{\textit{Left panel}: probability of generating observers in a given universe, at any point in time, as a function of $\Lambda$, assuming a direct proportionality to the stellar efficiency shown in Fig.~\ref{fig:ms-mb} (right panel) and imposing a flat prior on $\Lambda$. The peak occurs around $\Lambda\approx 800 \Lambda_{\rm obs}$. The observed value of $\Lambda$ (thin black vertical line) is an outlier in the distribution. \textit{Right panel}: cumulative probability function of the distribution shown in the left panel. The probability that $\Lambda \leq \Lambda_{\rm obs}$ is $\sim$\,0.5\%, meaning that this value is not anthropically favoured.} \label{fig:prob}
\end{figure*}

\section{Discussion}
\label{sec:discussion}

\subsection{Implications for anthropic arguments}
\label{sec:anthropics}

\subsubsection{Typicality and the `why now' problem}
\label{sec:why-now}

The first aspect of anthropic arguments that we will address concerns the so-called `why-now' problem, i.e. the fact that we live at an usual epoch when the density of matter and dark energy are of the same order of magnitude. The `weak' anthropic approach to this puzzle asserts that typical observers should live when the age of the universe corresponds
roughly to the lifetime of Sun-like stars, assuming these to be necessary for forming and sustaining habitable planets \citep{Carter_1974}. We can test this principle to some extent, by asking how typical we are as observers. Consider the fraction of stellar mass density produced up to present time in our own universe only (green line in Fig.~\ref{fig:rho*_frac}): this predicts that roughly 32\% of all the stars that will ever exist are already in place, so it appears that the present time is indeed typical, lying neither at the beginning nor at the end of the cosmic star formation history.

But this argument neglects the question of how long intelligent life may to take to emerge following the formation of a star. This is an issue on which cosmologists are unlikely to have a reliable opinion. But 
we can note that if the emergence of life is rare, then indeed an offset of the order of the typical stellar lifetime might be expected, and this is the basis of the standard anthropic argument for the current age of the universe. Alternatively, we can follow \cite{Peacock_2007} and eliminate the biological element by noting that the Sun formed at $z\approx0.45$. From our results, we find that $\sim\,$25\% of the asymptotic stellar density was in place at that point, so the Sun is a slight `early adopter' but by no means atypical.

The Sun thus formed almost exactly at the point of matter-$\Lambda$ equality, and the interesting question is whether this coincidence would still have held for a typical star if $\Lambda$ had taken a very different value. It is apparent from the predictions in Fig.~\ref{fig:rho*_frac} that the observed coincidence is not to be expected in all cases. For $\alpha_\Lambda=100$, the 50\% point of stellar production is reached at $t\approx 20\,\rm Gyr$. That corresponds to $1+z\approx 10^{-6}$ in that universe, at which point the density from $\Lambda$ exceeds that of matter by a factor of approximately $10^{14}$. We also note that the predicted CSFRD is not so very different from the observed one when we carry out the prediction for the EdS universe. In that case, there is still a well-defined typical era of star formation, which is clearly not dictated by the cosmological constant.

Our conclusion is then that the why-now puzzle is specific to the actual value of $\Lambda$. Weak anthropic selection that biases observers towards living at a special time in the universe would not have produced
such a coincidence in general.  We are therefore driven to ask if there is some more general selection effect that favours the observation of this particular value of $\Lambda$.

\subsubsection{Ensembles and the observed value of $\Lambda$}

The multiverse approach to  $\Lambda$ considers an ensemble of different universes, each with its own cosmological constant, and argues that only the universes that have a small enough value of $\Lambda$ would be able to host observers. As was emphasised by \cite{Efstathiou_1995}, this situation can be analysed in a Bayesian framework. If we consider the existence of observers as the `data', then the posterior on $\Lambda$ is 
\begin{equation}
\label{eq:bayes}
    p(\Lambda \vert \mathrm{O}) \propto p(\mathrm{O} \vert \Lambda) p(\Lambda) \, .
\end{equation}
In the equation above, $p(\rm O \vert \Lambda)$ is the likelihood, i.e. the probability of observers to emerge for a given value of $\Lambda$, and $p(\Lambda$) is the prior on $\Lambda$. 

Both the prior on $\Lambda$ and the likelihood are non-trivial to determine. The main conceptual difficulty is that one would need to choose a well-defined measure for weighting different members of the ensemble. But for an inflationary multiverse, the bubble universes are formally of infinite volume, and it is therefore not clear how to define the measure (see, e.g., \citealt{Wenmackers_2023}). But if $\Lambda$ is the sole parameter that varies across the ensemble, then the problem can be evaded following \cite{Weinberg_1987, Weinberg_1989, Weinberg_2000_talk, Weinberg_2000}, who pointed out that there is no known physical mechanism that preferences $\Lambda =0$. Therefore, the prior on $\Lambda$ should be flat at least in a neighbourhood of $\Lambda =0$. We will adopt Weinberg's convention of a flat prior, and extend it to the full range of values of $\Lambda$ considered in our work, i.e., $0 \leq \alpha_\Lambda \leq10^5$. 

Regarding the likelihood $p(\rm O \vert \Lambda)$, we will adhere to the view expressed in the previous section that stars are a precondition for observers. Thus, we will simply assume that the likelihood is proportional to the cosmic star formation efficiency $\varepsilon$. We stress that we are agnostic concerning the time at which observers appear, unlike in the previous section: $\varepsilon$ is an integrated quantity over the entire history of the universe.

We can now write the posterior probability density in the interval $(\Lambda,\, \Lambda +  d\Lambda)$ as:
\begin{equation}
    \frac{dp(\Lambda \vert \mathrm{O})}{d\Lambda} \propto \varepsilon(\Lambda) \, 
\end{equation}
or, more conveniently, as the distribution of $\ln \Lambda$:
\begin{equation}
     \frac{dp(\Lambda \vert \mathrm{O})}{d\ln \Lambda} \propto \alpha_{\Lambda} \varepsilon(\alpha_{\Lambda}) \, .
\end{equation}
This probability distribution is shown in the left panel of Fig.~\ref{fig:prob} (see also 
\citealt{Sorini_proceeding}). We notice immediately that the peak corresponds to $\alpha_{\Lambda}\approx 800$: the most likely order of magnitude of $\Lambda$, weighted by observer number across the multiverse, is vastly greater than the observed value. Furthermore, $\alpha_{\Lambda}=1$ (indicated with a vertical black line in the plot) appears to be an outlier of the distribution. To quantify this more precisely, we also show the corresponding cumulative probability distribution in the right panel of Fig.~\ref{fig:prob} -- emphasising once again that this neglects the entirely distinct case of recollapsing models. We then find that the probability of observing a cosmological constant equal to or below the value of $\Lambda$ observed in our universe is around 0.5\%. Thus, with our stated assumptions of (1) the extension to the SP21 model of cosmic star formation presented in this work; (2) a flat prior on $\Lambda$; (3) and assuming a direct proportionality between the cosmic star-forming efficiency and the likelihood of generating observers at a given $\Lambda$, the simplest multiverse framework is not able to account naturally for the small observed value of $\Lambda$.

\subsection{Alternative anthropic assumptions}

The above failure to account for the magnitude of $\Lambda$ can mean one of three things: (a) there is no multiverse; (b) members of the ensemble vary in a more complicated way than simply altering $\Lambda$; (c) some of our basic assumptions are inappropriate.

Undoubtedly, it is interesting to consider a more complex multiverse ensemble, where more fundamental parameters are varied at the same time (up to 31: \citejap{Tegmark_2006}). Indeed, other works have studied the simultaneous impact of varying two or three cosmological parameters on the star formation history \citep{Cline_2008, Bousso_2013}. For instance, a universe with both a larger value of $\Lambda$ and a larger amount of matter may well be evolving in a similar fashion as our own universe, as the increased acceleration of the expansion induced by a higher cosmological constant would be compensated by a stronger clustering of matter.  More radically, we may consider altering other fundamental parameters, such as the gravitational constant. Increasing $G$ would obviously favour the collapse of structures, and may counteract the effect of a larger $\Lambda$. At the same time, such a change would also affect the structure of haloes and subsequently the cooling rate. If one changes other coupling constants, such as the fine structure constant, then the picture becomes increasingly more complex. That would affect the very structure of matter, and modify cooling rates in unusual ways. A comprehensive discussion on how star formation would proceed in such exotic scenarios is provided by \cite{Adams_2019}.
We will explore these generalisations and other matters in future work. 

\begin{figure}
    \centering
    \includegraphics[width=\columnwidth]{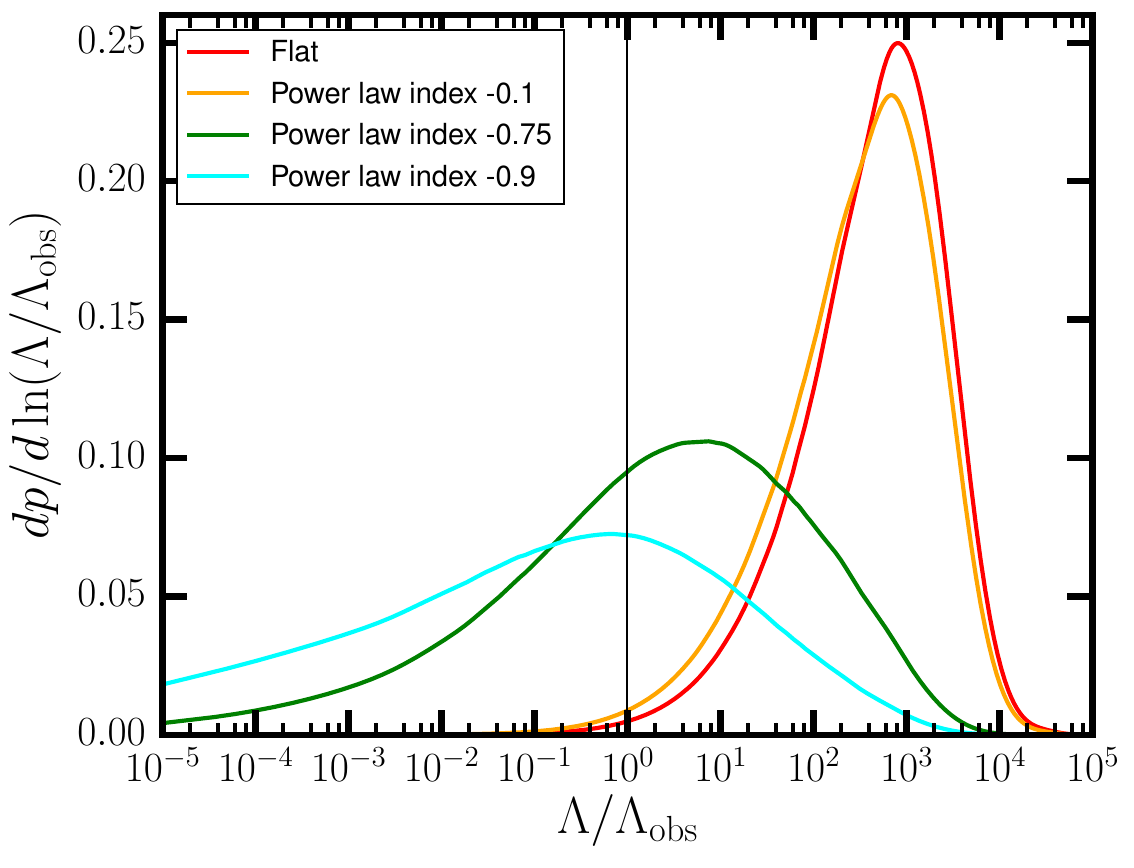}
    \caption{Same as in the left panel of Fig.~\ref{fig:prob}, but for different priors on $\Lambda$. The red line corresponds to the flat prior, and is therefore the same quantity plotted in the left panel of Fig.~\ref{fig:prob}. The other lines refer to a power-law prior, as reported in the legend. A power-law prior with an index in the range $(-1, -0.9]$ would strongly favour the observed value of $\Lambda$, but of course such a prior would need to be motivated by a theory and not chosen a posteriori in order to validate anthropic reasoning (see discussion in the main text).
    }
    \label{fig:power_law}
\end{figure}

The problem with all these more elaborate ensembles is the difficulty in deciding on the prior for the additional parameters. We have adopted Weinberg's prior on $\Lambda$, arguing that it is a special case. Nevertheless, it is interesting to consider what would happen if Weinberg's argument was in error, since a
prior that gives more weight to lower values of $\Lambda$ could radically change  the conclusions. A prior that is flat in $\ln(\alpha_{\Lambda})$ would yield by construction a posterior for  $\Lambda$ that is identical to the cosmic efficiency $\varepsilon$, properly renormalised. However, a logarithmic prior cannot be applied if we allow for $\Lambda=0$, let alone if we accept that $\Lambda<0$ is possible. A power-law prior on $\Lambda$ would be allowed if we insist that $\Lambda\geq0$, and we show posteriors on $\Lambda$ for various power-law slopes in Fig.~\ref{fig:power_law}. As the power law becomes steeper, the probability distribution becomes broader, and the peak moves towards smaller values of $\alpha_{\Lambda}$.

Of particular interest is the case of a prior in the form $\alpha_{\Lambda}^{-0.75}$. If one assumes that the cosmological constant is set by the energy of the vacuum $E_{\rm v}$, then $\Lambda \propto E_{\rm v}^4$. One might then opt for a flat prior on $E_{\rm v}$, corresponding to a prior on $\Lambda$ in the form $\alpha_{\Lambda}^{-0.75}$. In this case our model would yield a probability of observing $\Lambda \leq \Lambda_{\rm obs}$ of 39\%, very well consistent with observation. However, the prior should be motivated by a priori theoretical considerations, rather than being chosen in order to achieve by construction some preferred conclusion. At present, we do not have a compelling argument that would justify a uniform prior in $E_{\rm v}$, nor for the associated assumption that $\Lambda>0$.

There have been several attempts at solving the cosmological constant problem that predict a non-negative value. Supersymmetry generates a strictly positive vacuum energy density if it is spontaneously broken \citep[see e.g.][]{CC_problem_review}. \cite{Kaloper_2014} proposed a reformulation of General Relativity whereby all vacuum energy is sequestered from the matter sector. This prevents the vacuum energy from sourcing spatial curvature and predicts a positive cosmological constant, which is expressed as the space-time average of the trace of the  energy-momentum tensor. However, the model requires the universe to end in order to produce a positive $\Lambda$. This is not necessary in a later generalisation of the sequestering model by \cite{Lombriser_2019_seq}, who revisited the work of \cite{Kaloper_2014} by explicitly modelling the effect of collapsed structures. A positive $\Lambda$ is also predicted by models attempting to solve the cosmological constant problem by varying the Planck mass \citep{Lombriser_2019, Sobral-Blanco_2020, Sobral-Blanco_2021}, at least in their simplest incarnation. The basic idea is to vary the Einstein--Hilbert action with respect to both the metric and the Planck mass over a manifold. The resulting cosmological constant can be expressed in terms of the average energy-momentum  tensor, and can be interpreted as a `backreaction' of structures on the expanding universe. Assuming the typical size of collapsed structures, the model predicts $\Omega_{\Lambda}=0.704$, very close to the observed value. Recently, \cite{Gatzanaga_2021} proposed that $\Lambda$ represents a zero-action boundary term in the Einstein--Hilbert action. It can be  derived that $\Lambda$ is proportional to the average density of the universe and is thus a non-negative quantity. 

There are however other strong theoretical grounds for considering a negative cosmological constant. In supergravity, the vacuum energy can be negative \citep{CC_problem_review}. In the landscape of string theory, negative values of $\Lambda$ emerge naturally, and predicting a small positive value remains a challenge \citep[see e.g.][]{Feng_2021, Demirtas_2022}. These scenarios would then not be compatible with a power-law or logarithmic prior on $\Lambda$. 

Lastly, our conclusions would certainly be subject to change if we were to alter the model of cosmic star formation. We next discuss how the limitations of our present modelling could affect our conclusions in \S\,\ref{sec:limitations}. In \S\,\ref{sec:comparison}, we will then make a detailed comparison of our findings with other related works in the literature.

\subsection{Limitations of our study}
\label{sec:limitations}

This paper is based on the SP21 model of cosmological star formation. This aimed to retain the physical transparency of the original \cite{HS03} model, which comes at the price of some simplifications. Chief amongst these is the fact that the model formalism does not explicitly include AGN feedback, but only supernovae-driven winds. Nevertheless, the parameters underlying the stellar feedback model in the SP21 formalism are tuned to reproduce observations of the baryonic Tully-Fisher relationship \citep{bTFR} and of the Kennicutt-Schmidt relationship \citep{Kennicutt_1998} in our universe, which does harbour AGN feedback. Thus, the effect of AGN feedback is implicitly allowed for, as in the case of cosmological simulations that contain stellar feedback only, but which calibrate the parameters of their subgrid prescriptions to reproduce observed data. This approach was followed, for example, by \cite{BK_2022}, who analysed the impact of $\Lambda$ in a suite of hydrodynamical simulations that extended up to $t=97 \, \rm Gyr$, including only stellar feedback. They argued that any energy fed back into the environment by black holes at late times will happen on a time scale correlated to the feedback from star formation. Then, even without including an AGN feedback model, their feedback parameters would capture the global energy input that would be expected from both star formation and black holes. The empirical correlation of stellar mass and black hole mass in galaxies is commonly taken as evidence for co-evolution of these two populations \citep[see e.g.][]{Kormendy_2013, Heckman_2014}, and the neglect of explicit AGN feedback here is justifiable as long as co-evolution holds exactly.

However, several hydrodynamical simulations do indicate that AGN-driven jets are an important mechanism in the quenching of star formation after $z=2$ \citep[see, e.g.][]{Vogelsberger_2013, Weinberger_2017, IllustrisTNG2018, Sorini_2022, Scharre_2024}. While other groups found a more limited impact of AGN feedback on the low-redshift CSFRD (\citealt{McCarthy_2017, Salcido_2018}; see also the discussion in \citealt{Salcido_2020}), the general consensus is that AGN feedback should be explicitly included in realistic models of the star formation history. Indeed, the total energy input from feedback processes is not the sole important factor in determining the quenching of star formation. Cosmological simulations show that the way such energy propagates, for example through diffuse winds or collimated AGN-driven jets, has a potentially significant impact on the star formation history and the diverse gaseous phases within haloes \citep[e.g.][]{Weinberger_2017, Sorini_2022, Scharre_2024, Yang_2024}. But such considerations are beyond the scope of the present work, and their incorporation would require a major expansion of the SP21 model.

As SP21 pointed out, the introduction of AGN feedback in their formalism could plausibly diminish the low-$z$ CSFRD and hence also the long-term efficiency of star formation. Because most of the stellar mass in universes with large values of $\Lambda$ is formed in the far future (see Fig.~\ref{fig:rho*_frac}), AGN feedback may suppress the cosmic star formation efficiency at the high-$\Lambda$ end more rapidly. This would in turn increase the probability of observing a value of $\Lambda$ that is $\leq \Lambda_{\rm obs}$, reducing the discrepancy with observation. It remains to be seen whether the explicit inclusion of AGN feedback would significantly alter the conclusions in our work regarding the anthropically favoured value of $\Lambda$, or whether its impact would be sub-dominant. We leave this important analysis, which requires incorporating a first-principles model of AGN feedback within our formalism, for future work.

Other potentially important physical ingredients omitted from the SP21 model are chemical evolution and metal enrichment of the interstellar and intergalactic media. Following the evolution of the metallicity of stars and gas within and outside haloes would alter the cooling function over time. At later times, and reasonably also in the future of the universe, the metallicity would increase. For virial temperatures above $T=10^4 \K$, which are those of interest for the SP21 model, the cooling function becomes larger as the metallicity increases \citep[see][]{Sutherland_1993}. Hence, in the future of the universe the cooling rate would become higher, increasing the star formation efficiency.  

We also acknowledge that more exotic effects might become significant in the far future of the universe. For instance, gas cooling via inverse Compton scattering of CMB photons might become an important additional cooling mechanism in the far future \citep{Adams_1997, Bousso_2010, Bousso_2013}. As a further example, \cite{Adams_1997} argued that gravitational instabilities within galaxies could deviate the trajectory of stars from their orbits, so that stars either collapse towards the centre of the galaxy, or drift away from it, becoming isolated and unbound objects. In the former scenario, there can be a secondary spark of star formation, triggered by the clash between brown dwarfs around the galactic centre. Such exotic possible channels of star formation occur in the extreme far future,  $t\approx 10^{15}\,\rm Gyr$. The ejection of stars from the galaxy due to the accumulation of stellar encounters would occur on a smaller, but still colossal, time scale, $t\approx 10^{10} - 10^{11}\,\rm Gyr$ \citep{Adams_1997}. These processes occur well after the point where our calculations of star formation have converged, and there is no reason to believe that they will dominate over previous activity in terms of total stellar production. 

One can also ask whether it is reasonable to believe that `conventional star formation' proceeds all the way up to the $10^{10}\,\rm Gyr$ time scale represented in Fig.~\ref{fig:rho*_frac}. \cite{Adams_1997} argued that, based on the life time of typical red dwarfs, `conventional' star formation should halt on a time scale of $t\sim 10^4\,\rm Gyr$. If we imposed such a cutoff in the star formation history, that would lead to a lower star formation efficiency in universes with lower $\Lambda$. On the contrary, it would not affect as significantly universes with higher $\Lambda$, where the bulk of star formation occurs at earlier times (see Fig.~\ref{fig:rho*_frac}). Therefore, the posterior on $\Lambda$ would be more skewed towards higher values, hence making the anthropic prediction even less consistent with observation.
However, arguments on the time scales of different channels of star formation in the far future of the universe are themselves fairly uncertain. We therefore opted for not introducing an additional ill-defined parameter in our model, and simply integrated the star formation history up to $t \rightarrow \infty$. 

Throughout this work we have assumed that generation of observers is simply proportional to the cosmic star formation efficiency. More realistically, we might consider what is known about statistics of exoplanets as a function of stellar mass. This variation would not matter provided the IMF was invariant, but given that the typical density and temperature of star-forming gas is likely to change with epoch, there is ample scope for the IMF to alter. We have neglected this complication, as have most authors in this area. 

Another option could be to weight the CSFRD by the metallicity of the stars formed.
Indeed,  \cite{Dayal_2015} argued that the stellar mass, total metal mass and star formation rate all conspire in enhancing the habitability of galaxies such as metal-rich giant elliptical galaxies. Furthermore, accounting for the effect of radiation on the emergence of life on potentially habitable planets \citep[e.g.][]{Totani_SN, Gobat_2021} could bring in the structure of the galaxy as a relevant factor. Indeed, recent works defined a `galactic habitable zone' \citep{Gobat_2016}, in analogy to the  circumstellar habitable zone \citep{Huang_1959, Kasting_1993}. However, these variations in the generation of observers within and between galaxies are not a concern for the main focus of the present paper, which is on the global production of observers summed over the universe and over all cosmic time. Therefore it seems reasonable to neglect such issues for the present.

In conclusion, our model can certainly be improved in different ways, and we hope to address some of them in the future. But despite its simplicity the model is capable of capturing the main physical processes that are relevant for a sound description of the star formation history in the universe, both in the past and the future, and in different cosmologies.

\begin{table*}
    \centering
    \caption{Median, $16^{\rm  th}$ and $84^{\rm th}$ percentiles of the posterior distribution on $\alpha_{\Lambda}$ for different choices of the prior on $\Lambda$ or of the measure of observers in the multiverse, and for different models. The table also shows the probability of observers measuring a value of $\Lambda$ equal to or smaller than $\Lambda_{\rm obs}$. Unless otherwise explicitly stated in the main text, results from other works are quoted directly from the respective manuscripts, or derived from fitting formulae therein.
    }
    \label{tab:priors}
    \begin{tabular}{ccccccc}
         Star formation model & Observers model  & prior/measure &  $16^{\rm th}$ percentile & median & $84^{\rm th}$ percentile & $p(\Lambda \leq \Lambda_{\rm  obs})$  \\
         \hline
         This work & star formation & flat in $\alpha_{\Lambda}$ & 73 & 539 & 2306 & 0.52\% \\
         & & $\alpha_{\Lambda}^{-0.1}$ & 44 & 389 & 1868 & 1.0\%\\
         & & $\alpha_{\Lambda}^{-0.75}$ & 0.041 & 2.8 & 84 & 39\%\\
         & & $\alpha_{\Lambda}^{-0.9}$ & $6.9 \times 10^{-4}$ & 0.18 & 12 & 65\%\\ 
         \cite{BK_2022} & star formation & single-scale approximation & 2.3 & 12.6 & 43 & 8.2\%\\
          & & simulation & 1.4 & 8.0 & 34 & 12.9\% \\
         \cite{Barnes_2018} & star formation + delay & mass weighted & 10 & 59 & 194 & 1.9\%\\
          & & causal patch & 0.04 & 0.34 & 0.96 & 86\% \\
          & & causal diamond & 0.13 & 0.65 & 3.15 & 73\% \\
          & star formation + lifetime & mass weighted & 10 & 59 & 194 & 1.9\%\\
          & & causal patch & 0.009 & 0.089 & 0.849 & 90\%\\
          & & causal diamond & 0.01 & 0.25 & 0.96 & 86\% \\
          & star formation + metals & mass weighted & 8 & 45 & 163 & 2.5\% \\
          & & causal patch & 0.004 & 0.07 & 0.71 & 93\% \\
          & & causal diamond & 0.01 & 0.17 & 0.87 & 90\%\\
    \end{tabular}
\end{table*}

\subsection{Comparison with previous work}
\label{sec:comparison}

\subsubsection{Posterior on $\Lambda$ and anthropic considerations}

In this section, we will discuss our results on the posterior distribution on $\Lambda$ and the implications for anthropic arguments in the context of previous relevant work in the literature. 

\cite{BK_2022} ran a suite of hydrodynamical simulations based on the \texttt{Enzo} code \citep{Enzo_Bryan2014}, for an EdS universe and for \LCDM universes with different values of $\Lambda$, up to $100$ times the observed value. The simulations included a sub-grid model for stellar feedback \citep{BK_2020}, but no AGN feedback. This model was shown to yield physically sensible results even when run up to a cosmic time of $100\,\rm Gyr$ \citep{BK_2021}. Using the same assumptions adopted in the present work concerning the probability of generating observers in a given universe, \cite{BK_2022} found that the median of the posterior distribution on $\Lambda$ corresponds to $8.0 \, \Lambda_{\rm obs}$, with 95\% of the distribution spanning the range $\alpha_{\Lambda} = 0.32-105$, and thus consistent with the observed $\Lambda$. They find that the probability that $\Lambda \leq \Lambda_{\rm obs}$ is 12.9\%. \cite{BK_2022} also consider the traditional assumption for the anthropic weighting, based on the single-scale approximation for the collapsed fraction of matter in galaxy-scale haloes \citep[see][]{Peacock_2007}. Under this assumption, the median of the posterior distribution on $\Lambda$ rises to $12.6 \, \Lambda_{\rm obs}$, and the corresponding probability of an observer measuring $\Lambda \leq \Lambda_{\rm obs}$ is $8.2\%$. The overall conclusions from the detailed modelling in \cite{BK_2022} are thus not hugely different from the simple single-scale approach. These conclusions are qualitatively in accord with our own, to the extent that the observed value of $\Lambda$ is a low outlier in the ensemble. But we find an
unacceptably small probability of  $\Lambda \leq \Lambda_{\rm obs}$ -- whereas \cite{BK_2022} find a probability that, while small, does not strongly rule out the multiverse hypothesis.

An earlier example of a large `multiverse simulation' is given by \cite{Barnes_2018}. They ran a suite of variants of the \texttt{EAGLE} hydrodynamical simulation \citep{EAGLE_Schaye2015}, for values of $\alpha_{\Lambda}$ in the range $0-300$, until cosmic time $t=20.7\,\rm Gyr$. These simulations include recipes for both stellar feedback and AGN feedback. \cite{Barnes_2018} consider three possible models for the generation of observers:
\begin{enumerate}
    \item Observers are generated following the formation of stars, after a fixed delay of $5\,\rm Gyr$, which would represent the timescale for the evolution of intelligent life; 
    \item The rate of generation of observers at a given time around a certain stellar population is proportional to the fraction of stars that are still on the main sequence, to account for the fact that life appears to be a rare phenomenon in the universe, and therefore should occur towards the end of the main sequence lifetime of a star \citep{Carter_1983, Barrow_1986};
    \item The rate of generation of observers at a given time is proportional to the metallicity of the fraction of stars that are still on the main sequence for the stellar population considered, to reflect the argument that life would more easily emerge on rocky planets, which would be more likely to form around stars with higher metallicity (\citealt{Gonzalez_1997, Fischer_2005}; but see also \citealt{Buchhave_2005, Wang_2015}).
\end{enumerate}
The main impact of these different choices is to change when the observers form, rather than the total eventual numbers, which is our main concern. The results for all these models are reported in Table~\ref{tab:priors}, in correspondence of the cells in the second column, labelled `star formation + delay', `star formation + lifetime' and `star formation + metals', respectively. For each of the aforementioned models, \cite{Barnes_2018} further considered three possible ways of constructing a measure in the multiverse:
\begin{enumerate}
    \item a mass-weighted measure, whereby a given mass element in the universe can inhabit a region with any value of $\Lambda$ with equal probability;
    \item a causal-patch measure, in which regions of the universe that have equal co-moving volume of their causal patch at a certain cosmic time host the same number of observers, for any value of $\Lambda$ considered;
    \item a causal-diamond measure, analogous to the previous measure, except that the co-moving volume of the region (at given cosmic time $t$) that is enclosed by a photon that departs from a world line at reheating and returns at $t$ is considered instead of the volume of the causal patch.
\end{enumerate}
It is important to note that in all cases described above \cite{Barnes_2018} consider a flat prior on $\Lambda$. Thus their posterior on $\Lambda$ is the product of the measure times the weighting based on star formation. The mass-weighted measure is essentially what we adopt in the present work. 

The results that \cite{Barnes_2018} obtain for all observer generation models and all multiverse measures are listed in Table~\ref{tab:priors}. The mass-weighted measures yields a median of the posterior distribution on $\Lambda$ of 45 or 59, depending on the observer generation prescription. The 84$^{\rm th}$ percentile of the distribution can reach values as large as $194\, \Lambda_{\rm obs}$. On the other hand, the causal patch and causal diamond measures move the median $\Lambda$ to much smaller values, and always below $\Lambda_{\rm obs}$. Consequently, the mass-weighted measure would predict a significantly smaller probability of observing $\Lambda \leq \Lambda_{\rm obs}$ (1.9\%--2.5\%) than the other two measures (73\%--93\%). Thus, the choice of the measure appears to be crucial.

However, the measure is not a degree of freedom that can be adjusted at will (see the discussion in \citealt{Barnes_2018}). We confess that we find the motivation for the causal measures obscure: in our own universe there are definitely pairs of galaxies that are causally disconnected (those in opposite directions on the sky at high redshift), and yet each can potentially host observers. \cite{Barnes_2018} mention
theoretical motivations for choosing the causal patch and diamond measures in the context of the quantum xeroxing paradox \citep{Susskind_1993, Bousso_2006, Bousso_eternal_inflation} and holographic probability in eternal inflation \citep{Bousso_2006}. But
it seems to us that the results are biased, and systematically exclude potential observers at early times when the horizon is small. We will therefore compare our predictions with the results in \cite{Barnes_2018} arising from their mass-weighted measure only.

We can see that our results predict a median value of $\Lambda$ that is $9-12$ times larger than that of \cite{Barnes_2018}, and our probability for $\Lambda \leq \Lambda_{\rm obs}$ is about $\sim$\,$4-5$ times smaller. However, we integrate the CSFRD all the way to $t\rightarrow \infty$, whereas \cite{Barnes_2018} halt at $20.7\,\rm Gyr$ in all cases. Nevertheless, we verified that even if we stopped our integration at this same early time, the probability of measuring $\Lambda \leq \Lambda_{\rm obs}$ increases only slightly to 0.7\%. This increase does not qualitatively affect our conclusions.

Anthropic calculations have also been pursued using semi-analytic methods. \cite{Sudoh_2017} adopted the semi-analytic model of galaxy formation by \cite{Nagashima_2004}, coupled with a Monte Carlo calculation for the merger history of dark matter haloes. Adopting a flat prior on $\Lambda$ and using stellar mass density as a proxy for observer weighting, they obtained a median value of $\Lambda$ in the range $\alpha_{\Lambda} = 9.1-12$, depending on the exact details of their model. Overall, the distribution is broad, and their probability of observing $\Lambda \leq \Lambda_{\rm obs}$ is between 6.7\% and 9.7\%. Such values are in agreement with the results by \cite{BK_2022}, but are in tension with our conclusions and with \cite{Barnes_2018}. It is not surprising that \cite{Sudoh_2017} obtain a lower median $\alpha_{\Lambda}$ compared with us. Rather than integrating the mass of cool gas that is converted into stars over the entire history of the universe, they estimate the number of observers by calculating the total stellar mass density at the fixed cosmic time $t=15\,\rm Gyr$. The authors argue that this is a reasonable approximation, but it is not an assumption that is supported by our model: rather, we predict that in universes with a large value of $\Lambda$ the majority of stars form at $t \gg 15\,\rm Gyr$ (see e.g. Fig.~\ref{fig:rho*_frac}). 

To summarise, a common feature of all work on the simplest anthropic ensemble is that the posterior probability distribution on $\Lambda$ is broad, with a median that is substantially larger than $\Lambda_{\rm obs}$. The exact value is heavily dependent on the underlying star formation model and the probability measure, but in all works considered the observed value of the cosmological constant seems to be located in the tail of the posterior for $\Lambda$ (except for the causal patch and diamond measures in \citealt{Barnes_2018}). Nevertheless, in the previous literature discussed here the probability of measuring $\Lambda \leq \Lambda_{\rm obs}$ is still of the order of a few percent or even $\sim$\,10\%, thus failing to reject the multiverse model. In contrast, our results pose a challenge in this respect, given that this probability is only $0.5\%$. 
The viability of Weinberg's simple anthropic explanation for $\Lambda$ thus depends on which of these calculations is closest to a correct physical prediction of star formation in counter-factual universes. All existing studies have limitations, and ours is definitely no exception. But in yielding a result that is so far from the observed level of $\Lambda$, it raises the motivation for continuing to seek a more realistic treatment of this problem.

\subsubsection{Dependence of the cosmic stellar efficiency on $\Lambda$}
\label{sec:match_sims}

\begin{figure*}
    \centering
    \includegraphics[width=\textwidth]{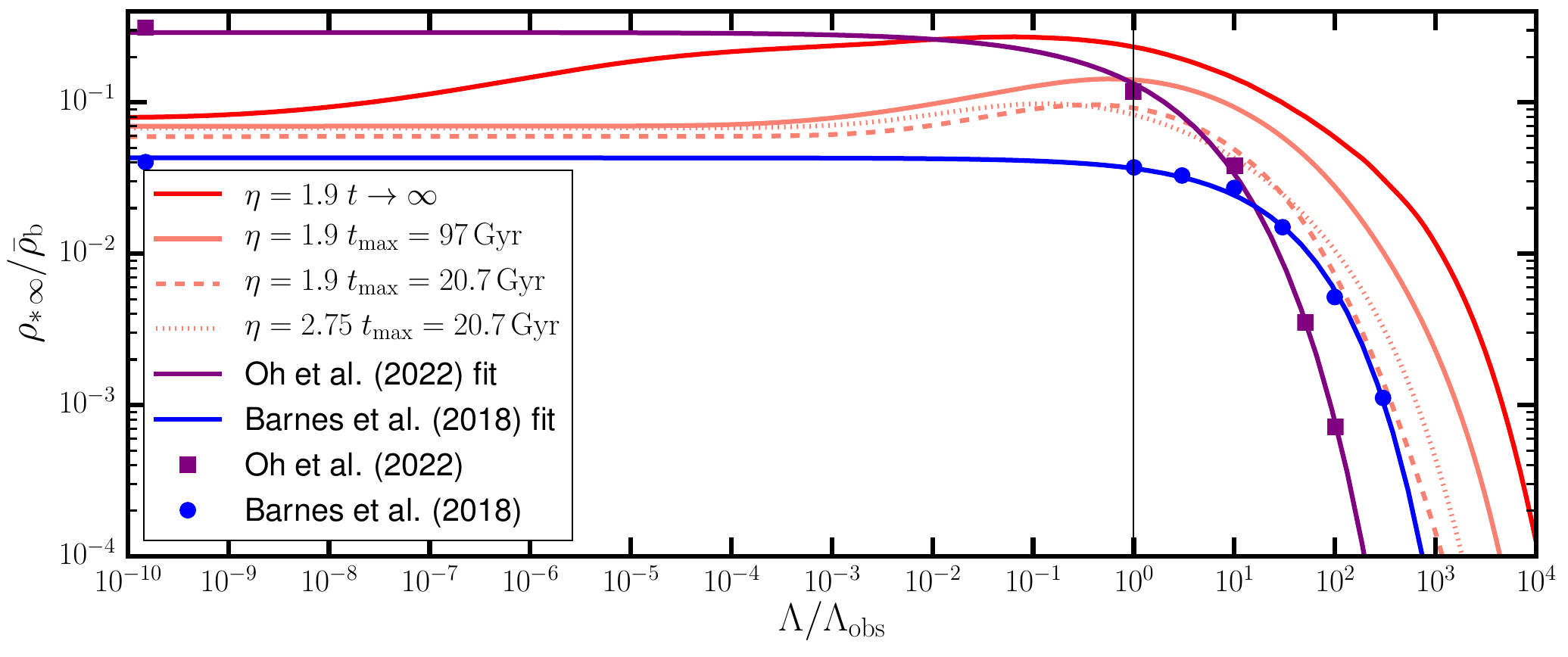}
    \caption{Same as Fig.~\ref{fig:ms-mb}, for the model presented in this work (red line), and the cosmological simulations by \protect\cite{Barnes_2018} and \protect\cite{BK_2022} (blue circles and purple squares, respectively). The corresponding fitting formulae to the numerical results, proposed by \protect\cite{BK_2022}, are represented with blue and purple solid lines, respectively. Integrating the cosmic star formation rate density given by our model up to the same finite cosmic time as in the simulations ($t_{\rm max}=20.7 \, \rm Gyr$ and $t_{\rm max}=97 \, \rm Gyr$ for \protect\citejap{Barnes_2018} and \protect\citejap{BK_2022}, respectively) improves the agreement with the numerical results (dashed and solid salmon lines, respectively). The match with \protect\cite{Barnes_2018} is further improved by properly changing the slope of the gas density profiles within haloes, $\eta$, which is a free parameter in our model (dotted salmon line; see discussion in \S~\ref{sec:match_sims} for details).}
    \label{fig:comparison}
\end{figure*}

The conclusions on the posterior distribution of $\Lambda$ reached by the different works discussed in the previous section are predicated on the dependence of the comic star formation efficiency on $\Lambda$. Analysing this key quantity is informative regardless of any anthropic consideration, as it explores the predictions of different cosmic star formation models outside the canonical values of the underlying cosmological parameters. In this section, we will discuss similarities and discrepancies of our results on the cosmic star formation efficiency (Fig.~\ref{fig:ms-mb}) with respect to the results from direct simulations.

Fig.~\ref{fig:comparison} shows the cosmic star formation efficiency predicted by our model, i.e. the same as in Fig.~\ref{fig:ms-mb}, in comparison with the efficiencies obtained with the \texttt{Eagle} simulations by \cite{Barnes_2018} and the \texttt{Enzo} simulations by \cite{BK_2022}.
For $\Lambda>\Lambda_{\rm obs}$, there is general qualitative agreement, with all works considered showing a suppression of the cosmic star formation efficiency for high values of $\Lambda$. Our model predicts a peak at $\Lambda \approx 0.1 \, \Lambda_{\rm obs}$, and lower efficiencies as we approach an EdS universe. By contrast, the fitting functions to the numerical results appear to indicate a monotonic increase in the asymptotic cosmic star formation efficiency as $\Lambda \rightarrow 0$. While this trend is qualitatively different from our result, no simulations  actually cover the range $0<\Lambda <\Lambda_{\rm obs}$. It would be interesting to run further simulations in this range in order to establish  whether the simulated dependence of the cosmic star formation efficiency on $\Lambda$ is genuinely monotonic.

Nevertheless, assuming that the fitting functions proposed by \cite{BK_2022} do represent the actual trend of the efficiency in the range $0<\Lambda <\Lambda_{\rm obs}$, it is worth speculating on the possible origins of the different result that we obtain. We should however first make sure that we make a fair comparison across all models considered. The simulations are inevitably stopped at some finite future time; for our cosmology and an EdS universe, this maximum time is $t_{\rm max}=97\, \rm Gyr$ and $t_{\rm max} = 20.7 \, \rm Gyr$ in \cite{Barnes_2018} and \cite{BK_2022}, respectively. To account for the star formation history at arbitrarily large times, \cite{Barnes_2018} extrapolate the time evolution of the CSFRD beyond $t_{\rm max}$ with a decaying exponential, and conclude that the contribution from times $t>t_{\rm max}$ to the total cosmic star formation efficiency is negligible. Therefore, for a fairer comparison between our model and the \cite{BK_2022} fitting formula to the \cite{Barnes_2018} simulations (blue line in Fig.~\ref{fig:comparison}), we should re-compute our cosmic star formation efficiencies by integrating the CSFRD up to $t_{\rm max}=20.7\, \rm Gyr$, and not in the limit $t\rightarrow \infty$.

The result is represented by the dashed salmon line in Fig.~\ref{fig:comparison}, which considerably improves the agreement with the fitting formula to the \cite{Barnes_2018} results. Our model recovers the cosmic efficiency in the  \cite{Barnes_2018} simulations within 50\% for an EdS universe, and yields an excellent match at the high-$\Lambda$ end. The largest discrepancy occurs for $\Lambda\approx 0.3 \Lambda_{\rm obs}$, and amounts to less than a factor of $\sim$\,3. This deviation is smaller than the discrepancy between \cite{Barnes_2018} and \cite{BK_2022}. However, the non-monotonic dependence of the cosmic star formation efficiency on $\Lambda$ for $\Lambda<\Lambda_{\rm obs}$ does not disappear by simply imposing a cutoff in the time integral of the CSFRD. But the peak becomes much less prominent, and the discrepancy with the numerical results is reduced.

If we apply the same reasoning for a fair comparison between our model and the \cite{BK_2022} simulations, we reach overall similar conclusions. To estimate the integral of the CSFRD in the limit $t\rightarrow \infty$, \cite{BK_2022} extrapolated their numerical results beyond $t_{\rm max}$ by fitting their CSFRD with the same functional shape used by \cite{Madau_rev}. This analytical formula allowed us to subtract the contribution from cosmic times $t> t_{\rm max}$ from the cosmic star formation efficiencies computed by \cite{BK_2022}, and we verified that the change would be in any case negligible. We can then simply cut the time integral of the CSFRD at $t_{\rm max}=97\, \rm Gyr$. The result is given by the salmon solid line in Fig.~\ref{fig:comparison}. The peak is less conspicuous, and the agreement with the fit to the numerical results (purple line) improves. However, the decay at large values of $\Lambda$ is less steep compared to \cite{BK_2022} work, and the efficiency in the EdS universe is recovered within a factor of $\sim$\,4.5.

Without adding any new physics to our model, we can check whether a change in the underlying parameters would yield a better match with cosmological simulations. This can be done by considering the late-time behaviour of the CSFRD. As mentioned earlier, \cite{BK_2022} showed that their CSFRDs can be well described with the same fitting formula adopted by \cite{Madau_rev}. We verified that this is the case for the \cite{Barnes_2018} CSFRDs too. At late times, the \cite{Madau_rev} curve can be in turn approximated with a power law scaling of the CSFRD with $(1+z)$. Our model also predicts a power-law scaling in an EdS universe at late times, with an index that depends on the slope of the gas density profile within haloes, $\eta$ (see Appendix~\ref{app:zero_L}). We can therefore find the value of $\eta$ that yields the same late-time power-law scaling in an EdS universe as in \cite{Barnes_2018} and \cite{BK_2022}.

In the first case, we find that the desired value is $\eta=2.75$. We then run our model for such value of $\eta$, while leaving all other parameters the same. We integrate the CSFRD for any value of $\Lambda$ up to $t_{\rm max} = 20.7 \, \rm Gyr$, and obtain the cosmic star formation efficiency. Moving from $\eta=1.9$ to $\eta=2.75$ further reduces the prominence of the peak in the cosmic star formation efficiency. However, the overall agreement with \cite{Barnes_2018} does not significantly improve. Also, the value $\eta=2.75$, while mathematically acceptable, would fall outside the range that provides a consistency within $3\sigma$ with observational constraints from the Kennicutt-Schmidt relationship and the baryonic Tully-Fisher relationship (see SP21 for details). If we apply the same logic to the \cite{BK_2022} simulations, we would find $\eta=3.21$. This is not even mathematically acceptable in our formalism, as values of $\eta>3$ would cause the gas mass of haloes to diverge (see equation~\ref{eq:dens_prof}). Therefore, we cannot fine tune $\eta$ to match the late-time behaviour of the CSFRD found by \cite{BK_2022} for an EdS universe.

To summarise, the qualitative discrepancy between the cosmic star formation efficiency predicted by our model and by analytical fits to cosmological simulations is considerably mitigated once a cutoff time is taken into consideration, for a fair comparison with the numerical results.  Where possible, the significance of the peak in the cosmic star formation efficiency predicted by our model can be further reduced by regulating the value of $\eta$ to match the late-time behaviour of the simulated CSFRD in an EdS universe. However, the peak does not disappear completely, therefore some change in the physics included in our model may be necessary in order to obtain a cosmic star formation efficiency that monotonically decreases with $\Lambda$.

The residual qualitative discrepancies between our model and numerical simulation occur in the range $0<\Lambda< \Lambda_{\rm obs}$. Within our formalism, the asymptotic stellar mass produced in a halo of a fixed virial mass $M\lesssim 10^{11}\, \rm M_{\odot}$ does not vary appreciably with the cosmological constant for $\Lambda<\Lambda_{\rm obs}$. But for the same cosmologies, the stellar mass produced in high-mass haloes ($M\gtrsim 10^{11}\, \rm M_{\odot}$) can vary by up to one order of magnitude (see Fig.~\ref{fig:efficiency}). Therefore, the treatment of star formation within high-mass haloes is probably the main reason behind the discrepancies between our model and simulations. In this respect, an obvious missing piece of physics in our model is AGN feedback, as discussed in \S~\ref{sec:limitations}. An explicit AGN feedback mechanism could reduce the contribution of high-mass haloes to the star formation history, and may bring the cosmic efficiency at $0< \Lambda < \Lambda_{\rm obs}$ closer to that of an EdS universe.

It is worth noting that \cite{BK_2022} predict a decreasing efficiency even in the absence of AGN feedback. There may therefore be other ways to achieve the same qualitative trend without (or in addition to) AGN feedback. For example, one may include chemical evolution models, or an explicit shutdown of star formation at very late times, as detailed in \S~\ref{sec:limitations}. We intend to address these issues in future work. At the same time, we point out that it would be very informative to run cosmological simulations in the range $0<\Lambda< \Lambda_{\rm obs}$, which is the range where our predictions differ from the fits to the numerical results by \cite{Barnes_2018} and \cite{BK_2022}.

\section{Conclusions and Perspectives}
\label{sec:conclusions}

We have improved the SP21 analytical model for cosmic star formation \citep{SP21}, which was itself based on the work by \cite{HS03}. We have then used this model to predict the cosmic star formation history in flat \LCDM cosmological models where the cosmological constant is varied between $\Lambda=0$ and $\Lambda=10^5 \, \Lambda_{\rm obs}$. This ensemble of universes was designed so that all members are indistinguishable at high redshift, but later part company as their different cosmological constants come to dominate. 

The cosmic star formation rate density is obtained by integrating the star formation rate within haloes over the halo mass function. We assumed that the same astrophysical processes operate in each member of the ensemble as in our own universe. At high redshift, the star-formation rate is determined by the typical time taken to convert cool gas into stars, whereas at low redshift it is set by the cooling time. We were then able to compute the total stellar mass generated by a given time, as a fraction of the total baryonic mass, per co-moving volume. We showed that this cumulative star-formation efficiency tends to a well-defined limit as $t\to\infty$, and we determined the dependence of this asymptotic efficiency on the value of $\Lambda$. We then investigated the implications of our results for anthropic attempts to explain the observed value of $\Lambda$. Our main conclusions are as follows:

\begin{enumerate}
    \item The SP21 formalism predicts a converging total comoving stellar mass density as the cosmic time tends to infinity. In our universe, about 32\% of the eventual stellar mass has been formed by $z=0$ (Fig.~\ref{fig:rho*_frac}). Thus, we are typical observers in terms of the cosmological epoch in which we live, which is marginally $\Lambda$-dominated. But we find that this coincidence between typical eras of star formation and $\Lambda$-domination does not hold for other values of $\Lambda$; hence single-universe anthropic arguments do not account for the `why-now' problem. 
    
    \item Our model predicts that the cosmic star formation efficiency peaks at around 27\% for $\Lambda \approx 0.1 \Lambda_{\rm obs}$ (Fig.~\ref{fig:ms-mb}). In an Einstein--de Sitter universe with $\Lambda=0$, the efficiency is approximately a factor of 3 lower than this; but the efficiency is much more strongly suppressed at high values of $\Lambda$. The decline is however relatively slow, and the efficiency only becomes negligibly small as $\Lambda$ approaches $10^5 \, \Lambda_{\rm obs}$. 
    
    \item For a given $\Lambda$, there is a characteristic halo mass above which the stellar efficiency within haloes is maximised (Fig.~\ref{fig:efficiency}). A halo with a given virial mass is smaller in universes with larger $\Lambda$, so that its typical density is higher. This in turn makes gas cooling, and subsequently star formation, more efficient. Thus, the suppression of the cosmic star formation efficiency at high $\Lambda$ must be cosmological in nature, and not astrophysical: it is induced by the cutoff in the halo mass function occurring at lower halo masses. Conversely, the reason for the reduced cosmic star formation efficiency at low $\Lambda$ is astrophysical: haloes are more diffuse as the cosmological model approximates an Einstein--de Sitter universe, so that the cooling rate is reduced. Thus, according to our model, astrophysics favours star formation in those haloes that exist when $\Lambda$ is high, whereas cosmology increases the abundance of haloes when $\Lambda$ is low. The observed value of the cosmological constant finds itself near the sweet spot of these two opposite trends (Fig.~\ref{fig:ms-mb}). However, that does not imply that the generation of observers is maximised around the same value of $\Lambda$.
    
    \item Assuming a flat prior on $\Lambda$ and that the observer weighting of a universe with a given $\Lambda$ is proportional to the cosmic star formation efficiency in that universe, the peak of the posterior on $\Lambda$ occurs at $\Lambda \approx 800 \, \Lambda_{\rm obs}$ (Fig.~\ref{fig:prob}). The median of the distribution is $\Lambda=539$, and the $16^{\rm th}-84^{\rm th}$ percentile range corresponds to $\Lambda = 73-2306$ (Table~\ref{tab:priors}). The probability that an observer measures a value of the cosmological constant $\Lambda \leq \Lambda_{\rm obs}$ is $\sim$\,0.5\%. Thus, Weinberg's anthropic reasoning would be disfavoured by the SP21 model, under the given assumptions.
\end{enumerate}
 
It is striking that the observed value of $\Lambda$ is within one order of magnitude of the optimal value that maximises the cosmic star formation efficiency. We also succeed in providing an explanation for why this is the case, based on calculations from first principles. However, we also found that this asymptotic efficiency of star formation declines rather slowly as $\Lambda$ increases, falling below 1/10 of the peak value only for $\Lambda>500\, \Lambda_{\rm obs}$ 
(Fig.~\ref{fig:ms-mb}). Hence, with Weinberg's assumption of a uniform prior on $\Lambda$, it is inevitable that typical observers would expect to experience values of $\Lambda$ much greater than the one we observe, and this is why we find that values of $\Lambda$ as small as observed are highly unlikely. In principle this rules out the simple anthropic ensemble considered here, subject to the assumptions of our model. But we have reviewed alternative calculations and found that some give a less severe disagreement -- although there is unanimity that the observed $\Lambda$ is unusually small at some level. 

Future progress on this question requires more detailed and realistic modelling: we will
endeavour to generalise our model of cosmic star formation by including features such as chemical evolution of the interstellar/intergalactic medium and AGN feedback. In particular, the absence of an explicit AGN feedback mechanism may lead to an overproduction of stars in massive haloes, which, according to our model, are especially efficient in universes with a high value of $\Lambda$.
Whether or not this is important depends on the applicability of exact co-evolution between black holes and stellar content in galaxies. If this concept holds exactly, the modelled effective energy feedback from star formation may be able to capture the effect of AGN feedback. But if co-evolution does not hold exactly, then
the inclusion of separate AGN feedback in our formalism might alter the resulting conclusions on the anthropically favoured range of $\Lambda$. We plan to address this question in future work.

One would also like to see further direct calculations via simulations.
But hydrodynamical simulations are expensive to run, especially beyond $z=0$, thus it is important to invest the computational resources to study models that promise to bring the most valuable physical insights. Our analytic model enables a fast exploration of the dependence of star formation history on the cosmological parameters, so it can aid in designing a suite of simulations exploring cosmological models that differ from the fiducial  \LCDM cosmology in especially interesting ways.

Eventually, we can hope for a convergence of these studies and a consensus on the viability of Weinberg's anthropic explanation for $\Lambda$. But even if the verdict is negative, this would be far from a rejection of anthropic reasoning in general. This manuscript focuses on changing only one fundamental parameter, i.e. the cosmological constant -- and even there we have restricted ourselves to the case of $\Lambda\geq0$, leaving the whole topic of recollapsing models for a further investigation. Beyond this, one should in principle vary multiple parameters simultaneously, and explore their joint impact on star formation history and the generation of observers. We are already working on this kind of study (Lombriser et al., in prep.).

\section*{Acknowledgements}

We thank the anonymous reviewers for helpful comments. The authors thank Sownak Bose and Boon Kiat Oh for useful discussions, and Ed Wheeler for comments on a draft of this manuscript. DS and LL acknowledge support from the Swiss National Science Foundation (SNSF) Professorship grant no.~202671. DS and JAP were supported by the European Research Council, under grant no.~670193 and by the STFC consolidated grant no.~RA5496. This work made extensive use of the NASA Astrophysics Data System and of the astro-ph preprint archive at arXiv.org. For the purpose of open access, the author has applied a Creative Commons Attribution (CC BY) licence to any Author Accepted Manuscript version arising from this submission.\\

\noindent
DS dedicates this work to the memory of his grandmother Linda.

\section*{Data availability}

The code that we developed to produce the results presented in this manuscript is undergoing further extension, and we plan to release it publicly as part of follow-up work. Until then, the code and the output data published in this manuscript may be released upon reasonable request to the corresponding author.




\bibliographystyle{mnras}
\bibliography{SFR_Lambda}




\appendix

\section{Derivation of the asymptotic behaviour of star formation history}
\label{app:math}

In \S\,\ref{sec:future} we showed that the stellar mass fraction in haloes with fixed mass $M$ asymptotes to a constant in the far future of the universe. This means that the corresponding nSFR, and consequently the CSFRD, tend to zero in the limit $t\rightarrow \infty$. These conclusions result from the numerical implementation of the iterative procedure that we introduced to extend the SP21 analytic model (see \S\,\ref{sec:generalise}). In this Appendix we will show that, under well motivated approximations, it is possible to derive the asymptotic behaviour of the nSFR and CSFRD analytically. Crucially, this will enable us to prove that the SMD, which is the time integral of the CSFRD, converges in the limit $t\rightarrow\infty$.

Since the CSFRD depends on the HMF and the nSFR at fixed $M$ (see equation~\ref{eq:CSFRD}), we will first need to study the limit of these two quantities for arbitrarily large cosmic times. Once we obtain the asymptotic behaviour of the CSFRD, it will be straightforward to study the convergence of the SMD. We will need to distinguish between the two qualitatively different cases of $\Lambda=0$ and $\Lambda>0$. All computations in this section have been verified with the software \texttt{Mathematica} \citep{Mathematica}. 

\subsection{Einstein--de Sitter universe}
\label{app:zero_L}

\subsubsection{Normalised SFR}
\label{app:nSFR_EdS}

The evolution of the nSFR of a halo with fixed virial temperature $T$ in the low-redshift regime is given by equation~\eqref{eq:nSFR_low}. In an EdS universe, the effective cooling time coincides with the dynamical time at any cosmic epoch (see equation~\ref{eq:tcool_smooth}). Therefore, equation~\eqref{eq:nSFR_low} can be simplified to
\begin{equation}
\label{eq:nSFR_EdS}
    s_{\rm low}(M,\,z) \approx \tilde{S}(T(M,\,z)) \left( \frac{H(z)}{H_0} \frac{f_{\rm gas}(T(M, \, z), \, z)}{f_{\rm b}} \right)^{\frac{3}{\eta}} \, ,
\end{equation}
switching from fixed virial temperature to fixed virial mass as the independent variable. Thus, the first term in the r.h.s of the equation above acquires a redshift dependence, which can be  determined by replacing the virial temperature in SP21 equation (25) with the expression given by equation~\eqref{eq:Mvir_T}. Recalling the definition of $\tilde{S}(T)$ from SP21,
\begin{equation}
\label{eq:S_tilde}
    \tilde{S}(T) =  \frac{1}{\eta} \left[ \sqrt{\frac{\Delta}{2}} \frac{(3-\eta) f_{\rm dyn} f_{\rm b} X^2 \mu H_0 \mathcal{C}(T)}{6 \pi G m_{\rm H}^2  k_{\rm B} T} \right]^{\frac{3}{\eta}} \frac{6 \pi G m_{\rm H}^2 k_{\rm B} T}{ f_{\rm dyn}^2 X^2 \mu \mathcal{C}(T)} \, ,
\end{equation}
it becomes obvious that this task is complicated by the presence of the cooling function $\mathcal{C}(T)$, which has a non-trivial dependence on temperature. However, we can reasonably approximate the \cite{Sutherland_1993} cooling function for a plasma of primordial composition, in the temperature range of our interest ($T>10^{4.5}\K$), with a piece-wise power law:
\begin{gather}
\label{eq:CoolFunc_appr}
    \mathcal{C}(T) \propto T^{\theta}\; , \textrm{where} \;  
        \begin{cases}
         \theta = 10.75 & \textrm{if} \;\; T < 10^{4.5} \, \mathrm{K} \\
         \theta = -1.5 &  \textrm{if} \;\;  10^{4.5} \, \mathrm{K} <T< 10^{4.7} \, \mathrm{K}\\
         \theta = 1.7  &  \textrm{if} \;\;  10^{4.7} \, \mathrm{K} <T< 10^5 \, \mathrm{K}\\
         \theta = -1  &  \textrm{if} \;\;  10^5 \, \mathrm{K} <T< 10^6 \, \mathrm{K}\\
         \theta = 0.5 &  \textrm{if} \;\;  T > 10^6 \, \mathrm{K}\\
        \end{cases} \; .
\end{gather}
Within this approximation, and considering that in an EdS universe  $H(z)\propto (1+z)^{3/2}$, it follows that
\begin{equation}
\label{eq:stilde_EdS}
    \tilde{S}(M(T, \, z)) \propto (1+z)^{\frac{2(\theta-1)(3-\eta)}{3\eta}} \, .
\end{equation}

Let us now turn to the second factor in the r.h.s. of equation~\eqref{eq:nSFR_EdS}. As discussed in \S~\ref{sec:fb_haloes}, the critical temperature decreases with cosmic time in an EdS universe. Recalling that for the computation of the CSFRD we only consider haloes above a certain virial temperature threshold (in this work, $10^{4.5} \K $), it is apparent that all haloes that contribute to the CSFRD will exceed the critical temperature at a sufficiently late time. Thus, $f_{\rm gas}$ will asymptotically reach a non-null constant, so that the only remaining redshift dependence in the r.h.s. of equation~\eqref{eq:nSFR_EdS} is $H(z)^{3/\eta} \propto (1+z)^{9/2 \eta}$. Combining it with equation~\eqref{eq:stilde_EdS}, it follows that $ s_{\rm low}(M,\,z) \propto (1+z)^{\zeta}$, with
\begin{equation}
    \zeta = \frac{27+4 (\theta-1) (3-\eta)}{6\eta} \, .
\end{equation}

The nSFR then converges as $z \rightarrow -1$ (or, equivalently, $t \rightarrow \infty$), as long as $\zeta > -1$. In the physically relevant range $3/2<\eta<3$, this condition is satisfied for all values of $\theta$ quoted in equation~\eqref{eq:CoolFunc_appr}.

\subsubsection{Halo mass function}
\label{sec:hmf_EdS}

Within the Sheth-Tormen formalism \citep{ST99, ST02}, the halo multiplicity function is
\begin{equation}
    \frac{dF}{d\ln M} = - \frac{2N}{\sqrt{\pi}} \frac{d \ln \sigma_0(M)}{d\ln M} y (1+ \sqrt{2}y)^{-\frac{3}{5}} \exp(-y^2) \, ,
\end{equation}
where $\sigma_0(M)$ is the linear-theory fractional variance of matter density fluctuations averaged over spheres containing a mass $M$ at $z=0$, and $N$ is a normalisation constant chosen such that all cosmic mass is contained in haloes. We also defined
\begin{equation}
    y = \frac{\delta_{\rm c}}{2^{\frac{3}{4}}D(z) \sigma_0(M)} \, ,
\end{equation}
where $D(z)$ is the growth function, which, in an EdS universe, scales as $D(z) \propto (1+z)^{-1}$. Therefore, the scaling of $y$ with redshift is $y\propto(1+z)$.

It follows that the redshift dependence of the halo multiplicity function is given by
\begin{equation}
    \frac{dF}{d\ln M} \propto (1+z) [1 + \sqrt{2} (1+z)]^{-\frac{3}{5}} \exp[-(1+z)^2] \, .
\end{equation}
In the limit $z \rightarrow -1$, the scaling above reduces to $dF/d\ln M \propto(1+z)$.

\subsubsection{Stellar mass density}

The convergence of the CSFRD depends on the asymptotic behaviour of the integrand in equation~\eqref{eq:CSFRD}. As explained in \S~\ref{sec:fb_haloes}, the critical temperature becomes arbitrarily low at large times, therefore the relevant regime in our piece-wise approximation of the cooling function is $T_{\rm crit} < 10^{4.5} \K$, which corresponds to $\theta=10.75$. If we further set $\eta=1.9$, we obtain that $\zeta \approx 6.1$. We saw in \S\,\ref{sec:hmf_EdS} that the asymptotic redshift dependence of the halo multiplicity function in equation~\eqref{eq:CSFRD} is $\propto (1+z)$. Therefore, the integrand scales approximately as $(1+z)^{\zeta+1} \approx (1+z)^{7.1}$. This is then the redshift-dependence of the CSFRD in the far future.

The SMD is simply the time integral of the CSFRD. Changing the integration variable from cosmic time to redshift, the SMD is defined by equation~\eqref{eq:rho*}. For an EdS model, the integrand in that equation is approximately proportional to $\dot{\rho}_*(z) (1+z)^{-5/2} \propto(1+z)^{4.6}$, which is convergent for $z\rightarrow -1$. This proves that the mass density of stars formed over the entire history of an EdS universe is finite. We therefore used the scaling $\rho_*(z) \propto(1+z)^{\zeta-1/2} \approx (1+z)^{5.6}$ to extrapolate the SMD in the far future and compute the asymptotic SMD. We verified that the analytic approximation matches the numerical results at $z\lesssim -0.99$.

\subsection{Positive cosmological constant}
\label{app:pos_L}

To study the convergence of the SMD in a universe with $\Lambda>0$, we will still follow a similar logic to the one outlined in the previous section. We will highlight the differences in the reasoning that stem from having a positive rather than null cosmological constant.

\subsubsection{Normalised SFR}

In the case $\Lambda>0$, it is more convenient to consider the asymptotic behaviour of the nSFR in terms of cosmic time rather than redshift. This can be straightforwardly obtained by adapting equation (15) in SP21 such that haloes are taken at a fixed virial mass instead of a fixed virial temperature:
\begin{multline}
\label{eq:cool_rate_gen}
    s_{\rm low}(T(M, \, z), z) = \frac{S(T(M, \, z))}{M} \left( \frac{f_{\rm gas}(T(M, \, z), \,z)}{f_{\rm b}} \right)^{\frac{3}{\eta}} \\
    \left(\frac{H(z)}{H_0} \right)^{\frac{6}{\eta}-3} 
    (H_0 t_{\rm cool})^{\frac{3}{\eta}-2} \, \frac{dt_{\rm cool}}{dt} \, .
\end{multline}
In a universe with $\Lambda>0$ the Hubble parameter asymptotes to a positive constant, so we can ignore any $H(z)$ factor in the equation above for our convergence study. From equation~\eqref{eq:Tvir}, it also follows that haloes with a fixed mass will have an approximately constant virial temperature in the far future, so we can disregard any redshift dependence originating from the relationship between virial temperature and virial mass. Finally, we notice that our prescription for the cooling time (equation~\ref{eq:tcool_smooth}) yields $t_{\rm cool} \sim t$ in the far future. Therefore, we can immediately remove the last factor on the r.h.s. of equation~\eqref{eq:cool_rate_gen}. Its asymptotic behaviour in the limit $t \rightarrow \infty$ is therefore
\begin{equation}
\label{eq:nSFR_posL}
     s_{\rm low}(T(M, \, z), z) \propto\left( \frac{f_{\rm gas}(T(M, \, z), \,z)}{f_{\rm b}} \right)^{\frac{3}{\eta}} t^{\frac{3}{\eta}-2} \, .
\end{equation}
We then need to study the scaling of $f_{\rm gas}$ with cosmic time in this limit.

Our numerical results in Fig.~\ref{fig:Tcrit} suggest that  the critical temperature increases with time once $\Lambda$ dominates. Therefore, there will eventually be a time when the virial temperature falls below the critical temperature. At this point, the scaling for $f_{\rm gas}$ is set by the critical temperature -- see the case $T<T_{\rm crit}$ in equation~\eqref{eq:fbh_T}. We then need to determine the asymptotic time dependence of the critical temperature. 

As mentioned in \S\,\ref{sec:rcool}, the critical temperature is the virial temperature above which haloes retain a baryon mass fraction equal to the cosmic value $f_{\rm b}$. At low redshift, this temperature is the one that satisfies the following condition (see SP21):
\begin{equation}
\label{eq:Tcrit_cond}
    \left( \frac{-\eta^2 + 4\eta -2}{\eta} \frac{GM}{f_{\rm w} \beta x R^2} \right)^{\frac{2\eta-3}{\eta-1}} = \frac{(3-\eta) (f_{\rm b}-f_{* \, \infty }) \mu X^2 M \mathcal{C}(T)}{6\pi k_{\rm B} T m_{\rm H}^2 R^3} t_{\rm cool}\, ,
\end{equation}
where we make the implicit assumption that the stellar mass fraction asymptotes to a constant in the far future. This assumption will be validated a posteriori, when we will show that the formalism does predict a converging stellar mass fraction.
 
Taking $t_{\rm cool} \approx t$ and approximating again the cooling function as a piece-wise power law (see equation~\ref{eq:CoolFunc_appr}), we obtain:
\begin{equation}
\label{eq:Tcrit_asym_pos}
    T_{\rm crit} (t) \propto t^{\frac{2 (\eta - 1)}{2 (2 - \theta) \eta + 2 \theta -5 }} \, .
\end{equation}
It can be verified that, if $1.5<\eta<3$, the exponent in the equation above is positive for $\theta \lesssim 1.4$. Our numerical results show that for $\eta=1.9$, the critical temperature is above $10^5 \K$ at late times in universes with $\Lambda>0$ (see Fig.~\ref{fig:Tcrit}). In this temperature range, equation~\eqref{eq:CoolFunc_appr} returns $\theta = -1$ or $\theta=0.5$. In either case, this satisfies the conditions for which the exponent in equation~\eqref{eq:Tcrit_asym_pos} is positive. Therefore, the critical temperature keeps increasing with time.

One might wonder whether this result is specific to the value of $\eta=1.9$ adopted here. But SP21 showed that, for $\Lambda = \Lambda_{\rm obs}$, a different value of $\eta$ would change the late-time critical temperature only within a factor of $\sim$\,$2-3$. Thus, our considerations on the asymptotic behaviour of $T_{\rm crit}$ in a universe with $\Lambda >0$ should not be altered if we changed the value of $\eta$.

Inserting the time scaling of the critical temperature above in equation~\eqref{eq:rcool_T}, it can be easily seen that the cooling radius also scales with $t$ as a power law in the far future, with a positive index smaller than unity. This proves that the cooling radius increases with cosmic time, but does so progressively more slowly, as its time derivative flattens to zero as $t\rightarrow \infty$ (see the discussion in \S\,\ref{sec:rcool}).

If we now insert the scaling given by equation~\eqref{eq:Tcrit_asym_pos} in the $T<T_{\rm crit}$ case of equation~\eqref{eq:fbh_T}, and use the resulting time dependence for $f_{\rm gas}$ in equation~\eqref{eq:nSFR_posL}, we deduce that $s_{\rm low}(T(M, \, z), z) \propto t^{\xi}$, with
\begin{equation}
   \xi = \frac{ 4( \theta -2) \eta^2 + 5(5 - 2 \theta) \eta + 6 (\theta - 4)} { \eta [2(2 -  \theta) \eta + 2 \theta -5]} \, .
\end{equation}
We find that, for the value $\eta=1.9$ that we are adopting, $\xi<-1$ for $-0.3 \lesssim \theta \lesssim 1.5$. This includes $\theta=0.5$, which corresponds to $T_{\rm crit}>10^6 \K$, a condition that is certainly satisfied asymptotically. Hence, our analysis would predict that the nSFR converges in the limit $t \rightarrow \infty$. 

Having thoroughly tested our results by a comparison of numerics with the approximate analytical approach presented in this Appendix, we are therefore confident of the robustness of our findings regarding the convergence of the total stellar mass within haloes.

\subsubsection{Stellar mass density}

In a universe with $\Lambda>0$, the growth function asymptotes to a constant, so that the halo multiplicity function follows the same asymptotic behaviour for a fixed halo mass (see also \S\,\ref{sec:hmf_EdS}). The time dependence of the CSFRD is then simply obtained by integrating the nSFR over all halo masses, weighted by the asymptotic multiplicity function. However, there is a complication: not all haloes lie below the critical temperature threshold at any fixed cosmic time. Some will therefore follow the time-dependence for the nSFR found in the previous section, which applies to the case $T<T_{\rm crit}$, while others will yet have to enter that regime. Therefore, the asymptotic time dependence of the nSFR cannot be simply extracted from the integral in equation~\eqref{eq:CSFRD}, and the late-time behaviour of the CSFRD will deviate from that.

We empirically verified that the a good analytical approximation of the late-time CSFRD is given by
\begin{equation}
    \ln \left(\frac{\dot{\rho}_*(t)}{\rm U} \right) =\ln  \left(\frac{\dot{\rho}_*(t_{\rm l})}{\rm U} \right) \left( \frac{t}{t_{\rm l}} \right)^{\phi} \, ,
\end{equation}
where $t_{\rm l}$ is a sufficiently late cosmic time and $\phi$ is the exponent that minimises the residuals with the CSFRD that we computed numerically. In the above, we defined $\rm U = 1 \rm M_{\odot} \, yr^{-1} \, cMpc^{-3}$, so that the argument of the logarithms is dimensionless. We verified that our fitting formula yields an accuracy of a few percent at $1+z \lesssim 10^{-5}$, for all cosmological models considered. The values of the fitting parameters depended very weakly on the choice of $t_{\rm l}$.

The fitting function allows us to calculate the asymptotic SMD analytically, as we shall now show. The SMD at any time $t>t_{\rm l}$ is given by:
\begin{equation}
    \rho_*(t)= \rho_* (t_{\rm l}) + \int _{t_{\rm l}} ^{t} \dot{\rho}(t') dt' \approx \rho_* (t_{\rm l}) + \mathrm{U} \int _{t_{\rm l}} ^{t}  \exp \left[ P \left(  \frac{t'}{t_{\rm l}} \right)^{\phi} \right] \, dt' \, ,
\end{equation}
where we defined $P = \ln (\dot{\rho}_*(t_{\rm l})/ \rm U)$. With the substitution $t' \rightarrow \tilde{t} = -P (t'/t_{\rm l})^{\phi}$, we obtain
\begin{align}
    \rho_*(t) &\approx
    \rho_* (t_{\rm l}) 
    +  \frac{t_{\rm l} \, \mathrm{U} }{\phi (-P)^{\frac{1}{\phi}}} \int _{-P} ^{-P \left(\frac{t}{t_{\rm l}} \right)^{\phi}} \tilde{t}^{\frac{1}{\phi}-1} \exp{(-\tilde{t})} \, d \tilde{t}  \nonumber \\
      &= \rho_* (t_{\rm l}) 
     + \frac{t_{\rm l} \, \mathrm{U} }{\phi (-P)^{\frac{1}{\phi}}}  \left[ \gamma_{\rm l} \left( \frac{1}{\phi}, \, -P \left(\frac{t}{t_{\rm l}} \right)^{\phi} \right) - \gamma_{\rm l} \left( \frac{1}{\phi}, \, -P \right) \right] \, ,
     \label{eq:asym_rho*}
\end{align}
where $\gamma_{\rm l}$ is the lower incomplete gamma function. We note that because the CSFRD decays with time, one can always choose a large enough $t_{\rm l}$ such that $P<0$. For $t>t_{\rm l}$, the logarithm of the CSFRD is negative and increases in absolute value, so that $\ln \dot{\rho}_*(t)/\ln \dot{\rho}_*(t_{\rm l})$ increases with time. Thus, $\phi>0$ and the lower incomplete gamma function in equation~\eqref{eq:asym_rho*} is always well defined. In the limit $t\rightarrow \infty$, the above expression converges to
\begin{equation}
    \rho_{* \, \infty}= \rho_* (t_{\rm l}) + \frac{t_{\rm l} \, \mathrm{U} }{\phi (-P)^{\frac{1}{\phi}}} \, \gamma_{
    \rm u} \left( \frac{1}{\phi}, \, -P \right)  \, ,
\end{equation}
where $\gamma_{\rm u}$ is the upper incomplete gamma function. This is the expression that we used to compute the asymptotic SMD in universes with $\Lambda >0$.


\bsp	
\label{lastpage}
\end{document}